\documentclass[prb,nobalancelastpage,twocolumn,nofootinbib,superscriptaddress,showpacs]{revtex4-1}

\usepackage{color,amsthm,amsmath,amsfonts,graphicx,bm}
\usepackage{bm,eufrak}
\usepackage{amsfonts}
\usepackage{amssymb}
\usepackage{amsmath,mathrsfs}
\usepackage{epsfig}
\usepackage{graphicx}
\usepackage{enumerate}
\usepackage{multirow}
\usepackage{verbatim}
\usepackage{subfigure}
\usepackage{bbold}
\usepackage{soul}
\usepackage{scrextend}
\usepackage{tikz}
\usetikzlibrary{decorations.pathreplacing}

\usepackage{dcolumn}
\usepackage{latexsym}
\usepackage{amssymb}
\usepackage{array}
\usepackage{framed}
\usepackage{braket}
\usepackage{comment}

\usepackage{tikz-cd}

\graphicspath{ {./figures/} }

\newcommand\otimesg{\otimes_{\mathfrak{g}}}
\newcommand\vk[1]{| #1 )}
\newcommand\vb[1]{( #1 |}

\newcommand{\mc}[1]{\mathcal{#1}}
\newcommand{\mr}[1]{\mathrm{#1}}
\newcommand{\dg}{\dagger}

\newcommand{\mb}{\mathbf}

\newcounter{notes}
\stepcounter{notes}

\DeclareFontFamily{OT1}{pzc}{}
\DeclareFontShape{OT1}{pzc}{m}{it}{<-> s * [1.10] pzcmi7t}{}
\DeclareMathAlphabet{\mathpzc}{OT1}{pzc}{m}{it}

\graphicspath{ {./figures/} }

\begin{document}

\title{Classification of symmetry-protected topological many-body localized phases in one dimension}

\date{\today}

\author{Amos Chan}
\affiliation{Rudolf Peierls Centre for Theoretical Physics, Clarendon Laboratory, Parks Road, Oxford OX1 3PU, United Kingdom}
\author{Thorsten B. Wahl}
\affiliation{Rudolf Peierls Centre for Theoretical Physics, Clarendon Laboratory, Parks Road, Oxford OX1 3PU, United Kingdom}




\begin{abstract}
We provide a classification of symmetry-protected topological (SPT)  phases of many-body localized (MBL) spin and fermionic systems in one dimension. For spin systems, using tensor networks we show that all eigenstates of these phases have the same topological index as defined for SPT ground states. For unitary on-site symmetries, the MBL phases are thus labeled by the elements of the second cohomology group of the symmetry group. A similar classification is obtained for anti-unitary on-site symmetries, time-reversal symmetry being a special case with a $\mathbb{Z}_2$ classification (cf. [Phys. Rev. B 98, 054204 (2018)]). 
For the classification of fermionic MBL phases, we propose a fermionic tensor network diagrammatic formulation. We find that fermionic MBL systems with an (anti-)unitary symmetry are classified by the elements of the (generalized) second cohomology group if parity is included into the symmetry group. However, our approach misses a $\mathbb{Z}_2$ topological index expected from the classification of fermionic SPT ground states. 
Furthermore, we show that all found phases are stable to arbitrary symmetry-preserving local perturbations. Conversely, different topological phases must be separated by a transition marked by delocalized eigenstates. 
Finally, we demonstrate that the classification of spin systems is complete in the sense that there cannot be any additional topological indices pertaining to the properties of individual \textit{eigenstates}, but there can be additional topological indices that further classify Hamiltonians. 

\end{abstract}

\maketitle

\tableofcontents

\section{Introduction} \label{sec:intro}


Many-body localization (MBL)~\cite{AltmanReview,Luitz2017,Abanin2017,ImbrieLIOMreview2017,Alet2017} is the interacting analogue of Anderson localization~\cite{anderson1958absence}. It refers to strongly disordered (isolated) quantum systems, which fail to thermalize, because all their eigenstates are localized. As a consequence, such systems retain a memory of their initial state for arbitrarily long observation times, violating the eigenstate thermalization hypothesis~\cite{1984Peres,deutsch1991quantum,srednicki1994chaos,srednicki1999,Rigol:2008bh,DAlessio2015aa,Borgonovi2016}. In the case of one-dimensional systems, this phenomenon is well-established both theoretically~\cite{gornyi2005interacting,basko2006metal,pal2010mb,oganesyan2007localization,imbrie2016many} and experimentally~\cite{Schreiber842}. While strongly disordered higher dimensional systems might not be strictly many-body localized~\cite{deRoeck2017Stability,Altman2018stability}, astronomically long relaxation times most likely lead to MBL-like behavior on all practically relevant time scales~\cite{Choi1547,2D_quantum_bath,chandran2016higherD,luitz2017smallbaths}. 

In MBL systems, all eigenstates fulfill the area law of entanglement and are thus very much alike ground states of local gapped Hamiltonians~\cite{Friesdorf2015}. The latter can be classified into different topological phases, where eigenstates within one topological phase can be connected by short-depth quantum circuits~\cite{Zoo}. For one-dimensional spin systems, distinct topological phases only exist if a symmetry is imposed on the system (and the connecting quantum circuits), known as \textit{symmetry-protected topological (SPT) phases}~\cite{Pollmann2010,2011Schuch,2011Chen}. Much of the interest in topological systems stems from their ability to protect quantum information against noise: Topologically non-trivial phases allow to encode quantum information in a global fashion, i.e., local perturbations do not affect it. In higher dimensions, certain topological systems even allow to carry out quantum computations in a fault-tolerant way~\cite{topQC}. As a result, there has been huge theoretical interest in classifying topological phases with and without imposed symmetries~\cite{Levin_Wen,Levin_Wen2,Chen_Gu,2015Gu,fermionic_string_net,Kitaev2012,Lan2014}. Much of this classification was carried out using \textit{tensor network states}~\cite{PerezGarcia2007,PEPS,verstraete2008matrix}, as they approximate ground states of local gapped Hamiltonians efficiently~\cite{Verstraete2006,Molnar2015}. 

Since the eigenstates of MBL systems fulfill the area law of entanglement, they all have the capacity to display topological features.  MBL eigenstates can be efficiently described by tensor network states: In one dimension, the corresponding tensor network states (matrix product states), have been shown to yield very high accuracies for the approximation of individual eigenstates~\cite{Khemani2016MPS,Yu2017}. The eigenstates to be approximated get in part selected by the optimization algorithm, and it is unclear to what extent they represent other eigenstates at the same energy density. This shortcoming can be circumvented by approximating the full set of eigenstates by quantum circuits - a specific type of tensor networks involving only unitary matrices~\cite{Pollmann2016TNS,Wahl2017PRX}. Numerical simulations using this approach have produced high accuracies for strongly disordered one-dimensional systems and the first quantitative theoretical results on two-dimensional MBL-like systems~\cite{2DMBL}. The one-dimensional calculations in Ref.~\onlinecite{Wahl2017PRX} were carried out using two-layer quantum circuits with wide gates. Based on numerical evidence and analytical considerations, the error of the approximation decreases exponentially with the width of the gates. As the computational cost is exponential in the width of the gates, the error decreases polynomially with computational effort. Hence, quantum circuits approximate MBL systems efficiently. As a result, they also constitute a valuable analytical tool for the classification of topological MBL phases. 

For MBL systems it is \textit{a priori} not clear whether all eigenstates are in the same topological phase (as defined for ground states)  or not. If they are in the same topological phase, quantum information can be protected at all energy scales, as all eigenstates involved in the dynamics offer the same type of topological protection~\cite{2013Bauer_Nayak,Huse2013LPQO,Chandran2014SPT,kjall2014many,2015Slagle,2015Potter}. For the one-dimensional disordered cluster model, numerical simulations have indicated that all eigenstates of that MBL system are indeed in the same SPT phase~\cite{bahri2015localization}. In the case of one-dimensional MBL systems with time-reversal symmetry, this has been shown rigorously to be the case using two-layer quantum circuits~\cite{Thorsten}. However, the extension to on-site symmetries and fermionic systems remained open problems. 

\begin{figure}[htb]
	\includegraphics[width=0.4\textwidth]{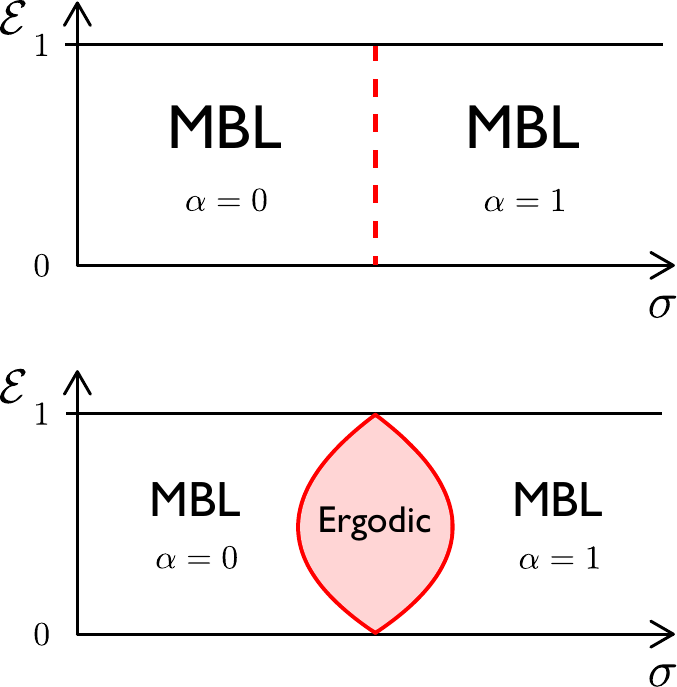}
	\caption{Two possible scenarios of transitions between two SPT MBL phases where $\sigma$ is some parameter of the Hamiltonian, $\alpha$ is a  topological label, and $\mathcal{E} = (E_n - E_{\mathrm{min}}) / (E_{\mathrm{max}} - E_{\mathrm{min}})$ ($E_n$ correspond to the energies of the eigenstates, $E_\mr{max}$ is the highest energy and $E_\mr{min}$ the ground state energy). \textit{Top:} As $\sigma$  increases, the system traverses from an SPT MBL phase labeled by $\alpha =0$ to another one labelled by $\alpha = 1$.  The transition is marked by a critical line (indicated in red), where the majority of eigenstates are volume law-entangled.  
\textit{Bottom:} In this scenario, there is an extended region with volume law-entangled eigenstates separating the two phases in the thermodynamic limit.  
}
	\label{fig:spt}
\end{figure}

In this work, we classify SPT MBL phases of one-dimensional spin and fermionic systems with (anti-~)unitary  on-site symmetries using quantum circuits. For spin systems, we show that these phases can be labeled by the elements of the generalized second cohomology group of the symmetry group and that the corresponding topological index is the same for all eigenstates. We show that those SPT MBL phases are robust to symmetry-preserving perturbations. Conversely, a system transiting  between two different SPT MBL phases (labeled by different topological indices) must at some point become ergodic. Two possible scenarios of transition are described in Fig.~\ref{fig:spt} with either a critical line or an extended region separating the two phases.  Furthermore, we demonstrate that the classification for spin systems is complete in the sense that there cannot be any additional topological labels which affect individual eigenstates (but there can be additional topological indices that further classify Hamiltonians.).

For fermionic systems, we first introduce a diagrammatic approach for fermionic tensor networks. We obtain that using (anti-)unitary symmetry groups the topological classes are given by the (generalized) second cohomology group of the overall symmetry group containing parity as a subsymmetry. We point out that a $\mathbb{Z}_2$ topological label is missing compared to the classification of fermionic SPT ground states, namely the index related to the  fractionalization of fermionic parity~\cite{pollmann2011ferm}. An important example is the case of time-reversal symmetry (a $\mathbb{Z}_2$ anti-unitary symmetry). In this case, we obtain two out of three $\mathbb{Z}_2$ topological indices using our method. On the example of the Kitaev chain, we illustrate that phases where the remaining topological $\mathbb{Z}_2$ label is non-trivial cannot be represented exactly by parity-preserving short-depth quantum circuits and are thus not captured by our classification. 

In Section~\ref{sec:introMBL}, we provide a brief introduction to MBL in one dimension, the SPT phases it can give rise to, and the formalism of tensor networks. 
In Section~\ref{sec:non-technical} we give an overview over the main results derived in this article in a non-technical manner. 
In Section~\ref{sec:underlying_technical} we specify the assumptions underlying our technical derivation. 
In Section~\ref{sec:on-site} the formalism for the classification of spinful MBL systems with unitary on-site symmetries is introduced. We demonstrate that such SPT MBL systems are labeled by the elements of the second cohomology group. We use the same approach for anti-unitary on-site symmetries in Section~\ref{sec:anti-unitary}. A special case thereof is time-reversal symmetry with a $\mathbb{Z}_2$ classification~\cite{Thorsten}, which is explicitly derived using the above formalism in Section~\ref{sec:TRS}. 
	In Section~\ref{sec:fermclass}, we propose a diagrammatic approach for fermionic tensor networks (Sec.\ref{sec:fermform}), obtain a classification of fermionic SPT MBL phases (Sec.~\ref{sec:extproof}), and explicitly derive two out of three topological invariants for time-reversal symmetry (Sec.~\ref{fermtrs}). 
 Finally, we show that the SPT MBL phases obtained with our approach are robust to symmetry-preserving perturbations (Section~\ref{sec:robustness}) and that the classifications in the spinful case is complete in the sense that there cannot be any additional topological indices pertaining to the properties of individual eigenstates (Section~\ref{sec:completeness}). However, there can be additional topological indices related to the overall Hamiltonian. 
Section~\ref{sec:conclusions} concludes the paper and points out open questions for future research. 
In Appendix~\ref{app:A}, we use our formalism to argue that FMBL systems invariant under a non-abelian symmetry are unstable. 


\section{Many-body localization and tensor networks}~\label{sec:introMBL}


\subsection{MBL and local integrals of motion}

The transition from the ergodic (thermal) phase to the MBL phase as a function of disorder strength cannot be captured by any theory of conventional phase transitions~\cite{VHA_MBLTransition,Potter:2015ab}. On the thermal side but close to the phase transition the system displays a mobility edge~\cite{Luitz2015,2DMBL}, which is the boundary of an energy window in the middle of the spectrum within which eigenstates are volume-law entangled, i.e., delocalized. (However, arguments have been put forward challenging this picture~\cite{DeRoeck2016}.) Eigenstates outside this energy window are area-law entangled and thus often referred to as many-body localized. Our analysis here is restricted to the \textit{fully many-body localized} (FMBL) regime, which is characterized by a complete set of \textit{local integrals of motion} (LIOMs)~\cite{serbyn2013local,chandran2015constructing,ros2015integrals,Inglis_PRL2016,Rademaker2016LIOM,Monthus2016,
Pekker2017,Goihl2018,Abi2018,Abi2019,Geraedts2017,ImbrieLIOMreview2017}. For spin-$1/2$ chains these are commonly denoted as  $\tau_z^i$ with site index $i = 1, 2, \ldots, N$, which commute with the Hamiltonian $H$ and with each other,
\begin{align}
[H,\tau_z^i] = [\tau_z^i, \tau_z^j] = 0.
\end{align}
They are effective spin degrees of freedom related by a quasi-local unitary transformation $U$ to the original spins. Thus, the former are exponentially localized around site $i$. The corresponding decay length is referred to as their \textit{localization length} $\xi_i$. Concretely, we define the FMBL regime by all $\xi_i$ being sub-extensive in the system size $N$ in the thermodynamic limit. 
 The unitary $U$ also diagonalizes the Hamiltonian,
\begin{align}
H &= U E U^\dagger, \\
\tau_\mu^i &= U \sigma_\mu^i U^\dagger,
\end{align}
where $E$ is a diagonal matrix containing the energies and $\sigma_\mu^i$ are the Pauli operators ($\mu = x,y,z$) acting on site $i$. Hence, the Hamiltonian can be written entirely in terms of the LIOMs,
\begin{align}
H &= c + \sum_{i = 1}^N c_i \tau_z^i + \sum_{i>j=1}^N c_{ij} \tau_z^i \tau_z^j + \sum_{i>j>k=1}^N c_{ijk} \tau_z^i \tau_z^j \tau_z^k \notag \\ &+ \ldots \label{eq:LIOM}
\end{align} 
$|c_{ijk\ldots}|$ decays exponentially with the largest difference  of its coefficients (site distance), where the decay length is tightly connected to the localization length. The eigenstates $|\psi_{l_1 \ldots l_N}\rangle$ are thus completely determined by the expectation values of the $\tau_z^i$ operators known as l-bits $l_i$,
\begin{align}
\tau_z^i |\psi_{l_1 \ldots l_i \ldots l_N}\rangle = (-1)^{l_i} |\psi_{l_1 \ldots l_i \ldots l_N} \rangle.
\end{align}
A classic example of an MBL system is the disordered Heisenberg model,
\begin{align}
H_\mr{Heisenberg} = J \sum_{i=1}^{N-1} \mb S_i \cdot \mb S_{i+1} + \sum_{i=1}^N h_i S_i^z,
\end{align}
where $h_i$ is chosen randomly between $-W$ and $W$, which is known as the \textit{disorder strength}. For $W \gtrapprox 3.5$, the system is in the FMBL regime~\cite{pal2010mb,Luitz2015}. 

\begin{figure}
\begin{picture}(70,120)
\put(0,0){\includegraphics[width=0.12\textwidth]{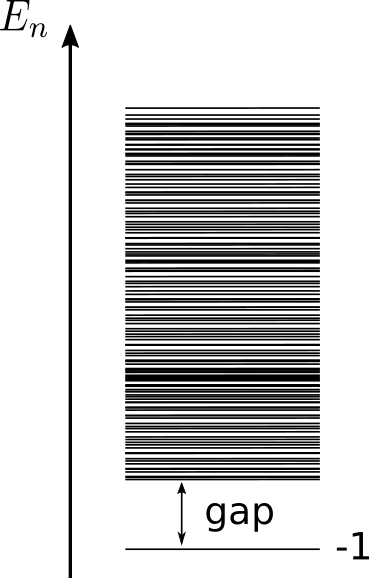}}
\put(-10,110){\textbf{a}}
\end{picture} \ \
\begin{picture}(70,120)
\put(0,0){\includegraphics[width=0.11\textwidth]{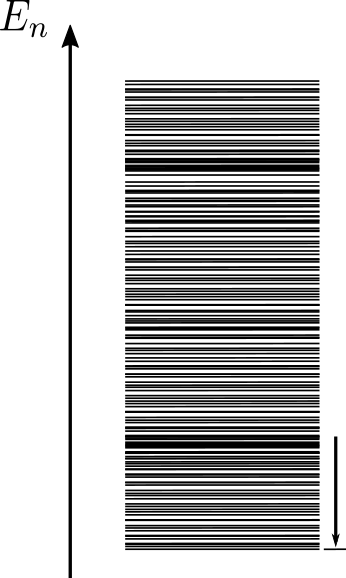}}
\put(-10,110){\textbf{b}}
\end{picture} \ \
\begin{picture}(70,120)
\put(0,0){\includegraphics[width=0.12\textwidth]{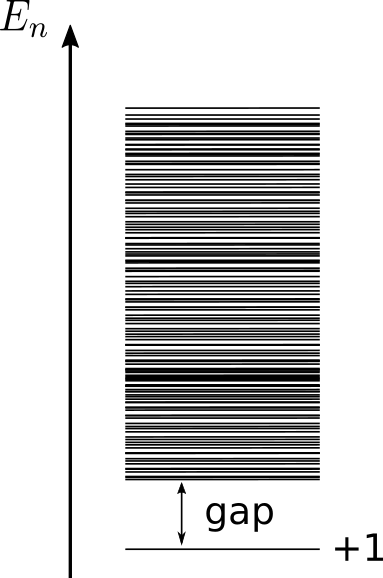}}
\put(-10,110){\textbf{c}}
\end{picture} \ \
\begin{picture}(200,170)
\put(0,0){\includegraphics[width=0.4\textwidth]{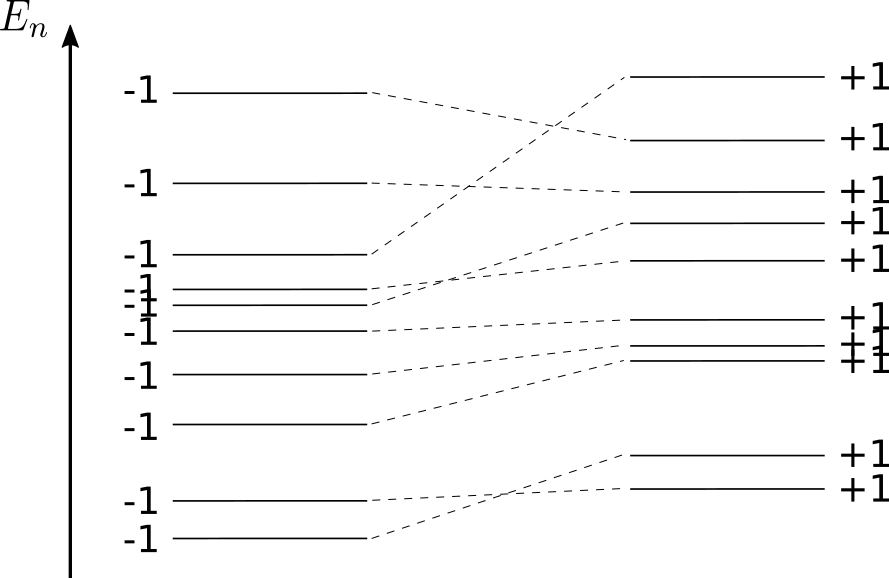}}
\put(-10,140){\textbf{d}}
\end{picture}
  \caption{(a-c): Transition from a topologically non-trivial ground state with topological index $-1$ to a trivial state. (a) denotes the initial state; as a parameter in the Hamiltonian is changed adiabatically, the gap closes (b) before reopening, leaving a topologically trivial ground state behind (c). Note that the excited states above the gap are volume law-entangled and thus cannot be assigned a topological index. (d) Transition between two topologically distinct MBL phases as a parameter of the Hamiltonian is adiabatically changed (indicated by dashed lines). In this case, all eigenstates are area law-entangled and can thus be assigned a topological index. As the index of all eigenstates has to be the same, level crossings do not lead to a change of any topological eigenstate index. Hence, in order to transit into a topologically distinct phase, the eigenstates must become delocalized (i.e., volume law-entangled) along the adiabatic evolution, breaking the FMBL condition.}
\label{fig:transition}
\end{figure}
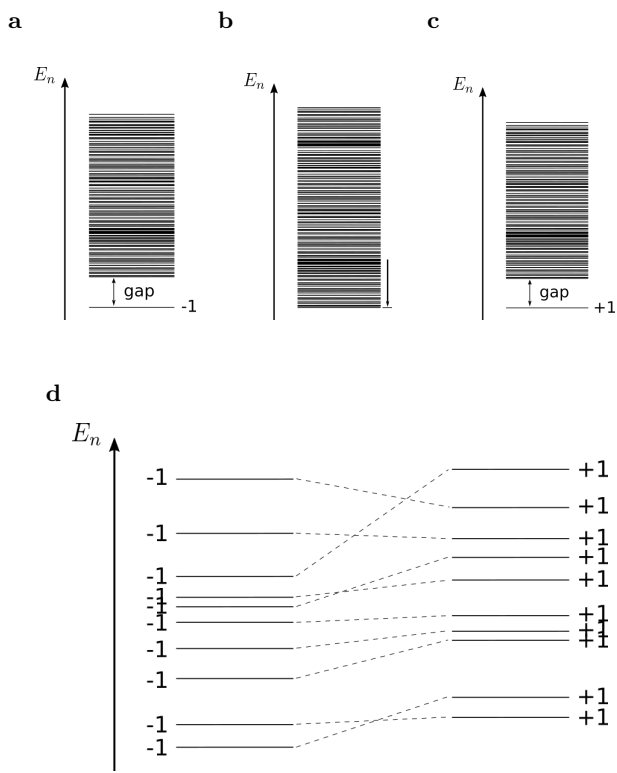

\subsection{Symmetry-protected topological MBL phases}

Ground states of gapped local Hamiltonians can be classified into different topological phases. A topological phase contains the set of local Hamiltonians (or alternatively, their ground states) which can be adiabatically connected with each other without closing the energy gap~\cite{huang2015}. In one dimension, gapped spin-Hamiltonians with a unique ground state lie all in the same topological phase~\cite{2011Schuch}. Fermionic one-dimensional systems without additional symmetries have two topological phases~\cite{Kitaev_chain,Bultnick2017}. 
In the case of time-reversal symmetry, there are two (eight) topologically distinct phases~\cite{Pollmann2010} for spins (for fermions~\cite{kitaevferm2010}). For on-site symmetries, the spinful  SPT phases are in one-to-one correspondence to the elements of the second cohomology group (cohomology classes) of the symmetry group~\cite{2011Schuch,2011Chen,Bultnick2017}.

In the field of MBL, one is interested in features shared between all eigenstates, as those features lead to constraints on the dynamics. Hence, a definition of an MBL topological phase should refer to the set of all eigenstates. We propose the following: Two local FMBL Hamiltonians $H_0$ and $H_1$ with a certain symmetry are said to be in the same SPT MBL phase if and only if they can be connected by a symmetry-preserving path $H(\lambda)$ (also assumed local), such that
\begin{align}
H(0) = H_0 \ \ \mr{and} \ \ H(1) = H_1
\end{align}
and FMBL is preserved along the path. Thus, the condition of a gapped path for ground states of local Hamiltonians has been replaced by the constraint of FMBL along the path. It is the natural extension to a full set of area-law entangled eigenstates, because ground states of local Hamiltonians are area-law entangled unless the gap closes, which can lead to delocalization, cf. Fig.~\ref{fig:transition}.

\begin{figure}
	\includegraphics[width=0.3\textwidth]{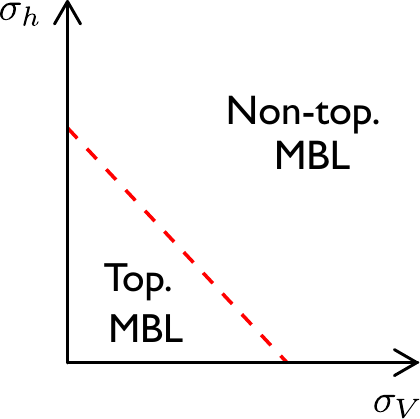}
	\caption{
	A schematic phase diagram of Eq.~\eqref{eq:top_Ham} for fixed non-zero  $\sigma_{\lambda}$, adapted from Ref.~\onlinecite{Eisert2019}. For $\sigma_\lambda \gg \sigma_h, \sigma_V$, there exists a topologically non-trivial fully many-body localized phase, where all eigenstates have four-fold degenerate entanglement spectra if periodic boundary conditions are imposed.
	}
	\label{fig:clusterphase}
\end{figure}

According to a numerical study carried out in Ref.~\onlinecite{bahri2015localization}, all eigenstates of FMBL systems are in the same ground state SPT phase. The authors considered the disordered cluster model given by the Hamiltonian
\begin{align}
H_\mr{cl} = \sum_{i=1}^N \left(\lambda_i \sigma^{i-1}_x \sigma_z^i \sigma^{i+1}_x+ h_i \sigma^i_z + V_i \sigma^i_z \sigma^{i+1}_z \right),
\label{eq:top_Ham}
\end{align}
where $\lambda_i, h_i$ and $V_i$ are chosen randomly according to a Gaussian probability distribution with mean 0 and standard deviation $\sigma_\lambda$, $\sigma_h$ and $\sigma_V$, respectively. A schematic phase diagram of the model for fixed non-zero $\sigma_{\lambda}$ is given in Fig.~\ref{fig:clusterphase}. For $\sigma_\lambda \gg \sigma_h, \sigma_V$, all eigenstates have four-fold degenerate entanglement spectra if periodic boundary conditions are imposed. Two symmetries independently protect this four-fold degeneracy: on the one hand, $\mathbb{Z}_2 \times \mathbb{Z}_2$ symmetry (represented by $\{\mathbb{1}, (\sigma_z \otimes \mathbb{1}_{2\times 2})^{\otimes N/2}, (\mathbb{1}_{2\times 2} \otimes \sigma_z)^{\otimes N/2}, \sigma_z^{\otimes N}\}$)~\cite{Son2011,bahri2015localization}; and on the other hand, time-reversal symmetry~\cite{Verresen2017}, for which it was proven in Ref.~\onlinecite{Thorsten} that all FMBL eigenstates necessarily have the same topological label. 
For open boundary conditions, all eigenstates are four-fold degenerate up to $\mathcal{O}(e^{-N})$ corrections in the topological phase. This is due effective spin-$1/2$ degrees of freedom, one at each boundary, which are completely decoupled from the remaining system~\cite{bahri2015localization}. Thus, the Hamiltonian written in terms of LIOMs~\eqref{eq:LIOM} must have the form
\begin{align}
H_\mr{cl} &= c + \sum_{i=2}^{N-1} c_i \tau_z^i + \sum_{i>j=2}^{N-1} c_{ij} \tau_z^i \tau_z^j + \sum_{i>j>k=2}^{N-1} c_{ijk} \tau_z^i \tau_z^j \tau_z^k \notag \\
&+ \ldots + c_{2 3 \ldots N-1} \tau_z^2 \tau_z^3 \ldots \tau_z^{N-1} + J_{1N} \tau_z^1 \tau_z^N.
\end{align}
The last term indicates that there is an exponentially small coupling between $\tau_z^1$ and $\tau_z^N$. We thus have 
\begin{align}
\| [H_\mr{cl}, \tau_x^1] \| = \mathcal{O}(e^{-N}), \ & \| [H_\mr{cl}, \tau_y^1] \| = \mathcal{O}(e^{-N}) \\
\| [H_\mr{cl}, \tau_x^N] \| = \mathcal{O}(e^{-N}), \ & \| [H_\mr{cl}, \tau_y^N] \| = \mathcal{O}(e^{-N})  
\end{align}
and of course $[H_\mr{cl}, \tau_z^1] = [H_\mr{cl}, \tau_z^N] = 0$. Hence, in the thermodynamic limit $N \rightarrow \infty$, the edge degrees of freedom $\tau_\mu^1$ and $\tau_\mu^N$ ($\mu = x,y,z$) can be used to encode qubits. It is challenging to address these boundary degrees of freedom in practise, as the $\tau_\mu^{1,N}$ are generally not known in actual experiments. For very large $\sigma_\lambda$, however, they are close to the pure cluster model case. Nevertheless, there is a fundamental advantage over non-topological MBL systems: By acting on several spins at the boundary, it is possible to retrieve quantum information after arbitrarily long times, whereas only the classical part of information can be recovered in the case of non-topological MBL systems with this protocol~\cite{Banuls2017}.

\begin{figure}
\centering
\begin{picture}(240,100)
\put(0,0){\includegraphics[width=0.45\textwidth]{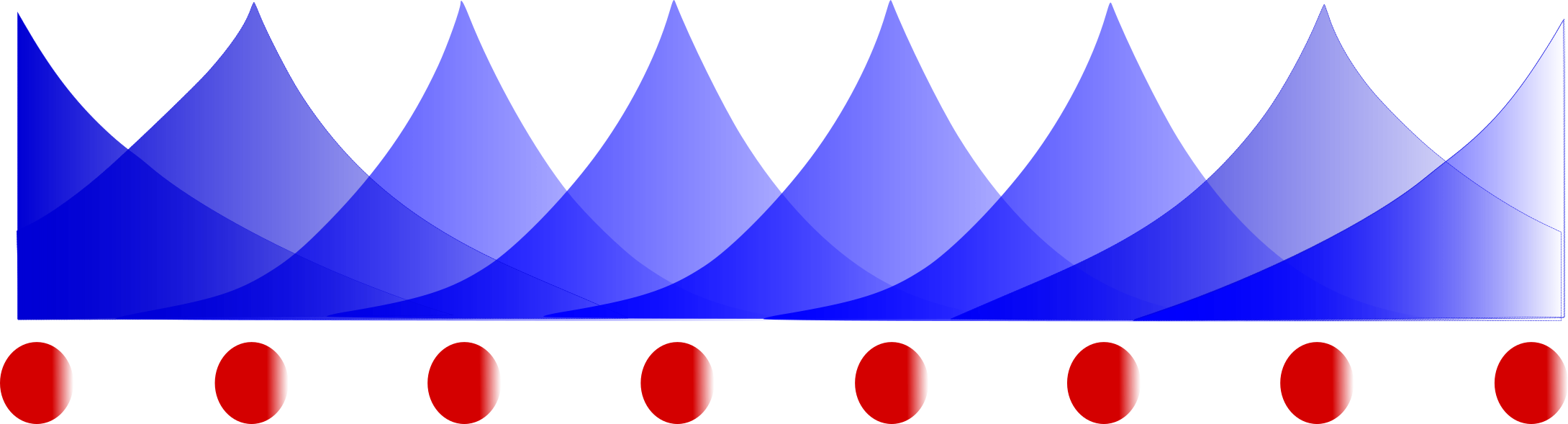}}
\put(-5,80){\textbf{a}}
\end{picture} \\
\begin{picture}(240,100)
\put(0,0){\includegraphics[width=0.45\textwidth]{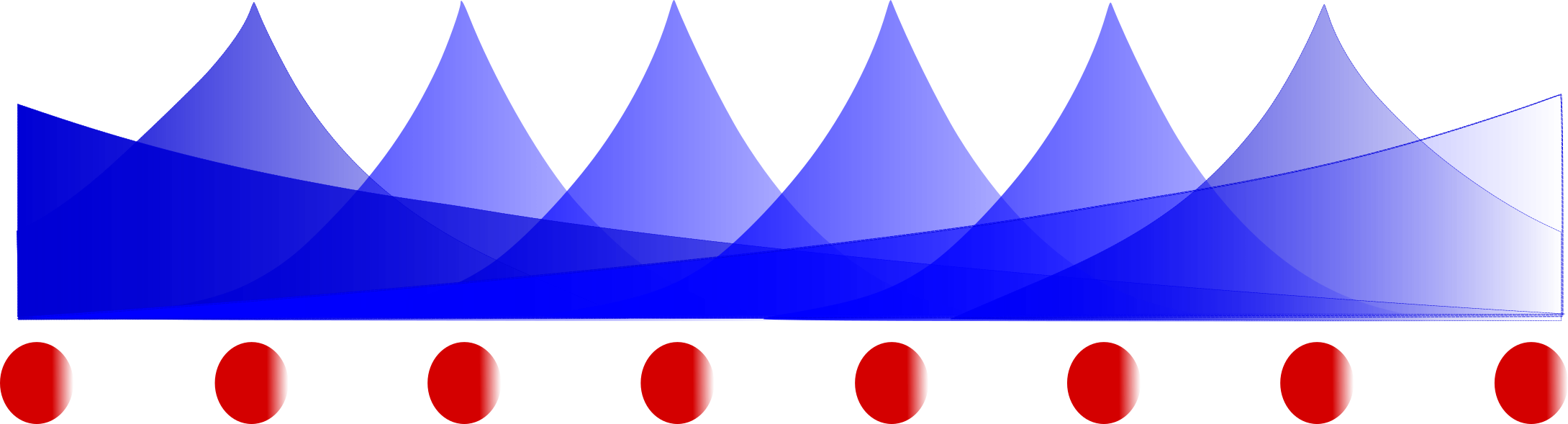}}
\put(-5,80){\textbf{b}}
\end{picture}
  \caption{(a) Local integrals of motion in a topologically non-trivial phase. (b) Integrals of motion at the phase transition to the topologically trivial phase: At least the boundary l-bits must completely delocalize, such that they have finite overlap in the thermodynamic limit, thus breaking the FMBL condition. This could be verified in ultracold atomic gas experiments initialized in a narrow domain-wall configuration near the edge~\cite{Choi1547}.}
\label{fig:taus}
\end{figure}

As one approaches the non-topological regime by increasing $\sigma_h$ and/or $\sigma_V$, the coupling between $\tau_z^1$ and $\tau_z^N$ must become finite even in the thermodynamic limit. That implies that at least the localization lengths of the operators $\tau_z^1$ and $\tau_z^N$ must diverge (leading to a $c_{1N}$ of order $\mathcal{O}(1)$), see Fig.~\ref{fig:taus}.  Hence, the FMBL condition must be violated when transiting between the two topologically distinct MBL phases. However, this is not sufficient to show that the eigenstates of the MBL system become volume law-entangled (delocalized) at the transition, as a complete set of LIOMs implies area law-entangled eigenstates~\cite{Friesdorf2015}, but not the other way around. Instead, the volume law-entanglement of an extensive number of eigenstates follows from the fact that the individual eigenstates cannot change their topological index while being area law-entangled.  
This is depicted in Fig.~\ref{fig:transition}d (cf. phase diagrams in Fig.~\ref{fig:spt}).

Below, we show that all eigenstates of FMBL systems with on-site symmetries also have to be in the same ground state SPT phase. This phase is labeled by an element of the second cohomology group of the symmetry group. Note that we only consider abelian symmetry groups, as FMBL systems with a non-abelian symmetry are unstable. This was first noted in Ref.~\onlinecite{2016Potter_Vasseur}, but we also provide an argument based on our formalism in the Appendix. We also demonstrate that the spinful classification is complete in the sense that there cannot be any additional topological indices which affect individual eigenstates (though there can be additional topological indices related to the overall Hamiltonian). 
The probably most important consequence of our derivation is that the topological properties of SPT MBL systems are robust to small symmetry-preserving perturbations (shown in Sec.~\ref{sec:robustness}).


\vspace{36pt}

\subsection{Tensor networks}\label{sec:tn}
	\begin{figure}[t!]
	\includegraphics[width=0.9\columnwidth]{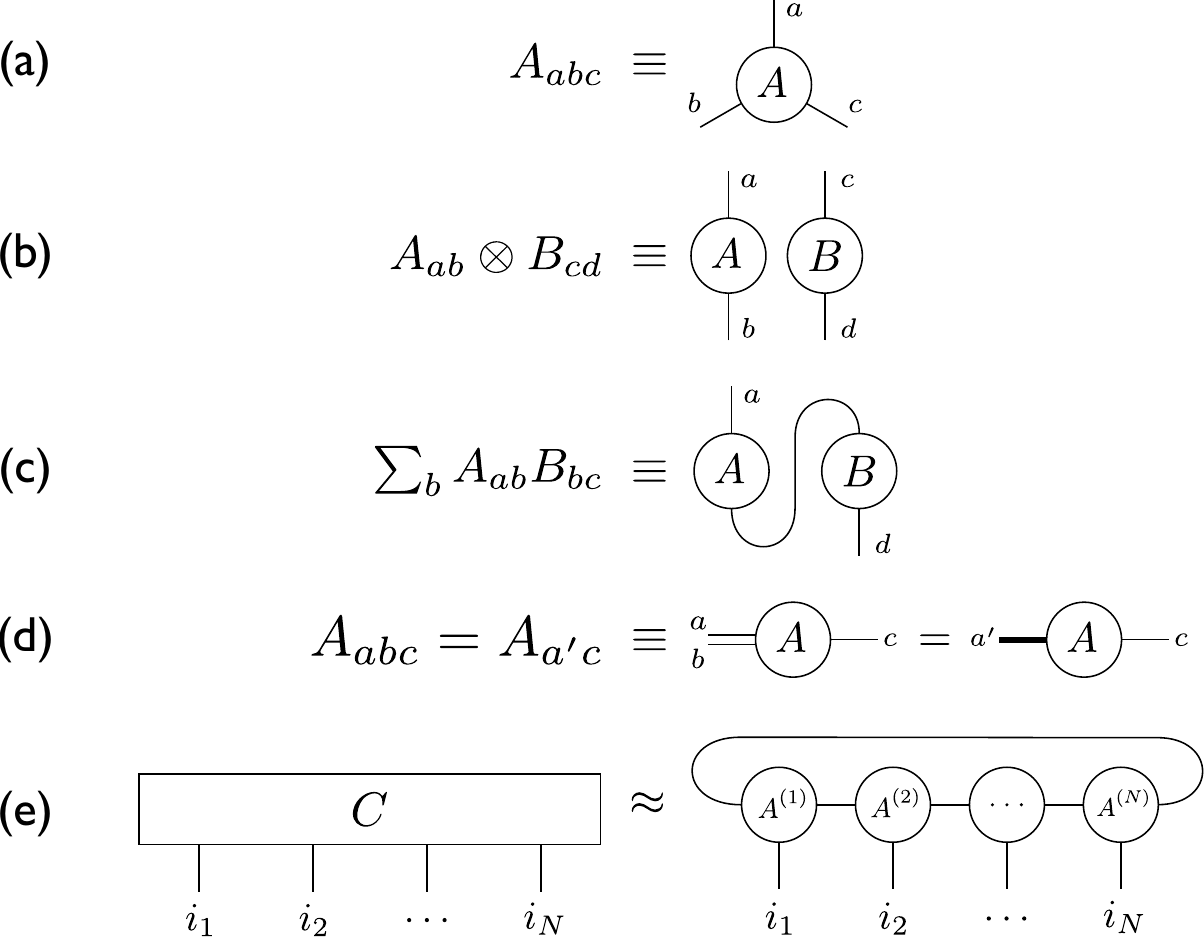}
	\caption{(a) to (d): Diagrammatic representation for tensors and tensor operations (see Sec.~\ref{sec:tn}). (e): An example of a matrix product state approximation with periodic boundary conditions.} \label{fig:tn}
\end{figure}
A central obstacle in the study of quantum many-body systems is the exponential growth of the Hilbert space as a function of system size. However, many physically interesting quantum states contain additional structures and exhibit special properties, like area law entanglement, which allows them to be efficiently represented by tensor networks. Tensor networks allow these states to be probed using variational methods numerically\cite{white1992a, white1992b}, and provides important analytical framework to understand universal properties of these states\cite{2011Chen, 2011Schuch}. Here we provide a brief review on the formulation of tensor networks, following Refs.~\onlinecite{bridgeman2017, orus2014}. 

A tensor is represented diagrammatically with a geometric shape with outgoing legs, each corresponding to an index. For example, a rank-three tensor $A_{abc}$ can be represented as in Fig.~\ref{fig:tn}a. Some important operations on tensors are represented diagrammatically as follows: (i) Tensor product of tensors $A$ and $B$ is represented by placing two tensor diagrams beside each other (Fig.~\ref{fig:tn}b).
(ii) Contraction of two indices of a given tensor is represented by connecting the two corresponding legs of the tensor (Fig.~\ref{fig:tn}c). (iii) Grouping and splitting of tensor indices can be represented by combining and splitting open legs (Fig.~\ref{fig:tn}d). 

A quantum state of a many-body system, consisting of $N$ $q$-level degrees of freedom, can be written as 
\begin{equation}
	\ket{\psi} = \sum_{i_1, i_2, \dots , i_N } C_{i_1, i_2, \dots, i_N}
	| i_1 i_2 \ldots i_N \rangle
	\; ,
\end{equation}
where $i_n = 1, \dots, q$, and $C$ is a rank-$N$ tensor specified by $q^N$ complex numbers. For quantum many-body states with area law entanglement (e.g. the eigenstates of an FMBL Hamiltonian, and the ground state of a gapped local Hamiltonian), the tensor $C$ (containing exponentially many degrees of freedom) can be approximated efficiently by a tensor network containing polynomially many degrees of freedom in $N$~\cite{MPS_faithful,Friesdorf2015}. An important example is the matrix product state (MPS) 
\begin{equation}
\ket{\psi} = \sum_{i_1, i_2, \dots , i_N } \mathrm{Tr} \left[ A^{(1)}_{i_1} A^{(2)}_{i_2} \dots A^{(N)}_{i_N} \right] 
| i_1 i_2 \ldots i_N \rangle
\; ,
\end{equation}
where $A^{(p)}_{i_p}$ is a $\chi^{(p)} \times \chi^{(p+1)}$ matrix, and its diagrammatic representation is given in Fig.~\ref{fig:tn}e. 
$\max_p \chi^{(p)}$ is called the bond dimension of the MPS. For a fixed bond dimension, the MPS is a computationally efficient representation of the original quantum state. In this article, we will classify the SPT MBL systems, using the fact that an FMBL Hamiltonian can be diagonalized by a unitary matrix that is efficiently represented as a quantum circuit -- a specific type of tensor network.


\vspace{36pt}
\section{Non-technical summary of results}~\label{sec:non-technical}

\subsection{Classification}

\subsubsection{Underlying assumptions}

Before introducing the main technique of this work, we briefly state the assumptions needed to demonstrate our claims: We consider only the fully many-body localized (FMBL) case, which we define as the regime where the probability of any LIOM $\tau_z^i$ having a localization length $\xi_i$ of order $\mathcal{O}(N)$ vanishes in the thermodynamic limit $N \rightarrow \infty$. That is, there are no thermal puddles of the order of the system size. This is expected to be the case for all disordered systems whose disorder strength is above the critical value~\cite{Abi2018}, i.e., it coincides with the standard definition of full many-body localization~\cite{Luitz2015}. Our central assumption is that in this regime, the LIOMs can be simultaneously efficiently approximated using a two-layer quantum circuit $\tilde U$ with gates of length $\ell \propto N$ (see below, Eq.~\eqref{eq:tildeU}). Concretely, that means that $\| \tau_z^i - \tilde U \sigma_z^i \tilde U^\dagger \|$ vanishes sufficiently fast in the thermodynamic limit~\cite{Thorsten} (for details, see next section).

We emphasize that gates whose length $\ell$ grows with the system size $N$ are required to efficiently approximate the FMBL system. The reason is that $\ell$ must be much larger than the largest localization length $\xi_i$ appearing in the system~\cite{Thorsten}. However, as $N \rightarrow \infty$, there are also arbitrarily large localization lengths $\xi_i$ in the system. Hence, a quantum circuit whose gate length $\ell$ grows with the system size is the approach which is physically best tailored to the problem at hand. 
More precisely, we define FMBL such that the longest localization length $\xi_i$ is of order less than $\mathcal{O}(N)$. Then, $\ell \propto N$ will be sufficient to allow for an approximation whose error converges to zero in the limit $N \rightarrow \infty$~\cite{Thorsten}. Note that for the classification of gapped ground states~\cite{Pollmann2010,2011Schuch,2011Chen,Bultnick2017}, one would not need to increase the range $\ell$ with $N$, since the correlation length of the system is finite.


 We assume that the symmetry of the system is abelian, i.e., it does not protect any exact degeneracies for finite system sizes. Note that non-abelian symmetries are incompatible with MBL~\cite{2016Potter_Vasseur}, as also shown in the Appendix. Note that while we only classify SPT MBL systems, our classification also applies in the \textit{presence} of spontaneous symmetry breaking.

\subsubsection{Spinful case}
In the case of an abelian on-site symmetry, there are generically no degeneracies in the energy spectra of disordered Hamiltonians. (We only consider periodic boundary conditions henceforth.) The case of accidental degeneracies can be remedied by adding infinitesimal symmetry-preserving perturbations to the Hamiltonian. In the absence of degeneracies and for finite $N$, eigenstates must be invariant under the symmetry and thus fulfill
\begin{align}
\mathpzc{v}^{\otimes N}_g |\psi_{l_1 \ldots l_N}\rangle = e^{i \varphi_{l_1 \ldots l_N}^g} |\psi_{l_1 \ldots l_N}\rangle ,
\end{align} 
where $\mathpzc{v}_g$ is the on-site action of the symmetry and represents the symmetry group $G \ni g$. The unitary matrix $U$ containing the eigenstates $|\psi_{l_1 \ldots l_N}\rangle$ thus fulfills
\begin{align}
\mathpzc{v}^{\otimes N}_g U = U \Theta_g, \label{eq:two-layer}
\end{align}
where $\Theta_g$ is the diagonal matrix with diagonal elements $e^{i \varphi_{l_1 \ldots l_N}^g}$. 

As elaborated on above, classifying MBL phases characterized by the unitary $U$ is equivalent to classifying two-layer quantum circuits $\tilde U$ if the range $\ell$ of the gates increases linearly with system size $N$,
\begin{equation}
\begin{aligned}
\includegraphics[width=0.4\textwidth]{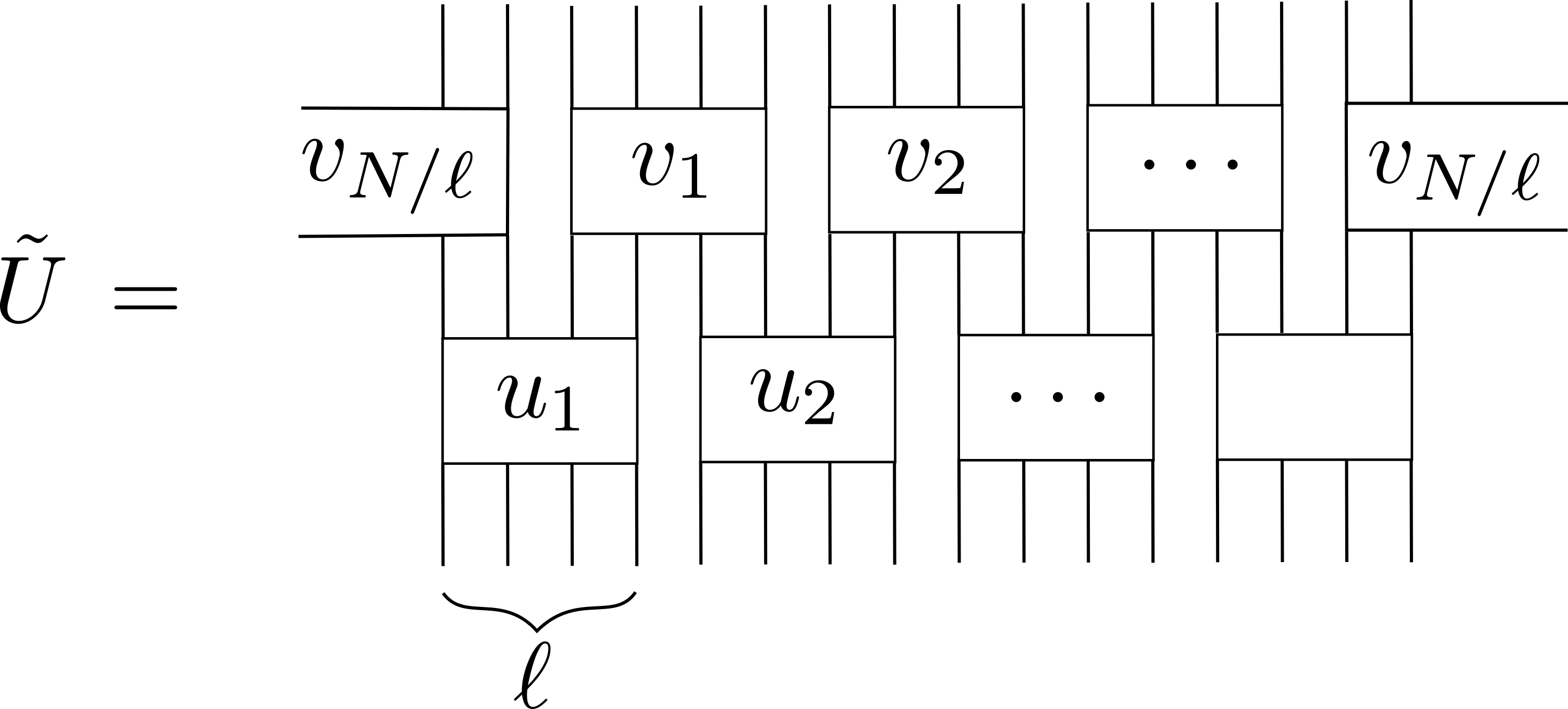} 
\end{aligned} \ . \label{eq:tildeU}
\end{equation}
In the diagram, lower legs represent l-bit indices, i.e., by fixing them, one obtains a matrix product state representation of the eigenstate $|\tilde \psi_{l_1 \ldots l_N}\rangle$  corresponding to those l-bits. Intuitively, the reason behind the efficiency of this approximation is that in the FMBL phase, for $N \rightarrow \infty$ the probability of finding a LIOM of localization length $\mc{O}(N)$ goes to zero. Therefore, in the thermodynamic limit, all of them can be captured exactly by a two-layer quantum circuit whose gates are of range $\ell = a N$ with $a$ fixed. 

Hence, we assume Eq.~\eqref{eq:two-layer} to be asymptotically true for $\tilde U$, that is
\begin{align} \label{eq:theta1}
\Theta_g = \tilde U^\dg v_g^{\otimes N} \tilde U. 
\end{align}
We now use the graphical notation to represent this equation, setting
\begin{equation}
\begin{tikzpicture}[scale=1.1,baseline=(current  bounding  box.center)]    
\coordinate[label=right:$\mathpzc{v}_g^{\otimes \ell/2} \ {=}$] (A) at (0,0);
\begin{scope}[shift={(2.7,0)}]
\draw[thick](-0.7,-0.7) -- (-0.7,0.7);
\draw[thick,fill=white] (-0.7,0) ellipse (0.3 and 0.3);
\coordinate[label=above:$g$] (A) at (-0.7,-0.25);
\end{scope} 
\end{tikzpicture}\ .
\end{equation}        
In the diagrammatic representation, multiplication order top to bottom corresponds to left to right in algebraic representation, such that Eq.~\eqref{eq:theta1} reads
\begin{equation}
\begin{aligned}
\includegraphics[width=0.47\textwidth]{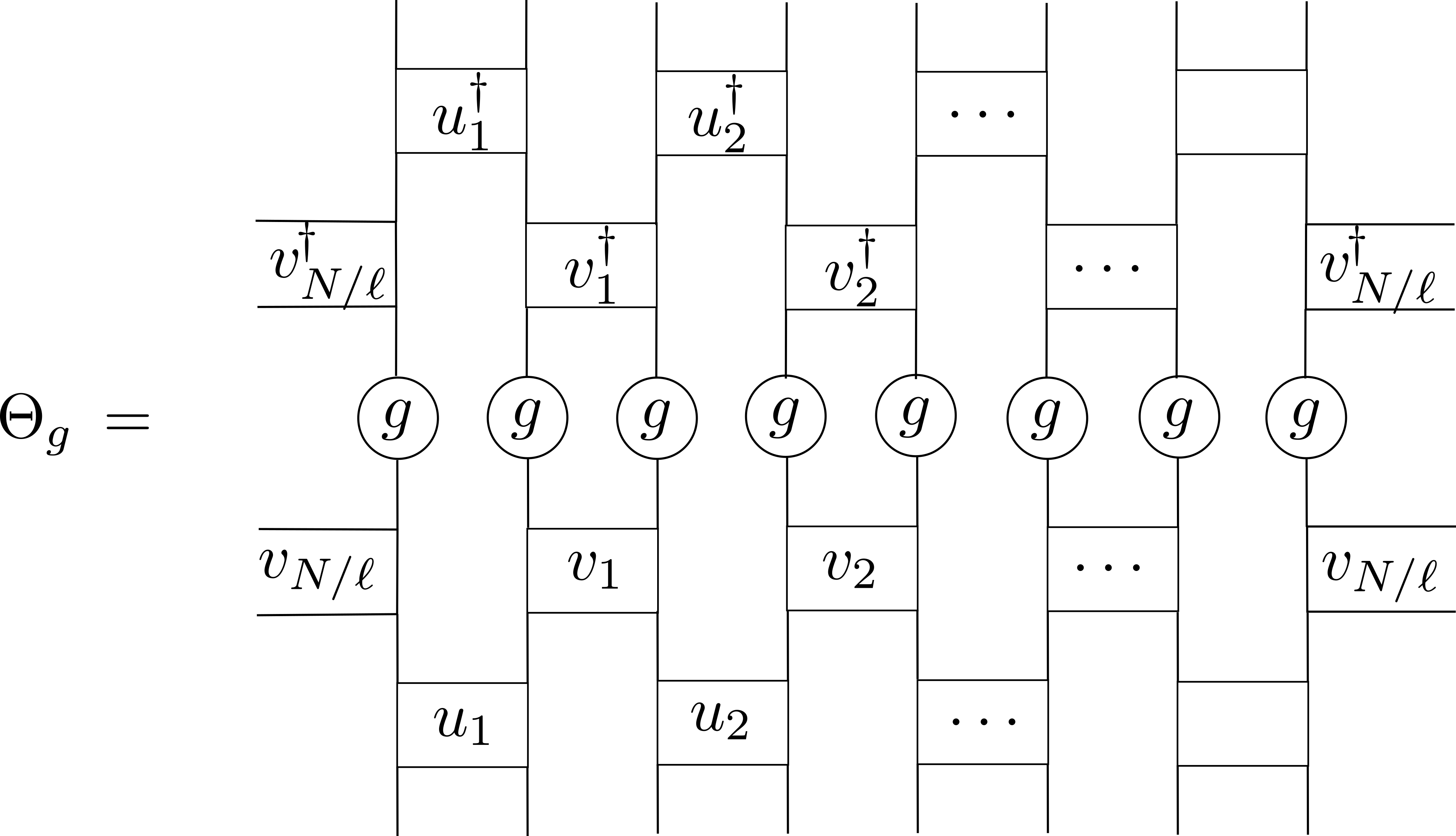}
 \end{aligned} \label{eq:theta_graph},
\end{equation}
where each leg represents $\frac{\ell}{2}$ legs in the previous diagram. By blocking unitaries together, it is possible to show that $\Theta_g$ can be written as a two-layer quantum circuit whose unitaries $\Theta_k^g$ are all diagonal,
\begin{equation} \label{eq:thetaqc}
\begin{aligned}
\includegraphics[width=0.4\textwidth]{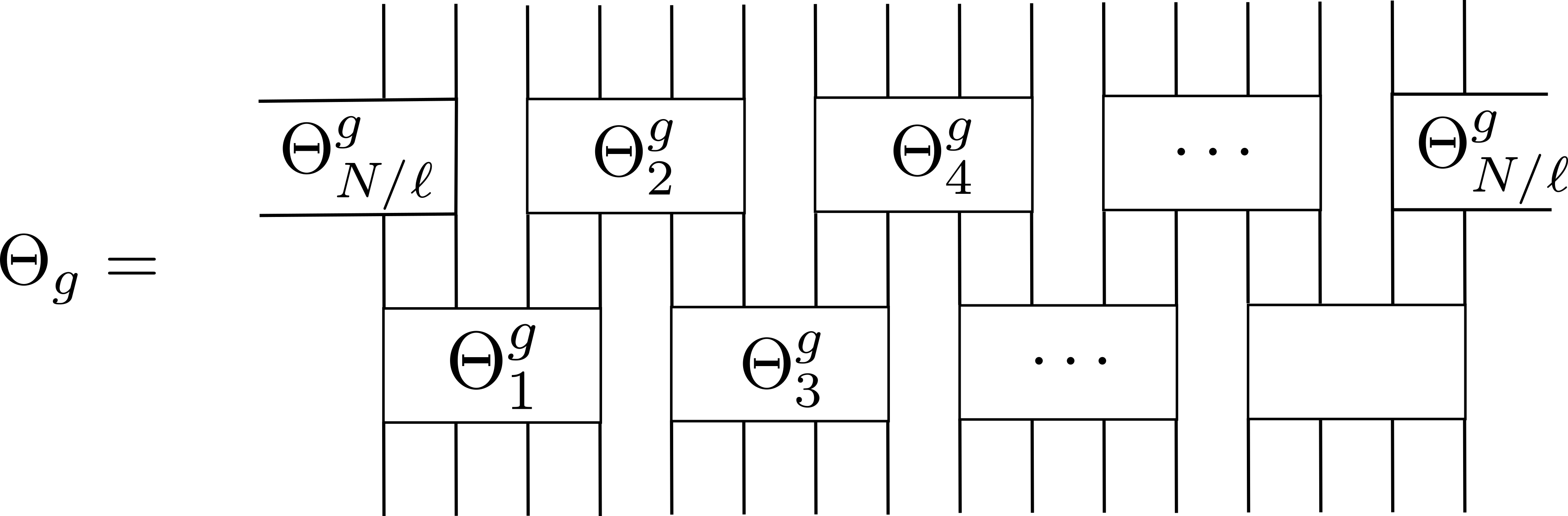}
 \end{aligned} .  
\end{equation}
Therefore, $\mathpzc{v}_g^{\otimes N} \tilde U = \tilde U \Theta_g$ is an equality of two two-layer quantum circuits if one blocks the unitaries of $\tilde U$ with $\mathpzc{v}_g^{\otimes N}$ on the left hand side and the ones of $\Theta_g$ as defined in Eq.~\eqref{eq:thetaqc} with those of $\tilde U$ on the right hand side.

We now use a trick from Ref.~\onlinecite{Thorsten}, but will obtain a slightly more compact result taking advantage of gauge degrees of freedom: If two two-layer quantum circuits are equal,
%
%
\begin{equation}
\begin{tikzpicture}[scale=1.1,baseline=(current  bounding  box.center)]    
\foreach \x in {0,1.6,3.2,4.8}{
\draw[thick](-0.4+\x,-0.6) -- (-0.4+\x,1.4);
\draw[thick](0.4+\x,-0.6) -- (0.4+\x,1.4);
\draw[thick,fill=white] (-0.4+\x,-0.25) rectangle (0.4+\x,0.25);		
}
\foreach \x in {0.8,2.4,4}{
\draw[thick,fill=white] (-0.4+\x,0.55) rectangle (0.4+\x,1.05);		
}
\draw[thick] (-0.8,0.55) -- (-0.4,0.55) -- (-0.4,1.05) -- (-0.8,1.05);
\draw[thick] (5.6,0.55) -- (5.2,0.55) -- (5.2,1.05) -- (5.6,1.05); 
\coordinate[label=right:$U_1$] (A) at (-0.3,0);
\coordinate[label=right:$U_2$] (A) at (1.3,0);
\coordinate[label=right:$\ldots$] (A) at (2.9,0);
\coordinate[label=right:$U_n$] (A) at (4.5,0);
\coordinate[label=right:$V_1$] (A) at (0.5,0.8);
\coordinate[label=right:$V_2$] (A) at (2.1,0.8);
\coordinate[label=right:$\ldots$] (A) at (3.7,0.8);
\coordinate[label=right:$V_n$] (A) at (5.15,0.78);
\coordinate[label=right:$V_n$] (A) at (-1,0.78);

\begin{scope}[shift={(0,-2.8)}]
\coordinate[label=right:${=}$] (A) at (-2,0.8);
\foreach \x in {0,1.6,3.2,4.8}{
\draw[thick](-0.4+\x,-0.6) -- (-0.4+\x,1.4);
\draw[thick](0.4+\x,-0.6) -- (0.4+\x,1.4);
\draw[thick,fill=white] (-0.4+\x,-0.25) rectangle (0.4+\x,0.25);		
}
\foreach \x in {0.8,2.4,4}{
\draw[thick,fill=white] (-0.4+\x,0.55) rectangle (0.4+\x,1.05);		
}
\draw[thick] (-0.8,0.55) -- (-0.4,0.55) -- (-0.4,1.05) -- (-0.8,1.05);
\draw[thick] (5.6,0.55) -- (5.2,0.55) -- (5.2,1.05) -- (5.6,1.05); 
\coordinate[label=right:$U_1'$] (A) at (-0.3,0);
\coordinate[label=right:$U_2'$] (A) at (1.3,0);
\coordinate[label=right:$\ldots$] (A) at (2.9,0);
\coordinate[label=right:$U_n'$] (A) at (4.5,0);
\coordinate[label=right:$V_1'$] (A) at (0.5,0.8);
\coordinate[label=right:$V_2'$] (A) at (2.1,0.8);
\coordinate[label=right:$\ldots$] (A) at (3.7,0.8);
\coordinate[label=right:$V_n'$] (A) at (5.15,0.78);
\coordinate[label=right:$V_n'$] (A) at (-1,0.78);
\end{scope}

\end{tikzpicture}  \label{eq:qu_circuits},
\end{equation}        
we can multiply both sides from the top by $\bigotimes_k V_k^\dg$ and from the bottom by $\bigotimes_k U_k'^\dg$, which results in
%
%
\begin{equation}
\begin{tikzpicture}[scale=1.1,baseline=(current  bounding  box.center)]    
\foreach \x in {0,1.6,3.2,4.8}{
\draw[thick](-0.4+\x,-0.75) -- (-0.4+\x,1.75);
\draw[thick](0.4+\x,-0.75) -- (0.4+\x,1.75);
\draw[thick,fill=white] (-0.4+\x,-0.25) rectangle (0.4+\x,0.25);		
}
 
\coordinate[label=right:${U_1'}^\dg$] (A) at (-0.35,0);
\coordinate[label=right:${U_2'}^\dg$] (A) at (1.25,0);
\coordinate[label=right:$\ldots$] (A) at (2.9,0);
\coordinate[label=right:${U_n'}^\dg$] (A) at (4.45,0);

\foreach \x in {0,1.6,3.2,4.8}{
\draw[thick,fill=white] (-0.4+\x,0.75) rectangle (0.4+\x,1.25);		
}

\coordinate[label=right:$U_1$] (A) at (-0.3,1);
\coordinate[label=right:$U_2$] (A) at (1.3,1);
\coordinate[label=right:$\ldots$] (A) at (2.9,1);
\coordinate[label=right:$U_{n}$] (A) at (4.5,1);


\begin{scope}[shift={(0,-4.2)}]
\coordinate[label=right:${=}$] (A) at (-2,1.4);

\foreach \x in {0,1.6,3.2,4.8}{
\draw[thick](-0.4+\x,0) -- (-0.4+\x,1.4);
\draw[thick](0.4+\x,0) -- (0.4+\x,1.4);
}
\foreach \x in {0.8,2.4,4}{
\draw[thick,fill=white] (-0.4+\x,0.55) rectangle (0.4+\x,1.05);		
}
\draw[thick] (-0.8,0.55) -- (-0.4,0.55) -- (-0.4,1.05) -- (-0.8,1.05);
\draw[thick] (5.6,0.55) -- (5.2,0.55) -- (5.2,1.05) -- (5.6,1.05); 
\coordinate[label=right:${V_1'}$] (A) at (0.5,0.8);
\coordinate[label=right:${V_2'}$] (A) at (2.1,0.8);
\coordinate[label=right:$\ldots$] (A) at (3.7,0.8);
\coordinate[label=right:${V_n'}$] (A) at (5.15,0.78);
\coordinate[label=right:${V_n'}$] (A) at (-1,0.78);

\begin{scope}[shift={(0,-0.6)}]
\foreach \x in {0,1.6,3.2,4.8}{
\draw[thick](-0.4+\x,2) -- (-0.4+\x,3.4);
\draw[thick](0.4+\x,2) -- (0.4+\x,3.4);
}
\foreach \x in {0.8,2.4,4}{
\draw[thick,fill=white] (-0.4+\x,2.35) rectangle (0.4+\x,2.85);		
}
\draw[thick] (-0.8,2.35) -- (-0.4,2.35) -- (-0.4,2.85) -- (-0.8,2.85);
\draw[thick] (5.6,2.35) -- (5.2,2.35) -- (5.2,2.85) -- (5.6,2.85); 

\coordinate[label=right:$V_1^\dg$] (A) at (0.5,2.61);
\coordinate[label=right:$V_2^\dg$] (A) at (2.1,2.61);
\coordinate[label=right:$\ldots$] (A) at (3.7,2.58);
\coordinate[label=right:$V_{n}^\dg$] (A) at (5.15,2.65);
\coordinate[label=right:$V_{n}^\dg$] (A) at (-1.1,2.6);
\end{scope}

\end{scope}
\end{tikzpicture} \ . \label{eq:tensor_product} 
\end{equation}        
Since the left and the right hand side of this equation are tensor products with respect to different partitions, they must both further subdivide into tensor products of tensors acting on blocks consistent with both partitions, 
%
%
\begin{equation} 
\begin{tikzpicture}[scale=1.1,baseline=(current  bounding  box.center)]    

\draw[thick](-0.4,-1.4) -- (-0.4,1.4);
\draw[thick](0.4,-1.4) -- (0.4,1.4);
\draw[thick,fill=white] (-0.4,0.3) rectangle (0.4,0.8);		
\draw[thick,fill=white] (-0.4,-0.8) rectangle (0.4,-0.3);		

\coordinate[label=right:$U_k$] (A) at (-0.3,0.55);
\coordinate[label=right:${U_k'}^\dg$] (A) at (-0.4,-0.55);

\coordinate[label=right:${=}$] (A) at (1,0);

\begin{scope}[shift={(3.4,0)}]
\draw[thick](-0.7,-1.4) -- (-0.7,1.4);
\draw[thick](0.7,-1.4) -- (0.7,1.4);
\draw[thick,fill=white] (-0.7,0) ellipse (0.55 and 0.55);
\draw[thick,fill=white] (0.7,0) ellipse (0.55 and 0.55);
\coordinate[label=right:$W_{2k-1}'$] (A) at (-1.3,0);
\coordinate[label=right:$W_{2k}'$] (A) at (0.2,0);

\end{scope}

\end{tikzpicture} 
\end{equation}        
and
\begin{equation}
\begin{tikzpicture}[scale=1.1,baseline=(current  bounding  box.center)]    

\draw[thick](-0.4,-1.4) -- (-0.4,1.4);
\draw[thick](0.4,-1.4) -- (0.4,1.4);
\draw[thick,fill=white] (-0.4,0.3) rectangle (0.4,0.8);		
\draw[thick,fill=white] (-0.4,-0.8) rectangle (0.4,-0.3);		

\coordinate[label=right:$V_k^\dg$] (A) at (-0.3,0.55);
\coordinate[label=right:$V_k'$] (A) at (-0.3,-0.55);

\coordinate[label=right:${=}$] (A) at (1,0);

\begin{scope}[shift={(3.4,0)}]
\draw[thick](-0.7,-1.4) -- (-0.7,1.4);
\draw[thick](0.7,-1.4) -- (0.7,1.4);
\draw[thick,fill=white] (-0.7,0) ellipse (0.55 and 0.55);
\draw[thick,fill=white] (0.7,0) ellipse (0.55 and 0.55);
\coordinate[label=right:$W_{2k}'$] (A) at (-1.2,0);
\coordinate[label=right:$W_{2k+1}'$] (A) at (0.1,0);

\coordinate[label=right:$e^{i \phi_k}$] (A) at (1.6,0.1);

\end{scope} 
\end{tikzpicture}.
\end{equation}        
Since the $U_k$'s and $V_k$'s are unitaries, the $W_j$'s are unitaries, too. \textit{A priori}, the factor $e^{i \phi_k}$ ($\phi_k \in [0,2\pi)$) has to be included, as decomposing equations of tensor products is unique up to prefactors (which have to be of magnitude one due to unitarity). However, it can be absorbed into the definition of $W_j$ by redefining
\begin{align}
W_1' = W_1,& \ \ \ W_2' = W_2 \\
W_{2k-1}' = W_{2k-1},& \ \ \ W_{2k}' = W_{2k} e^{-i \sum_{m = 1}^{k-1} \phi_m}
\end{align}
for $k > 1$. 
This is consistent with the constraint (following from Eq.~\eqref{eq:tensor_product}) 
\begin{align}
\sum_{k=1}^{\frac{N}{\ell}} \phi_k \mod 2\pi = 0 .
\end{align}  
Eqs.~\eqref{eq:gauge1},~\eqref{eq:gauge2} are a \textit{gauge transformation}, since they leave the overall quantum circuit invariant. Therefore,
%
%
\begin{equation} 
\begin{tikzpicture}[scale=1.1,baseline=(current  bounding  box.center)]    

\draw[thick](-0.4,-1.4) -- (-0.4,1.4);
\draw[thick](0.4,-1.4) -- (0.4,1.4);
\draw[thick,fill=white] (-0.4,0.3) rectangle (0.4,0.8);		
\draw[thick,fill=white] (-0.4,-0.8) rectangle (0.4,-0.3);		

\coordinate[label=right:$U_k$] (A) at (-0.3,0.55);
\coordinate[label=right:${U_k'}^\dg$] (A) at (-0.4,-0.55);

\coordinate[label=right:${=}$] (A) at (1,0);

\begin{scope}[shift={(3.4,0)}]
\draw[thick](-0.7,-1.4) -- (-0.7,1.4);
\draw[thick](0.7,-1.4) -- (0.7,1.4);
\draw[thick,fill=white] (-0.7,0) ellipse (0.55 and 0.55);
\draw[thick,fill=white] (0.7,0) ellipse (0.55 and 0.55);
\coordinate[label=right:$W_{2k-1}$] (A) at (-1.3,0);
\coordinate[label=right:$W_{2k}$] (A) at (0.2,0);

\end{scope}

\end{tikzpicture} \label{eq:gauge1}
\end{equation}        
and
\begin{equation}
\begin{tikzpicture}[scale=1.1,baseline=(current  bounding  box.center)]    

\draw[thick](-0.4,-1.4) -- (-0.4,1.4);
\draw[thick](0.4,-1.4) -- (0.4,1.4);
\draw[thick,fill=white] (-0.4,0.3) rectangle (0.4,0.8);		
\draw[thick,fill=white] (-0.4,-0.8) rectangle (0.4,-0.3);		

\coordinate[label=right:$V_k^\dg$] (A) at (-0.3,0.55);
\coordinate[label=right:$V_k'$] (A) at (-0.3,-0.55);

\coordinate[label=right:${=}$] (A) at (1,0);

\begin{scope}[shift={(3.4,0)}]
\draw[thick](-0.7,-1.4) -- (-0.7,1.4);
\draw[thick](0.7,-1.4) -- (0.7,1.4);
\draw[thick,fill=white] (-0.7,0) ellipse (0.55 and 0.55);
\draw[thick,fill=white] (0.7,0) ellipse (0.55 and 0.55);
\coordinate[label=right:$W_{2k}$] (A) at (-1.2,0);
\coordinate[label=right:$W_{2k+1}$] (A) at (0.1,0);


\end{scope} \label{eq:gauge2}
\end{tikzpicture} \ \ .
\end{equation}        

If one writes the quantum circuit as a matrix product operator with tensors $A^k$, 
\begin{equation}
\begin{aligned}
\includegraphics[width=0.4\textwidth]{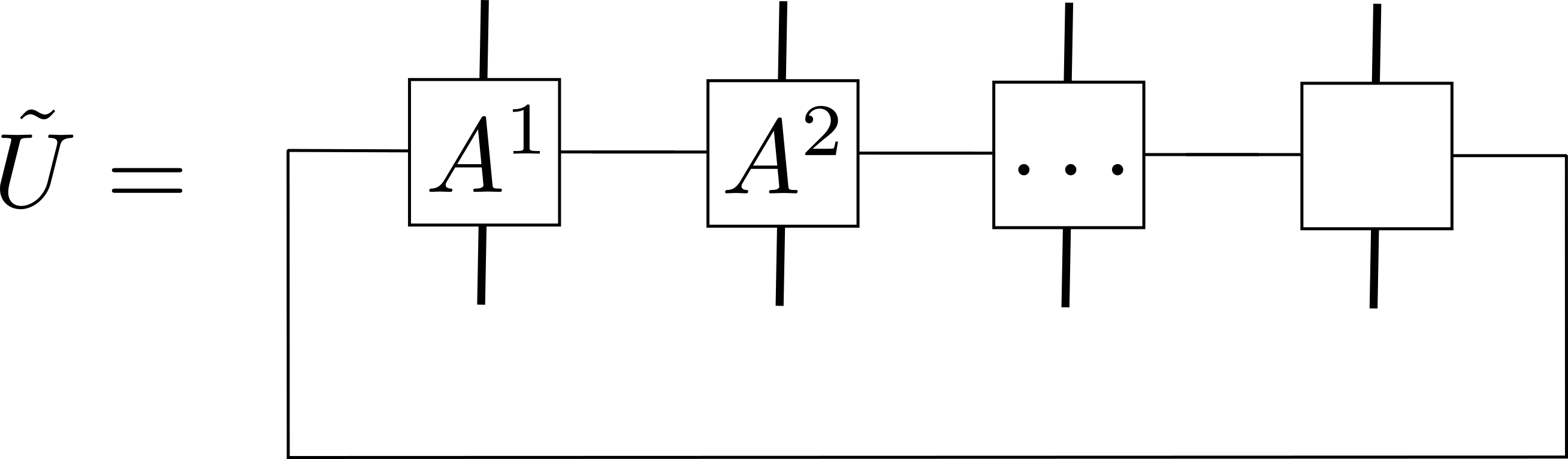} 
\end{aligned} \ ,
\end{equation}
they schematically fulfill the symmetry (cf. Eq.~\eqref{eq:symMPS} below)
\begin{equation}
\begin{aligned}
\includegraphics[width=0.4\textwidth]{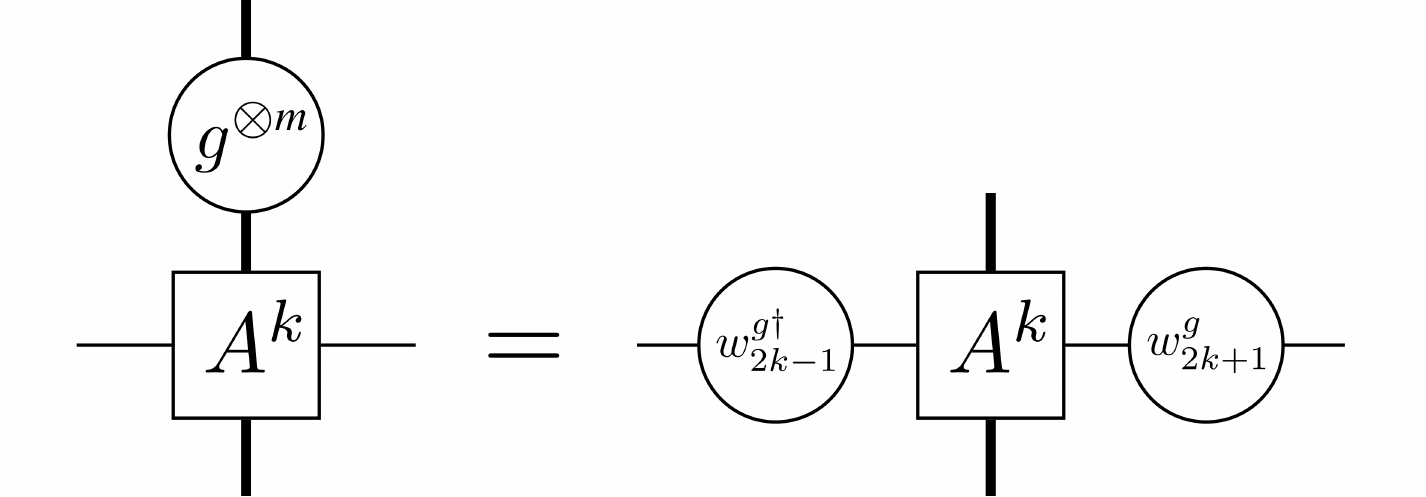}
\end{aligned} \label{eq:fundeq}
\end{equation}
with unitaries $w_j^g$. 
Each thick line represents $m = O(1)$ thin lines (which each have dimension $2^{\ell/2}$).

Consecutive application of this equation for two group elements $g$ and $h$ shows that the $w$-unitaries obey a set of equations (derived below as Eq.~\eqref{eq:cohomology1} and \eqref{eq:cohomology2}) schematically represented as 
\begin{align}
w_{2k-1}^{gh} &= w_{2k-1}^{g} w_{2k-1}^{h} e^{i \beta^{g,h}} \label{eq:beta1} \\
w_{2k+1}^{gh} &= w_{2k+1}^{g} w_{2k+1}^{h} e^{i \beta^{g,h}}. \label{eq:beta2}
\end{align}
Hence, they are \textit{projective representations} of the group $G$ and
correspond to the same element of the second cohomology group (cohomology class). 
Importantly, since Eq.~\eqref{eq:beta1} and \eqref{eq:beta2} have no dependence on the l-bit indices, all eigenstates have the same (ground state) topological label.

In particular, for time-reversal symmetry, 
it can be shown that the corresponding gauge transformation matrices  $w^{z}_{2k+1}$ fulfill (asterisk denoting complex conjugation)
\begin{align}
w^{\text{z}}_{2k+1} w^{\text{z} *}_{2k+1} &= (-1)^\kappa \mathbb{1} \label{eq:TRS_simple}
\end{align}
for all $k$ and for all l-bit labels, i.e. the topological label $\kappa = 0,1$ is shared by all eigenstates. $\kappa = 1$ corresponds to the topologically non-trivial phase. 

We note that it appears as if one could use the fact that the full set of eigenstates can be efficiently approximated by a short-depth quantum circuit $\tilde U$ to directly show that all of them must have the same topological label: $\tilde U$ approximately diagonalizes the Hamiltonian, i.e., with high accuracy $E = \tilde U^\dagger H \tilde U$ is a diagonal matrix. One can easily define a short-depth quantum circuit (e.g., of single-site unitaries) $V'(\lambda)$ with $\lambda \in [0,1]$ and $V'(0) = \mathbb{1}$ which continuously connects any two basis states, say $\tilde V(1) |l_1 \ldots l_N \rangle = |l_1' \ldots l_N' \rangle$. If we now set $V(\lambda) = \tilde U V'(\lambda) \tilde U^\dagger$, $V(\lambda)$ connects the corresponding eigenstates, i.e., $V(1) |\psi_{l_1 \ldots l_N} \rangle = |\psi_{l_1' \ldots l_N'}\rangle$. Hence, both eigenstates seem to correspond to the same SPT phase. However, for this to be the case the quantum circuit $V(\lambda)$ must be invariant under the symmetry, i.e.,
\begin{align}
v_g^{\otimes N \dagger} V(\lambda) v_g^{\otimes N} = V(\lambda),
\end{align}
which according to Eq.~\eqref{eq:theta1} implies
\begin{align}
\theta_g^* \tilde V(\lambda) \theta_g = \tilde V(\lambda).
\end{align}
Hence, $\tilde V(\lambda)$ can only act non-trivially in subspaces corresponding to equal $\theta_g$ diagonal elements. These subspaces correspond to eigenstates in the same symmetry-broken phase. This is consistent with the fact that only eigenstates belonging to the same SPT phase and the same symmetry-broken phase can be connected with each other via a short-depth quantum circuit~\cite{Chen_Gu} $V(\lambda)$. In other words, this argument can only be used to show that all eigenstates in the same symmetry-broken phase have the same topological label.

\subsubsection{Fermionic case}

The classification of fermionic  SPT MBL phases	can be obtained from the spinful one by introducing a diagrammatic formulation of fermionic tensor networks (Sec.~\ref{sec:fermform}), which faithfully represents the anti-commuting nature of the fermionic degrees of freedom. This approach allows us to derive the analogues of Eqs.~\eqref{eq:beta1} and~\eqref{eq:beta2}, and hence conclude that all eigenstates in a fermionic SPT MBL phase correspond to the same element of the (generalized) second cohomology group of the symmetry group combined with parity symmetry.  
However, compared to the classification of SPT fermionic ground states, this comes short of one $\mathbb{Z}_2$ topological label associated with the fractionalization of fermionic parity~\cite{pollmann2011ferm,Bultnick2017}. We attribute this to the inability of our ansatz to exactly represent eigenstates for which the this topological index  is non-trivial. 
For concreteness, we now review briefly the $\mathbb{Z}_8$ classification of one-dimensional interacting fermionic phases~\cite{kitaevferm2010, kitaevferm2011, pollmann2011ferm} and sketch the derivation in the MBL case of two out of three of the corresponding $\mathbb{Z}_2$ topological indices, which are shared by all eigenstates.

	Recall that the $\mathbb{Z}_8$ classification for one-dimensional interacting fermionic systems corresponds to three topological invariants defined as follows\cite{vers1610}: (i) 
	An index $\kappa = 0,1$ which arises from the fact that the time reversal operator on the virtual level of the matrix product state  representation of the ground state squares to positive or negative identity; 
	(ii) an index $\mu = 0,1$ 
	which indicates how the time reversal operator on the virtual level commutes with the parity operator on the virtual level;
	and 
	(iii) the fractionalization of the parity of the ground state~\cite{pollmann2011ferm}, i.e., whether there appears a decoupled  Majorana mode at each edge for open boundary conditions or after tracing out part of the system.  

	

For fermionic SPT MBL systems, (i) is derived analogously to the spin case (cf. Eq.~\eqref{eq:TRS_simple}). 
(ii) can be obtained as follows: By noting that the time-reversal operator and parity operator commute, we arrive at the following equations between the gauge transformation matrices of the time-reversal operator, $w^{z}_{i}$, and of the parity operator,  $w^{p}_{i}$, 
\begin{align}	
&
w^{z \dagger}_{2k-1} w^{p \dagger}_{2k-1} = w^{p \dagger}_{2k-1} w^{z \dagger}_{2k-1} e^{-i \pi \mu_k} 
\\
&
w^{p}_{2k+1} w^{z}_{2k+1} = w^{z}_{2k+1} w^{p}_{2k+1} e^{+i \pi \mu_k} 
\end{align} 
where $k$ labels a group of physical sites. After some simple algebraic manipulations, we show that
\begin{align}	
&
w^{z}_{i} w^{p}_{i} = (-1)^\mu w^{p}_{i} w^{z}_{i}
\end{align}  
where $\mu$ is a topological label, independent of site labels $k$ and $l$-bit labels. This implies that this topological label is shared by all eigenstates. The topological labels (i) and 
(ii) give rise to the $\mathbb{Z}_4$ classification of fermionic SPT MBL systems in the presence of time-reversal symmetry.  

	The current approach cannot be used to detect the presence (or absence) of the topological invariant associated with (iii), because the classification uses two-layer quantum circuits that locally \textit{preserve} fermionic parity and consequently cannot capture a topological label which requires long-range fermionic parity-preserving quantum circuits to be exactly represented. 
We illustrate this fact with the example of the Kitaev chain below. 	

\subsubsection{Completeness of classification for individual eigenstates}

The spinful classification derived in this article is complete in the sense that  there cannot be any additional topological indices which affect the properties of individual eigenstates (such as degeneracies in the entanglement spectra).
This does not rule out the possibility that there are topological obstructions to connecting different Hamiltonians with the same eigenstate topological index. However, we show that if there are Hamiltonians disconnected by such a topological obstruction, their topological distinctness cannot be visible on their individual eigenstates. 
 Furthermore, for both spin and fermionic systems we show that the topological index of the SPT MBL phase cannot change along the adiabatic evolution in the thermodynamic limit unless the symmetry or FMBL condition is broken.

Finally, Ref.~\onlinecite{2016Potter_Vasseur} provided a ``no-go theorem'' stating that MBL is not possible for symmetries that protect degeneracies, and in particular, for non-abelian symmetries (as also shown in the Appendix using our formalism). One way for a non-abelian symmetry and MBL to be compatible is for the system to spontaneously break the non-abelian symmetry while preserving an abelian subsymmetry. In this case, one might be able to use similar tools as the ones introduced here to classify the different SPT MBL phases with the corresponding abelian subsymmetry. 

\subsection{Example: The cluster model}

We can use these insights to show that the four-fold degenerate entanglement spectra of the disordered cluster model~\eqref{eq:top_Ham} are protected both by $G = \mathbb{Z}_2 \times \mathbb{Z}_2$ on-site symmetry and time-reversal symmetry: The Hamiltonian is invariant under the unitary transformations~\cite{bahri2015localization} $(\sigma_z \otimes \mathbb{1})^{\otimes \frac{N}{2}}$, $(\mathbb{1} \otimes \sigma_z)^{\otimes \frac{N}{2}}$ and consequently $\sigma_z^{\otimes N}$, which together with $\mathbb{1}$ represent the group $\mathbb{Z}_2 \times \mathbb{Z}_2$. On the other hand, it is also invariant under time-reversal symmetry defined by $\mc{T} = \sigma_z^{\otimes N} K$, $K$ carrying out complex conjugation. The unitary matrix $U$ diagonalizing the Hamiltonian has an exact representation in terms of a two-layer quantum circuit~\cite{Thorsten} for $\sigma_V = \sigma_h = 0$. We can use this representation to show that the Hamiltonian for $\sigma_V, \sigma_h \ll \sigma_\lambda$ is topologically non-trivial with respect to both symmetries. The unitaries act on $\ell = 2$ sites and are given by
\begin{align}
u_k = v_k  =  \frac{1}{2} \left(\begin{array}{cccc}
1&-1&-1&-1\\
-1&1&-1&-1\\
-1&-1&1&-1\\
-1&-1&-1&1
\end{array}\right). \label{eq:u_cluster}
\end{align}
This results in the following properties~\cite{Thorsten} (setting $X = \sigma_x$, $Y = \sigma_y$, $Z = \sigma_z$)
%
%
\begin{equation}
\begin{tikzpicture}[scale=0.95,baseline=(current  bounding  box.center)]

\draw[thick](-6.4,-0.55) -- (-6.4,0.55);
\draw[thick](-5.6,-0.55) -- (-5.6,0.75);
\draw[thick,fill=white] (-6.4,-0.25) rectangle (-5.6,0.25);		
\draw[thick,fill=white] (-5.6,0.75) rectangle (-4.8,1.25);	
	
\draw[thick] (-6.8,0.55) -- (-6.4,0.55);
\draw[thick] (-4.4,0.55) -- (-4.8,0.55) -- (-4.8,0.75);
\draw[thick] (-4.8,2.4) -- (-4.8,1.25) -- (-5.6,1.25) -- (-5.6,2.4);
\draw[thick,fill=white] (-5.6,1.85) circle (0.25);
\coordinate[label=above:$Z$] (A) at (-5.6,1.6);

\coordinate[label=right:$u_k$] (A) at (-6.3,0);
\coordinate[label=right:$v_k$] (A) at (-5.5,1);

\coordinate[label=right:${=} \ \  $] (A) at (-3.6,0.6);
\draw[thick](-1.4,-1.35) -- (-1.4,0.55);
\draw[thick](-0.6,-1.35) -- (-0.6,0.75);
\draw[thick,fill=white] (-1.4,-0.25) rectangle (-0.6,0.25);		
\draw[thick,fill=white] (-0.6,0.75) rectangle (0.2,1.25);		

\draw[thick] (-2.6,0.55) -- (-1.4,0.55);
\draw[thick,fill=white] (-2.05,0.55) circle (0.25);
\coordinate[label=left:$X$] (A) at (-1.75,0.55);
\draw[thick] (1.4,0.55) -- (0.2,0.55) -- (0.2,0.75);
\draw[thick] (0.2,1.6) -- (0.2,1.25) -- (-0.6,1.25) -- (-0.6,1.6);
\draw[thick,fill=white] (0.85,0.55) circle (0.25);
\coordinate[label=right:$X$] (A) at (0.55,0.55);
\draw[thick,fill=white] (-0.6,-0.8) circle (0.25);
\coordinate[label=below:$Z$] (A) at (-0.6,-0.55);

\coordinate[label=right:$u_k$] (A) at (-1.3,0);
\coordinate[label=right:$v_k$] (A) at (-0.5,1);

\end{tikzpicture} \  , \label{eq:Zeven}
\end{equation}
%
%
\begin{equation}
\begin{tikzpicture}[scale=0.95,baseline=(current  bounding  box.center)]

\draw[thick](-6.4,-0.55) -- (-6.4,0.55);
\draw[thick](-5.6,-0.55) -- (-5.6,0.75);
\draw[thick,fill=white] (-6.4,-0.25) rectangle (-5.6,0.25);		
\draw[thick,fill=white] (-5.6,0.75) rectangle (-4.8,1.25);	
	
\draw[thick] (-6.8,0.55) -- (-6.4,0.55);
\draw[thick] (-4.4,0.55) -- (-4.8,0.55) -- (-4.8,0.75);
\draw[thick] (-4.8,2.4) -- (-4.8,1.25) -- (-5.6,1.25) -- (-5.6,2.4);
\draw[thick,fill=white] (-4.8,1.85) circle (0.25);
\coordinate[label=above:$Z$] (A) at (-4.8,1.6);

\coordinate[label=right:$u_k$] (A) at (-6.3,0);
\coordinate[label=right:$v_k$] (A) at (-5.5,1);

\coordinate[label=right:${=} \ \  $] (A) at (-3.6,0.6);
\draw[thick](-1.4,-1.35) -- (-1.4,0.55);
\draw[thick](-0.6,-1.35) -- (-0.6,0.75);
\draw[thick,fill=white] (-1.4,-0.25) rectangle (-0.6,0.25);		
\draw[thick,fill=white] (-0.6,0.75) rectangle (0.2,1.25);		

\draw[thick] (-2.6,0.55) -- (-1.4,0.55);
\draw[thick,fill=white] (-2.05,0.55) circle (0.25);
\coordinate[label=left:$Z$] (A) at (-1.8,0.55);
\draw[thick] (1.4,0.55) -- (0.2,0.55) -- (0.2,0.75);
\draw[thick] (0.2,1.6) -- (0.2,1.25) -- (-0.6,1.25) -- (-0.6,1.6);
\draw[thick,fill=white] (0.85,0.55) circle (0.25);
\coordinate[label=right:$Z$] (A) at (0.6,0.55);
\draw[thick,fill=white] (-1.4,-0.8) circle (0.25);
\coordinate[label=below:$Z$] (A) at (-1.4,-0.55);

\coordinate[label=right:$u_k$] (A) at (-1.3,0);
\coordinate[label=right:$v_k$] (A) at (-0.5,1);

\end{tikzpicture} \ , \label{eq:Zodd}
\end{equation}
and
%
%
\begin{equation}
\begin{tikzpicture}[scale=0.95,baseline=(current  bounding  box.center)]

\draw[thick](-6.4,-0.55) -- (-6.4,0.55);
\draw[thick](-5.6,-0.55) -- (-5.6,0.75);
\draw[thick,fill=white] (-6.4,-0.25) rectangle (-5.6,0.25);		
\draw[thick,fill=white] (-5.6,0.75) rectangle (-4.8,1.25);	
	
\draw[thick] (-6.8,0.55) -- (-6.4,0.55);
\draw[thick] (-4.4,0.55) -- (-4.8,0.55) -- (-4.8,0.75);
\draw[thick] (-4.8,2.4) -- (-4.8,1.25) -- (-5.6,1.25) -- (-5.6,2.4);
\draw[thick,fill=white] (-5.6,1.85) circle (0.25);
\draw[thick,fill=white] (-4.8,1.85) circle (0.25);
\coordinate[label=above:$Z$] (A) at (-4.8,1.6);
\coordinate[label=above:$Z$] (A) at (-5.6,1.6);

\coordinate[label=right:$u_k$] (A) at (-6.3,0);
\coordinate[label=right:$v_k$] (A) at (-5.5,1);

\coordinate[label=right:${=} \ \  $] (A) at (-3.6,0.6);
\draw[thick](-1.4,-1.35) -- (-1.4,0.55);
\draw[thick](-0.6,-1.35) -- (-0.6,0.75);
\draw[thick,fill=white] (-1.4,-0.25) rectangle (-0.6,0.25);		
\draw[thick,fill=white] (-0.6,0.75) rectangle (0.2,1.25);		

\draw[thick] (-2.6,0.55) -- (-1.4,0.55);
\draw[thick,fill=white] (-2.05,0.55) circle (0.25);
\coordinate[label=left:$Y$] (A) at (-1.75,0.55);
\draw[thick] (1.4,0.55) -- (0.2,0.55) -- (0.2,0.75);
\draw[thick] (0.2,1.6) -- (0.2,1.25) -- (-0.6,1.25) -- (-0.6,1.6);
\draw[thick,fill=white] (0.85,0.55) circle (0.25);
\coordinate[label=right:$Y$] (A) at (0.6,0.55);
\draw[thick,fill=white] (-1.4,-0.8) circle (0.25);
\draw[thick,fill=white] (-0.6,-0.8) circle (0.25);
\coordinate[label=below:$Z$] (A) at (-1.4,-0.55);
\coordinate[label=below:$Z$] (A) at (-0.6,-0.55);

\coordinate[label=right:$u_k$] (A) at (-1.3,0);
\coordinate[label=right:$v_k$] (A) at (-0.5,1);

\end{tikzpicture} \ . \label{eq:Z}
\end{equation}
The tensors of the matrix product state representation of the eigenstate $|\psi_{l_1 \ldots l_N}\rangle$ are given by
\begin{equation}
\begin{tikzpicture}[scale=0.95,baseline=(current  bounding  box.center)]

\draw[thick] (-5.1,1.4) -- (-5.1,1);
\draw[thick] (-5.9,1) -- (-5.9,1.4);


\draw[thick] (-6.6,0.55) -- (-6.2,0.55);
\draw[thick] (-4.8,0.55) -- (-4.4,0.55);
\draw[thick,fill=white] (-6.2,0.2) rectangle (-4.8,1);

\coordinate[label=right:$A^k_{l_{2k-1} l_{2k}}$] (A) at (-6.3,0.6);

\coordinate[label=right:${=} \ \  $] (A) at (-3.5,0.6);
\draw[thick](-1.4,-0.55) -- (-1.4,0.55);
\draw[thick](-0.6,-0.55) -- (-0.6,0.75);
\draw[thick,fill=white] (-1.4,-0.25) rectangle (-0.6,0.25);		
\draw[thick,fill=white] (-0.6,0.75) rectangle (0.2,1.25);		

\draw[thick] (-1.8,0.55) -- (-1.4,0.55);
\draw[thick] (0.6,0.55) -- (0.2,0.55) -- (0.2,0.75);
\draw[thick] (0.2,1.6) -- (0.2,1.25) -- (-0.6,1.25) -- (-0.6,1.6);

\coordinate[label=below:$l_{2k-1}$] (A) at (-1.45,-0.55);
\coordinate[label=below:$l_{2k}$] (A) at (-0.5,-0.55);

\coordinate[label=right:$u_k$] (A) at (-1.3,0);
\coordinate[label=right:$v_k$] (A) at (-0.5,1);

\end{tikzpicture}\ . \label{eq:Ak}
\end{equation}
The corresponding projective representation of $\mathbb{Z}_2 \times \mathbb{Z}_2$ (whose elements we label by $g = II, ZI, IZ, ZZ$) is thus $w_{II} = \mathbb{1}$, $w_{ZI} = \sigma_x$, $w_{IZ} = \sigma_z$ and $w_{ZZ} = \sigma_y$. The Pauli matrices anticommute, which cannot be changed by modifying their overall phases: They represent a non-trivial element of the second cohomology group. Hence, the system is topologically non-trivial with respect to $\mathbb{Z}_2 \times \mathbb{Z}_2$. 

Time-reversal symmetry given by $\mc{T} = \sigma_z^{\otimes N} K$ corresponds to the symmetry~\eqref{eq:Z} with $w^z = Y$, i.e., $w^z w^{z*} = -\mathbb{1}$ since $u_k$ and $v_k$ are real. Note that $\mathbb{Z}_2$ symmetry alone (in the absence of complex conjugation) would not suffice to protect the four-fold degeneracy of the entanglement spectra. 

Hence, the four-fold degenerate entanglement spectra are also stable with respect to weak perturbations of the form 
\begin{align}
\sum_i t_i \sigma_x^{i-1} \sigma_y^{i+1}
\end{align}
with $t_i \in \mathbb{R}$ of small magnitude and chosen from a random distribution. In this case, time-reversal symmetry is broken, but $\mathbb{Z}_2 \times \mathbb{Z}_2$ is preserved. On the other hand, perturbations of the form
\begin{align}
\sum_i y_i \sigma_y^i
\end{align}
(with small random $y_i \in \mathbb{R}$), 
break $\mathbb{Z}_2 \times \mathbb{Z}_2$ (and the $\mathbb{Z}_2$ subgroups), but preserve time-reversal symmetry. Consequently, those perturbations also do not affect the four-fold degeneracy of the entanglement spectra.


\section{Classification of spinful SPT MBL phases} \label{sec:clbos}
\subsection{Underlying assumptions}\label{sec:underlying_technical}

Here we state the assumptions underlying our derivation, which will not take error bounds into account. However, we believe that it is possible to include them into the derivation, making it mathematically rigorous, similarly to Ref.~\onlinecite{Thorsten}. 

We consider a disordered Hamiltonian $H$ invariant under an abelian on-site (anti-)unitary symmetry defined on a spin or fermionic chain of length $N$. (The specific cases are considered in the following subsections.) The disorder is assumed to be sufficiently strong such that the system is in the FMBL regime, which we define as the realm where all LIOMs $\tau_z^i$ have subextensive localization lengths $\xi_i$. That is, in the limit $N \rightarrow \infty$, none of the localization lengths is of order $\mathcal{O}(N)$. Furthermore, we assume that for sufficiently large $N$, there exists a unitary $U$ exactly diagonalizing the Hamiltonian and a two-layer quantum circuit $\tilde U$ with $\ell = a N$ ($a > 0$ fixed) such that $\tau_z^i = U \sigma_z^i U^\dg$ and $\tilde \tau_z^i = \tilde U \sigma_z^i \tilde U^\dg$ fulfill $\|\tau_z^i - \tilde \tau_z^i \|_\mr{op} < c \, e^{-\frac{\ell}{\xi_i}}$ with constant $c > 0$. $\| \cdot \|_\mr{op}$ denotes the operator norm. This is a very reasonable assumption given that the unitaries of the quantum circuit act on $\ell$ sites. 
In words, these assumptions mean that the systems we consider can be described by a complete set of \textit{local} integrals of motion (whose number is equal to $N$), which can be efficiently approximated by two-layer quantum circuits with long gates. 
Since the error of the approximation vanishes in the thermodynamic limit $N \rightarrow \infty$, we do not keep track of the error in this paper for conciseness. The full derivation will associate SPT MBL systems with an integer-valued topological invariant up to an error which vanishes in the limit $N \rightarrow \infty$.  
For our derivation we will also need the fact that due to FMBL, such systems remain localized (and approximable in the above sense) under small local perturbations. 

Note that the gate length $\ell$ must grow with the system size $N$, since the error of the approximation is expected to be of order~\cite{Thorsten} $\sum_i e^{-\frac{\xi_i}{\ell}}$, where $\xi_i$ is the localization length of $\tau_z^i$, and with increasing $N$, larger and larger localization lengths are represented. We define FMBL as the phase where no localization length $\xi_i$ is of the order of the system size $N$. That is, if $\ell$ grows proportionally to $N$, the above error will converge to zero in the limit $N \rightarrow \infty$. Hence, having a gate length $\ell$ which increases with the system size is the physically most sensible approach. This is different from the classification of ground states of gapped local Hamiltonians~\cite{Pollmann2010,2011Schuch,2011Chen,Bultnick2017}, where due to the finite correlation length a constant gate length $\ell$ can be used.


Note that we use a very weak notion of locality; in practise only eight blocks of unitaries ($a = 1/8$) would be sufficient for our classification of phases. Such quantum circuits allow for basically arbitrary transformations of $1/8$ of the overall system (but not of the full system).  Physically this means that our approximate unitaries allow for integrals of motion which can be as big as 1/4 of the whole system size, and, therefore, thermal puddles of up to 1/4 of the whole system size are allowed, but the system is nevertheless MBL as a whole.
Yet, even under this weak notion of locality, we will show that there are topologically distinct phases, which cannot be continuously connected with our quantum circuits. Even if thermal puddles can extend over $1/4$ of the whole system, it is impossible to continuously connect MBL phases with different topological indices.  This has been shown mathematically rigorously in Ref.~\onlinecite{Thorsten} for the case of time-reversal symmetry.

This weak notion of locality allows our approximate eigenstates to have volume law-entanglement (though with a smaller coefficient than maximally entangled states whose half-chain entanglement entropy is $S = \frac{N}{2}\log(2)$): This can be seen from the fact that the matrix product operator corresponding to our quantum circuit has bond dimension $D = 2^{\ell/2} = 2^{N/16}$ for $\ell = N/8$. The entanglement between two halves of the chain it is able to represent is thus $S \leq \frac{N}{16} \log(2)$. Note that for typical tensor network states (such as matrix product states), an exponentially growing bond dimension would be detrimental to any classification of phases, as this would allow for volume-law entanglement between arbitrarily distant regions. However, for quantum circuits, $\ell = N/8$ can only lead to entanglement between regions which are at most $N/4$ separated, while regions with bigger separation cannot be entangled. 

 We also point out that the restriction to \textit{two-layer} unitaries is not a significant limitation for the classification in the case of $\ell = N/8$, since $n$ layers of unitaries of fixed length $L$ can be encoded into a two-layer quantum circuit with unitaries of length $\ell \approx n L/2$. Hence, the case of gates of length $\ell = N/8$ includes the case of quantum circuits with unitaries of fixed length and depth of order of the system size or less. 
 Thus, $\ell = N/8$ cannot lead to additional topological restrictions compared to a fixed length/increasing depth quantum circuit; it can possibly only lead to fewer restrictions. The remaining restrictions are truly topological and due to the fact that $\ell = N/8$ still describes an MBL system for any finite $N$, since thermal puddles cannot extend over the whole system. How the limit $N \rightarrow \infty$ is taken is thus crucial for obtaining meaningful results (cf. Ref.~\onlinecite{Thorsten}).

\subsection{MBL systems with a unitary on-site symmetry}\label{sec:on-site}

Assume the FMBL Hamiltonian $H$ is invariant under a local unitary $\mathpzc{v}_g$, which is a linear representation of the abelian symmetry group $G \ni g$. That is,
\begin{align} 
H = \mathpzc{v}_g^{\otimes N} H (\mathpzc{v}_g^\dg)^{\otimes N}. \label{eq:Ham_sym}
\end{align}
Following the line of reasoning of Ref.~\onlinecite{Thorsten}, it is easy to derive the action of the symmetry on the unitary $U$ which diagonalizes the Hamiltonian: $H = U E U^\dg$ implies
\begin{align}
E = U^\dg \mathpzc{v}_g^{\otimes N} U E U^\dg (\mathpzc{v}_g^\dg)^{\otimes N} U. \label{eq:insert_diagonalisation}
\end{align}
For finite system size $N$, $E$ cannot have any symmetry-enforced degeneracies, as the symmetry group is abelian. For the moment, we remove any other degeneracies by infinitesimal symmetry-preserving  local perturbations, which does not violate the FMBL condition~\cite{Thorsten}.  The case of accidental degeneracies is explicitly treated in Section~\ref{sec:robustness}, where the stability with respect to symmetry-preserving perturbations is shown. 
We hence assume $E$ to be non-degenerate, such that Eq.~\eqref{eq:insert_diagonalisation} implies
\begin{align} 
\Theta_g = U^\dagger \mathpzc{v}_g^{\otimes N} U 
\label{eq:theta}
\end{align}
with $\Theta_g$ being a diagonal matrix whose diagonal elements have magnitude 1.

One can use the same argument as in Ref.~\onlinecite{Thorsten} in order to show that $\Theta_g$ can be written as a two-layer quantum circuit whose unitaries are also diagonal, which we repeat here for the sake of completeness: Let $\mb l_{k}$ denote the l-bit indices (lower legs in Eq.~\eqref{eq:tildeU}) $l_{(k-1) \ell+1},l_{(k-1)\ell+2}, \ldots, l_{k \ell}$. Eq.~\eqref{eq:theta_graph} thus implies for the diagonal elements $\theta_{g,\mb l_1 \mb l_2 \ldots \mb l_{N/\ell}}$ of $\Theta_g$ that
\begin{equation}
\begin{aligned}
\includegraphics[width=0.48\textwidth]{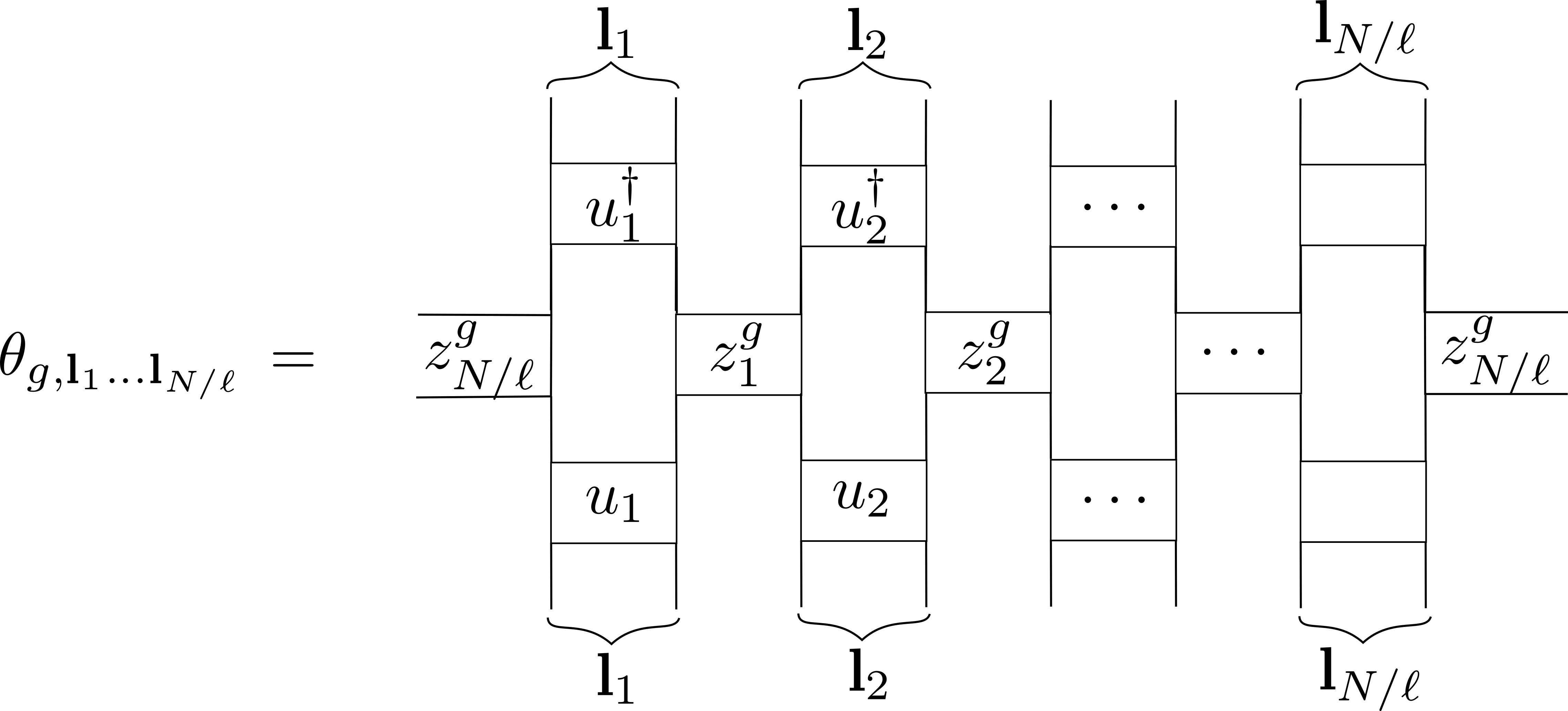}
 \end{aligned} \label{eq:theta_elements}  ,
\end{equation}
where we defined the unitaries $z_k^g = v_k^\dg \left(\mathpzc{v}_g^{\otimes \ell}\right) v_k$. Hence, the product $\theta_{g,\mb l_1 \ldots \mb l_k \ldots \mb l_{N/\ell}}^* \theta_{g,\mb l_1 \ldots \mb l_k' \ldots \mb l_{N/\ell}}$ can be written as 
\begin{equation}
\begin{aligned}
\includegraphics[width=0.46\textwidth]{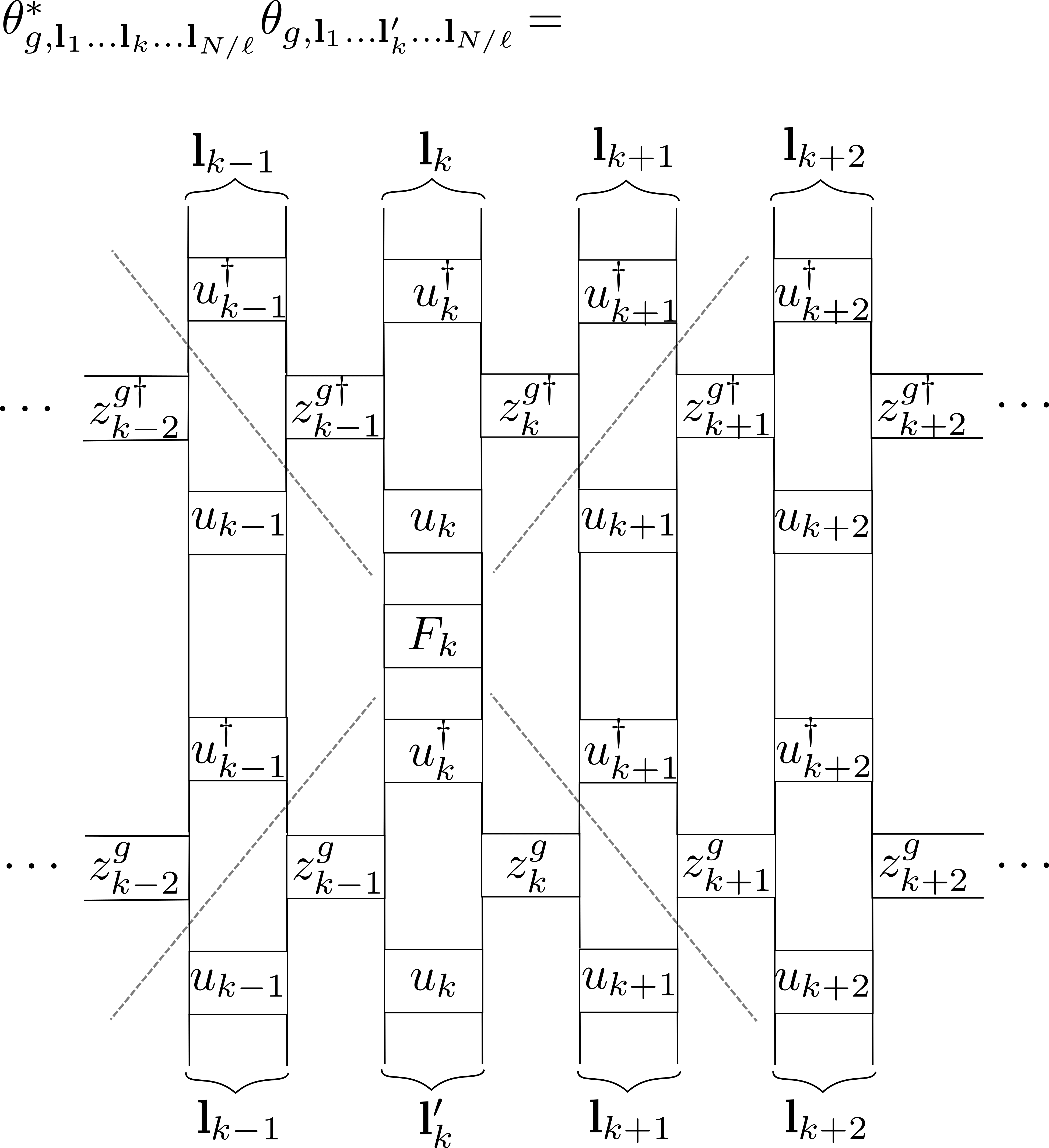}
 \end{aligned} \label{eq:theta_product}
\end{equation}
where we set $F_k = |\mb l_k \rangle \langle \mb l_k' |$ (and use cyclic indices). All unitaries outside the ``causal cone'' marked by dashed lines cancel, i.e., $\theta_{g,\mb l_1 \ldots \mb l_k \ldots \mb l_{N/\ell}}^* \theta_{g,\mb l_1 \ldots \mb l_k' \ldots \mb l_{N/\ell}}$ depends only on $\mb l_{k-1}, \mb l_k, \mb l_k', \mb l_{k+1}$. Therefore, we have 
\begin{align}\label{eq:lc1}
\theta_{g,\mb l_1 \ldots \mb l_k \ldots \mb l_{N/\ell}}^* \theta_{g,\mb l_1 \ldots \mb l_k' \ldots \mb l_{N/\ell}} = e^{-i p_k^g(\mb l_{k-1}, \mb l_k, \mb l_k', \mb l_{k+1})}
\end{align}
with unknown (discrete) functions $p_k^g \in \mathbb{R}$. We similarly define $\theta_{g,\mb l_1 \ldots \mb l_k \ldots \mb l_{N/\ell}} = e^{i f_g(\mb l_1, \ldots, \mb l_k, \ldots, \mb l_{N/\ell})}$, wherefore
\begin{align}
&f_g(\mb l_1, \ldots, \mb l_{k-1}, \mb l_k,\mb l_{k+1}, \ldots) - f_g(\mb l_1, \ldots, \mb l_{k-1}, \mb l_k',\mb l_{k+1},\ldots) \notag\\
&= p_k^g(\mb l_{k-1}, \mb l_k, \mb l_k', \mb l_{k+1}) \mod 2 \pi \label{eq:k} \\
&f_g(\mb l_1, \ldots, \mb l_{k-1}, \mb l_k',\mb l_{k+1}, \ldots) - f_g(\mb l_1, \ldots, \mb l_{k-1}, \mb l_k',\mb l_{k+1}',\ldots) \notag\\
&= p^g_{k+1}(\mb l_{k}', \mb l_{k+1}, \mb l_{k+1}', \mb l_{k+2}) \mod 2 \pi \label{eq:k+1}\\
&\ldots \notag\\
&f_g(\mb l_1', \ldots, \mb l_{k-1}, \mb l_k',\mb l_{k+1}', \ldots) - f_g(\mb l_1', \ldots, \mb l_{k-1}' ,\mb l_k',\mb l_{k+1}',\ldots) \notag\\
&=p_{k-1}^g(\mb l_{k-2}', \mb l_{k-1}, \mb l_{k-1}', \mb l_{k}') \mod 2 \pi. \label{eq:k-1}
\end{align}
In Eqs.~\eqref{eq:k+1} to~\eqref{eq:k-1} we consecutively flipped l-bits from $\mb l_m$ to $\mb l'_m$. 
Adding Eqs.~\eqref{eq:k} to~\eqref{eq:k-1} together yields
\begin{align}
&f_g(\mb l_1, \ldots, \mb l_{k-1}, \mb l_k,\mb l_{k+1}, \ldots) - f_g(\mb l_1', \ldots, \mb l_{k-1}', \mb l_k',\mb l_{k+1}',\ldots) \notag\\
&= p_k^g(\mb l_{k-1}, \mb l_k, \mb l_k', \mb l_{k+1}) \notag \\
&+ \sum_{m \in \{k+1, \ldots, \frac{N}{\ell}, 1, \ldots, k-2\}} p_m^g(\mb l_{m-1}', \mb l_m, \mb l_m', \mb l_{m+1}) \notag \\
&+ p_{k-1}^g(\mb l_{k-2}',\mb l_{k-1}, \mb l_{k-1}', \mb l_k') \mod 2 \pi .
\end{align}
We set $\mb l_1' = \mb l_2' = \ldots = \mb l_{N/\ell}' = \mb 0$, i.e.,
\begin{align}
&f_g(\mb l_1, \ldots, \mb l_{N/\ell}) - f_g(\mb 0, \ldots, \mb 0) = p_k^g(\mb l_{k-1}, \mb l_k, \mb 0, \mb l_{k+1}) \notag \\
&+ \sum_{m \in \{k+1, \ldots, \frac{N}{\ell}, 1, \ldots, k-2\}} p_m^g(\mb 0, \mb l_m, \mb 0, \mb l_{m+1}) \notag \\
&+ p_{k-1}^g(\mb 0,\mb l_{k-1}, \mb 0, \mb 0) \mod 2 \pi .
\end{align}
 Since $k$ is arbitrary, it follows that $f_g(\mb l_1, \ldots, \mb l_{N/\ell})$ can be written as a sum of real functions $\overline p_m^g$, which depend only on two consecutive blocked l-bits $\mb l_m, \mb l_{m+1}$ each, 
 \begin{align}
 f_g(\mb l_1, \ldots, \mb l_{N/\ell}) = \sum_{m=1}^{N/\ell} \overline p_m^g(\mb l_{m}, \mb l_{m+1}). 
 \end{align}
 Therefore, if we define diagonal matrices $\Theta^g_m$ whose diagonal elements are given by $e^{i \overline p_m^g(\mb l_m, \mb l_{m+1})}$, we arrive at the claimed two-layer quantum circuit representation
\begin{equation}
\begin{aligned}
\includegraphics[width=0.4\textwidth]{Theta_quantum_circuit.png}
 \end{aligned} \  \label{eq:theta_quantum_circuit}.
\end{equation}
We now insert this equation into $\tilde U \Theta_g = \mathpzc{v}_g^{\otimes N} \tilde U$, which leads to an equality of two two-layer quantum circuits if unitaries are blocked as indicated by dashed lines (we assume $N$ to be a multiple of $4 \ell$):
\begin{equation}
\begin{aligned}
\includegraphics[width=0.47\textwidth]{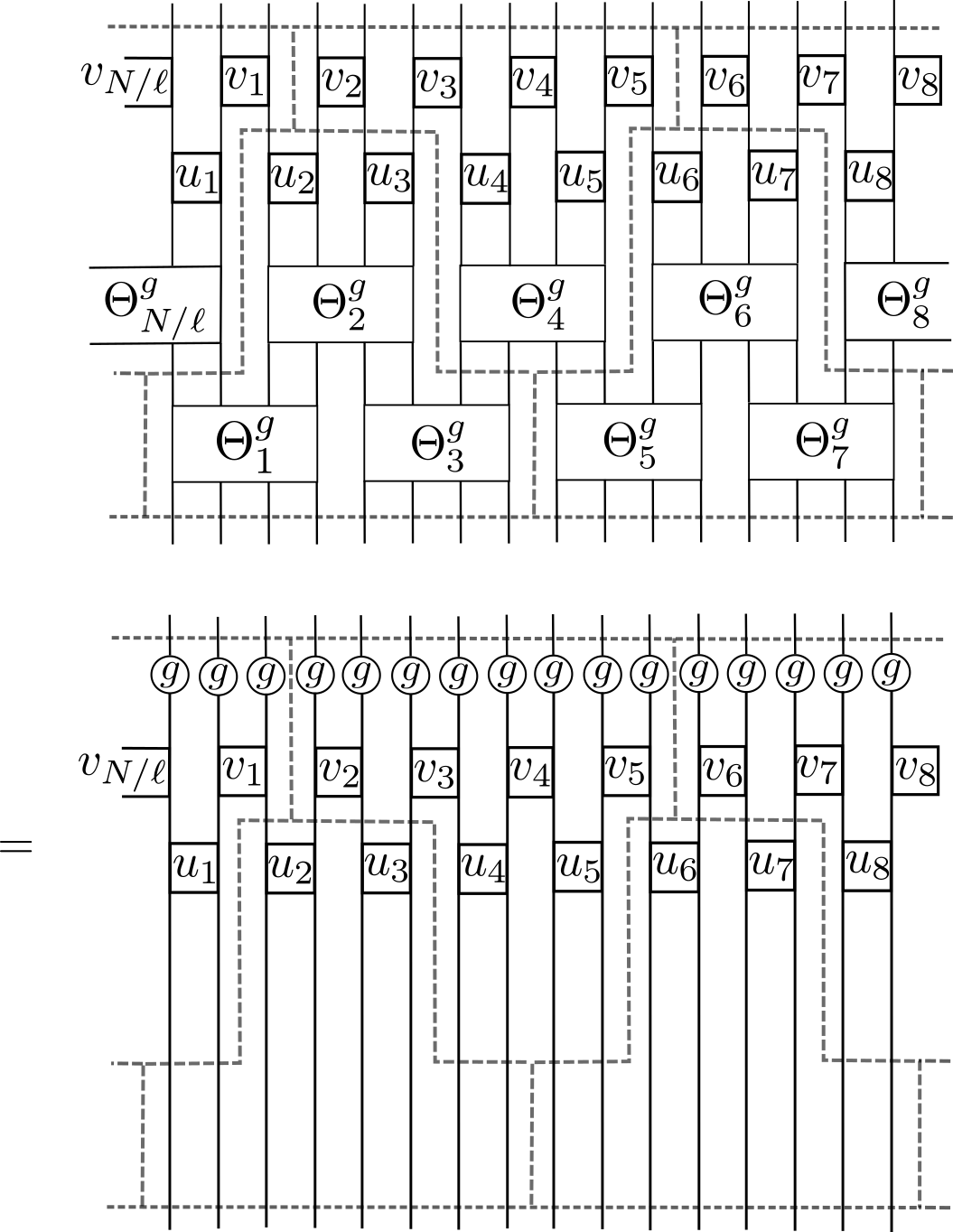}
 \end{aligned}   \label{eq:theta_equality}
\end{equation}
is equivalent to
\begin{equation}
\begin{aligned}
\includegraphics[width=0.43\textwidth]{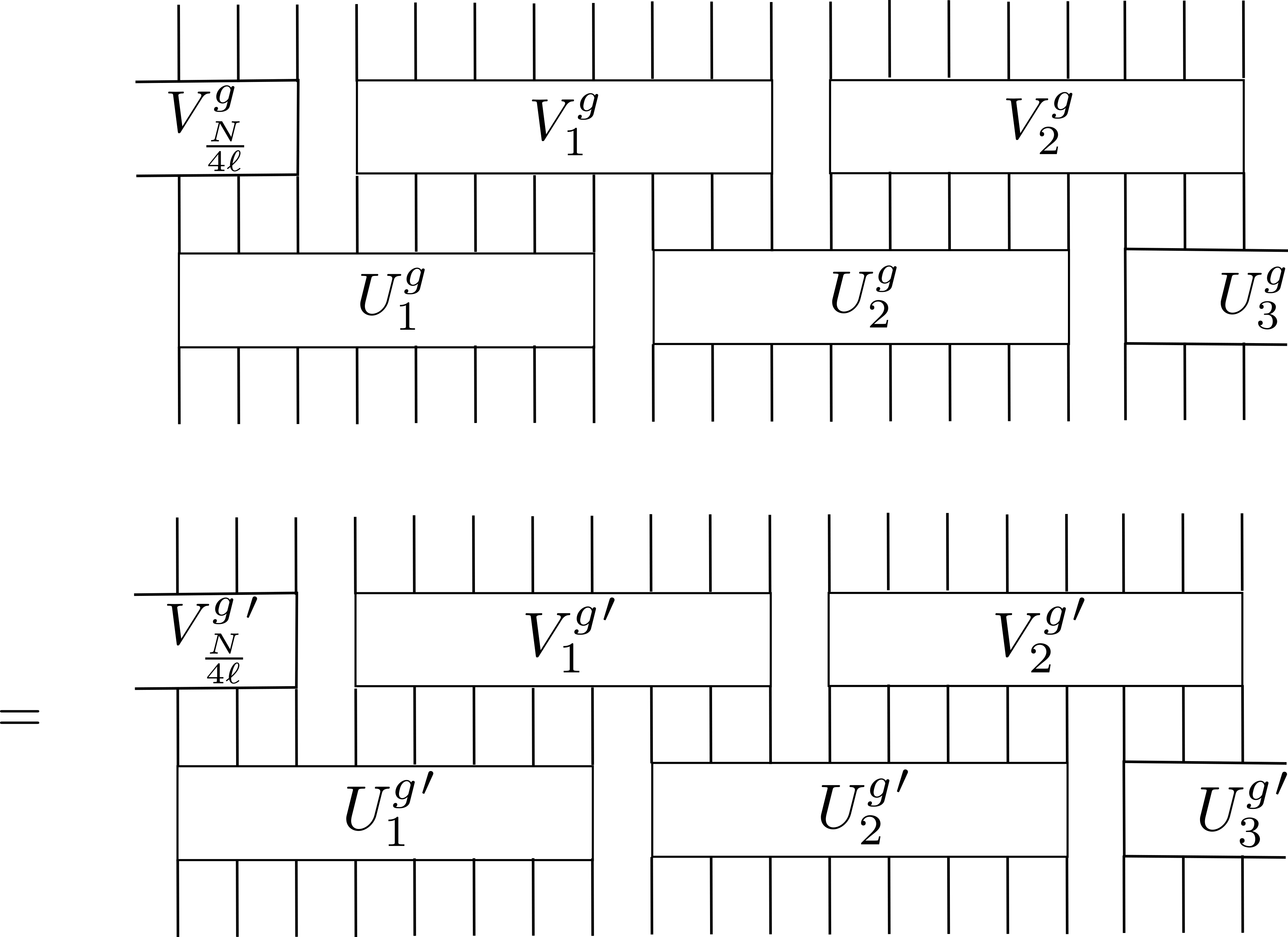}
 \end{aligned}   \label{eq:big_equality}
\end{equation}
if we define
\begin{equation}
\begin{aligned}
\includegraphics[width=0.45\textwidth]{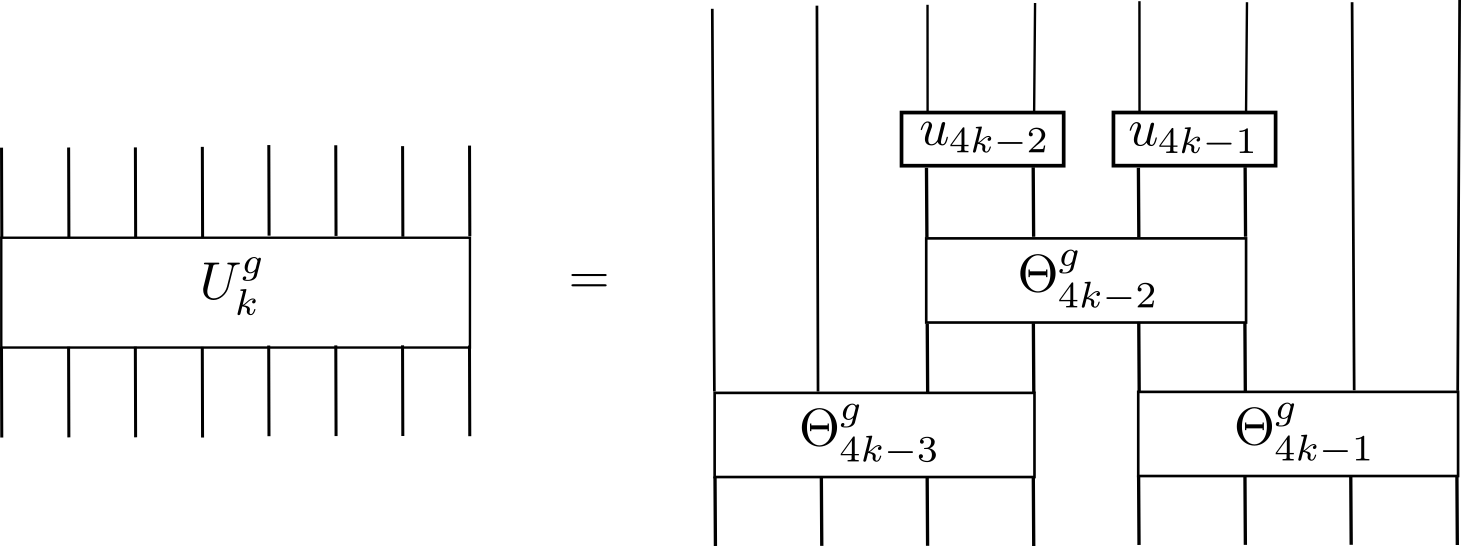}
 \end{aligned}  \  , \label{eq:Ugk}
\end{equation}
\begin{equation}
\begin{aligned}
\includegraphics[width=0.45\textwidth]{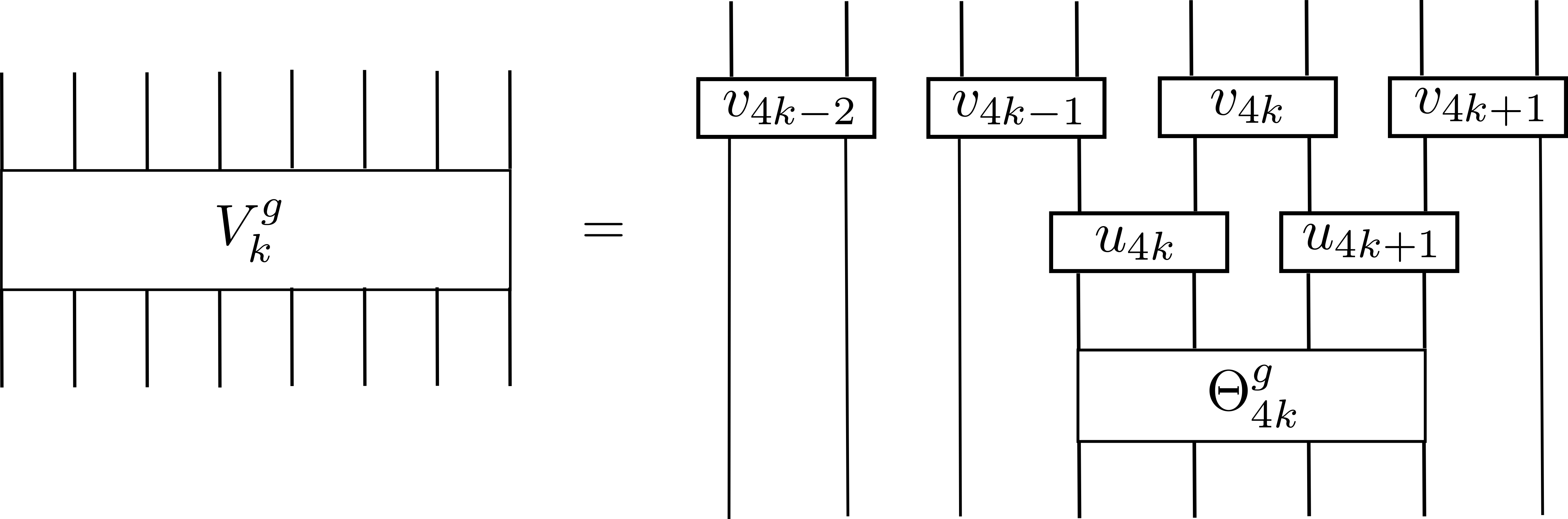}
 \end{aligned}  \  , \label{eq:Vgk}
\end{equation}
\begin{equation}
\begin{aligned}
\includegraphics[width=0.45\textwidth]{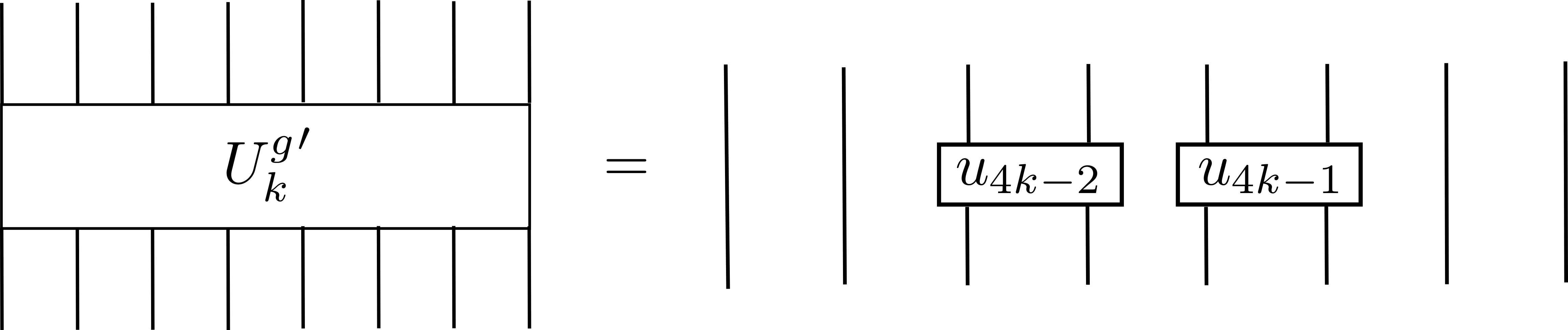}
 \end{aligned}  \  , \label{eq:Ugk2}
\end{equation}
and
\begin{equation}
\begin{aligned}
\includegraphics[width=0.45\textwidth]{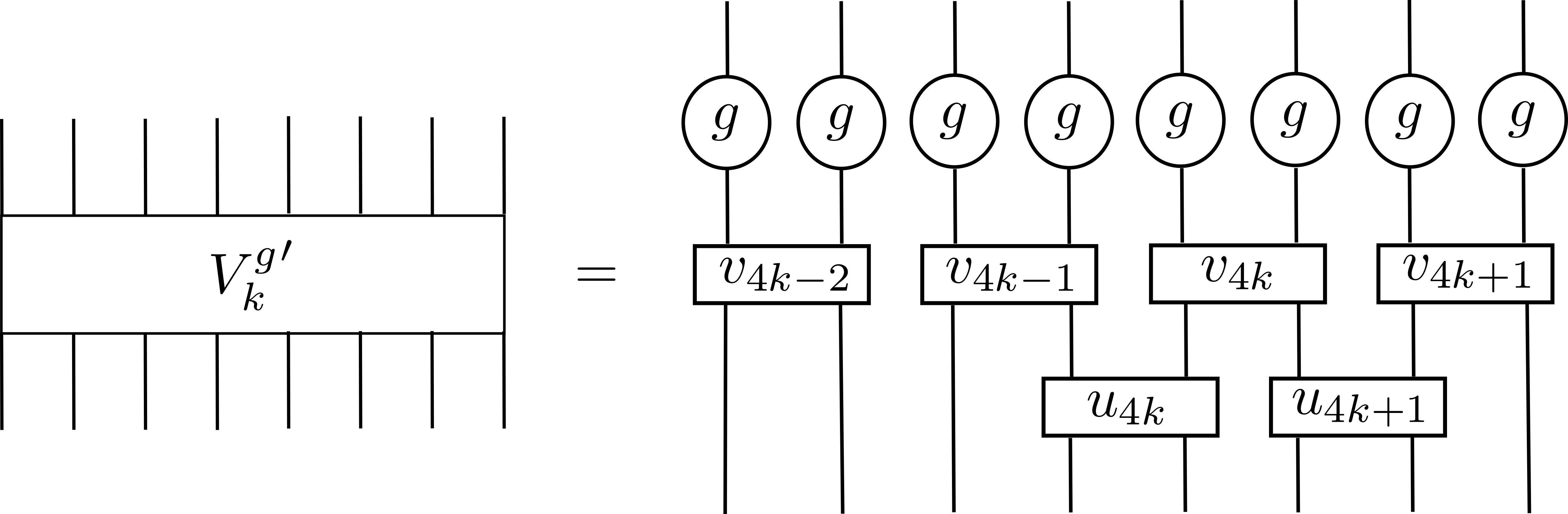}
 \end{aligned}  \ . \label{eq:Vgk2}
\end{equation}
The gauge transformations Eqs.~\eqref{eq:gauge1} and~\eqref{eq:gauge2} thus require
\begin{equation}
\begin{aligned}
\includegraphics[width=0.32\textwidth]{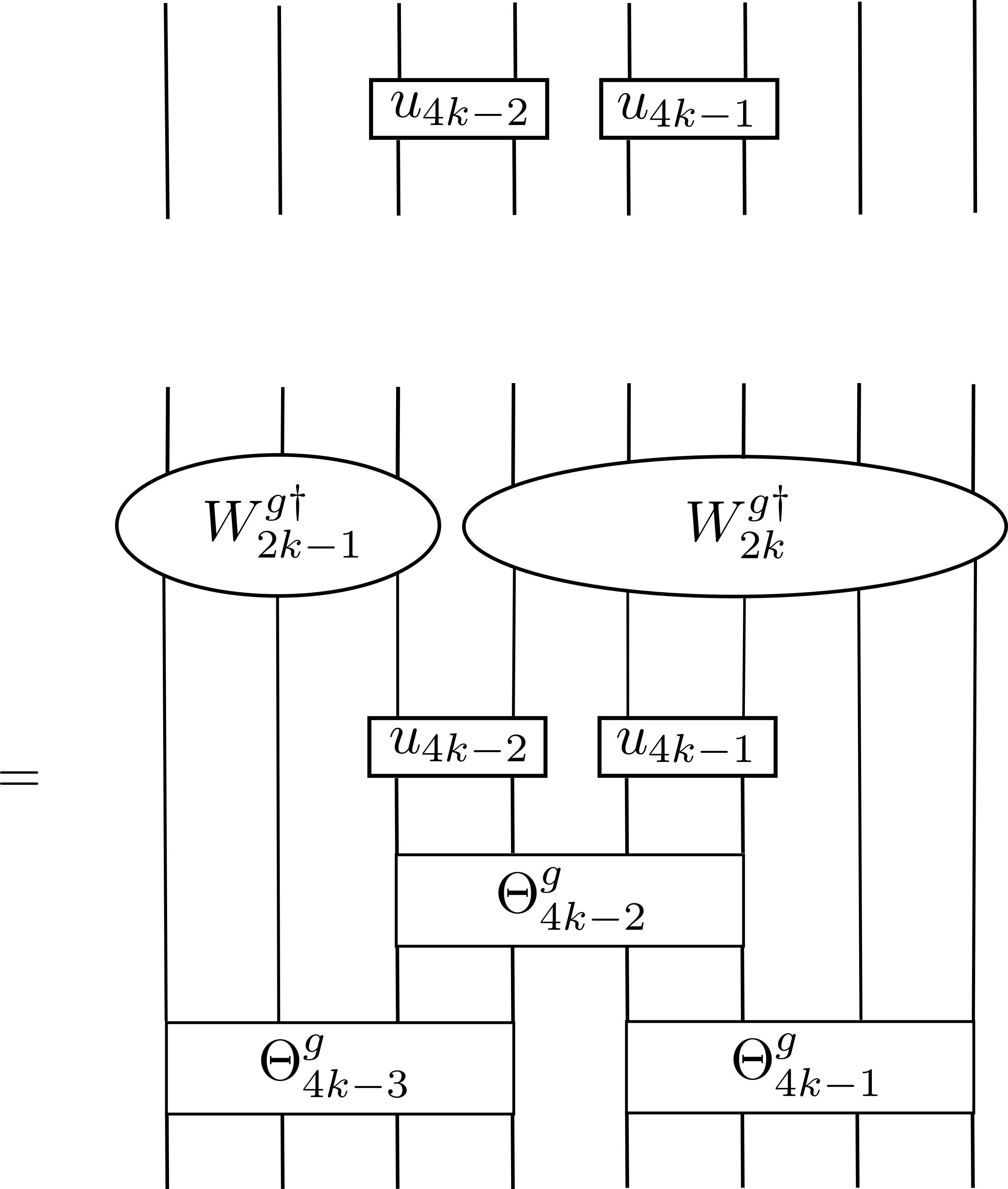}
 \end{aligned}  \label{eq:Ugk_trafo}
\end{equation}
and
\begin{equation}
\begin{aligned}
\includegraphics[width=0.38\textwidth]{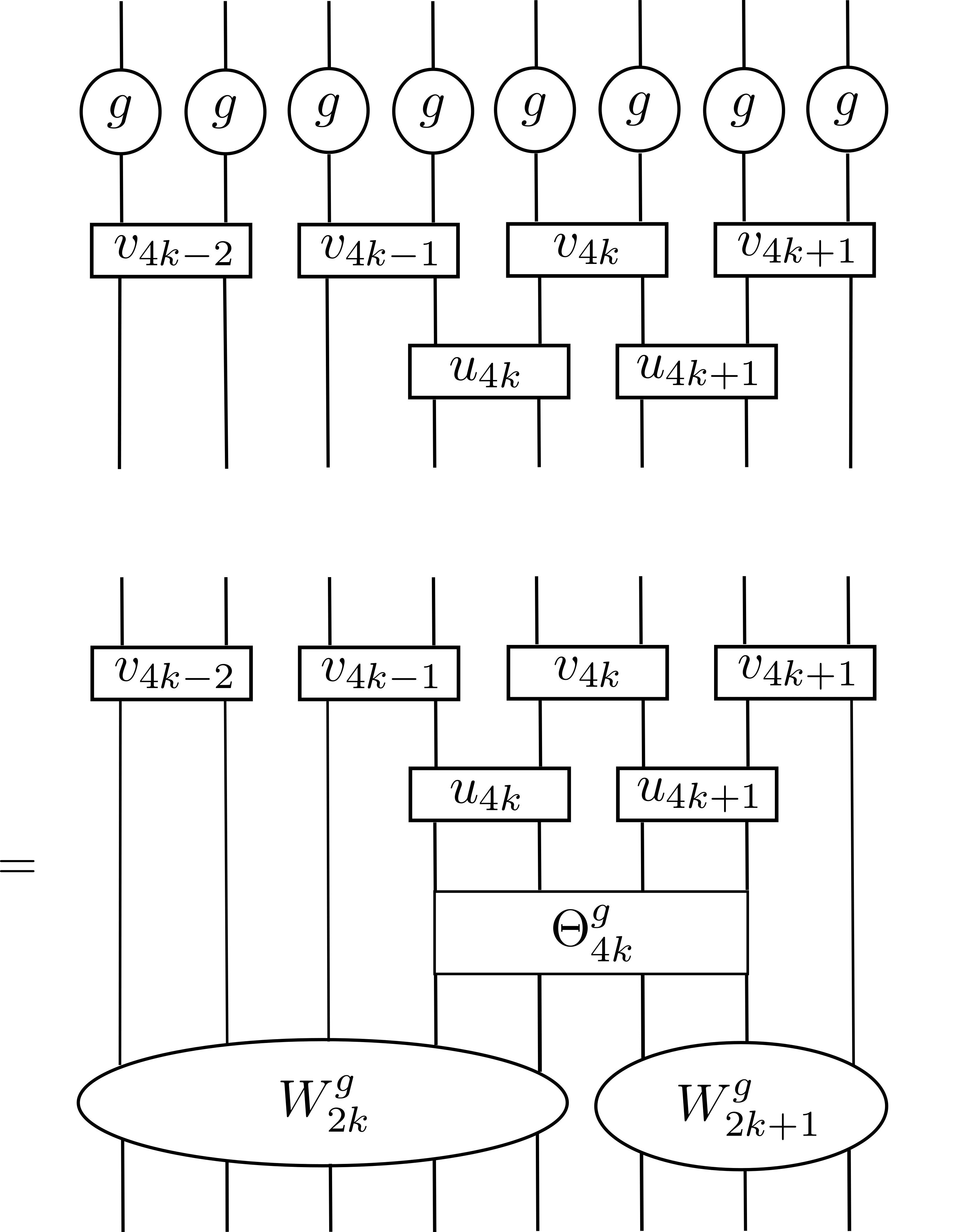}
 \end{aligned}  . \label{eq:Vgk_trafo}
\end{equation}
Eqs.~\eqref{eq:Ugk_trafo} and~\eqref{eq:Vgk_trafo} combined yield
\begin{equation}
\begin{aligned}
\includegraphics[width=0.45\textwidth]{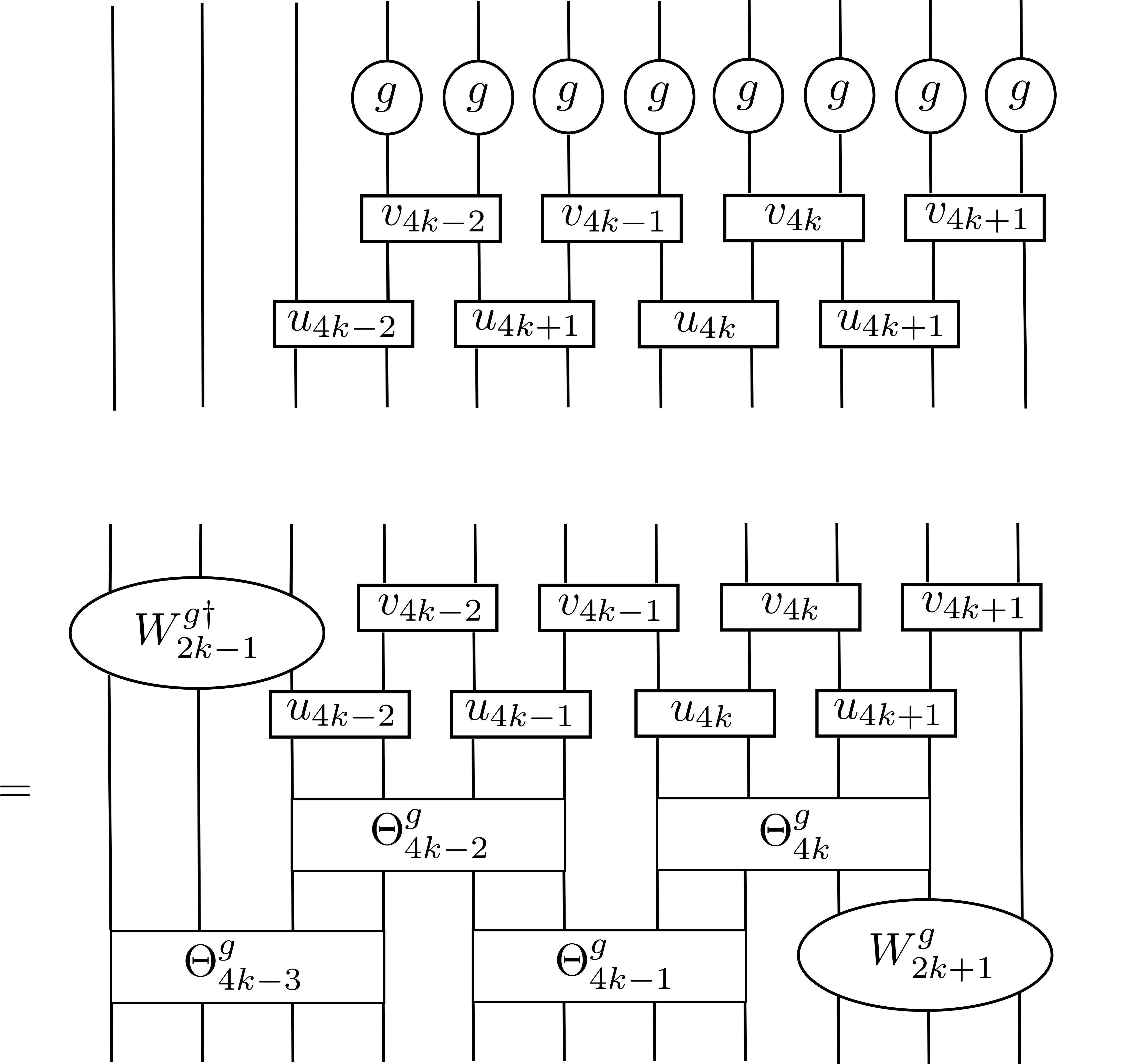}.
 \end{aligned}  \label{eq:combined_trafo}
\end{equation}

As can be seen from this equation, $W_{2k-1}^{g}$ is diagonal in the indices corresponding to its left two legs, i.e., diagrammatically
\begin{equation}
\begin{aligned}
\includegraphics[width=0.25\textwidth]{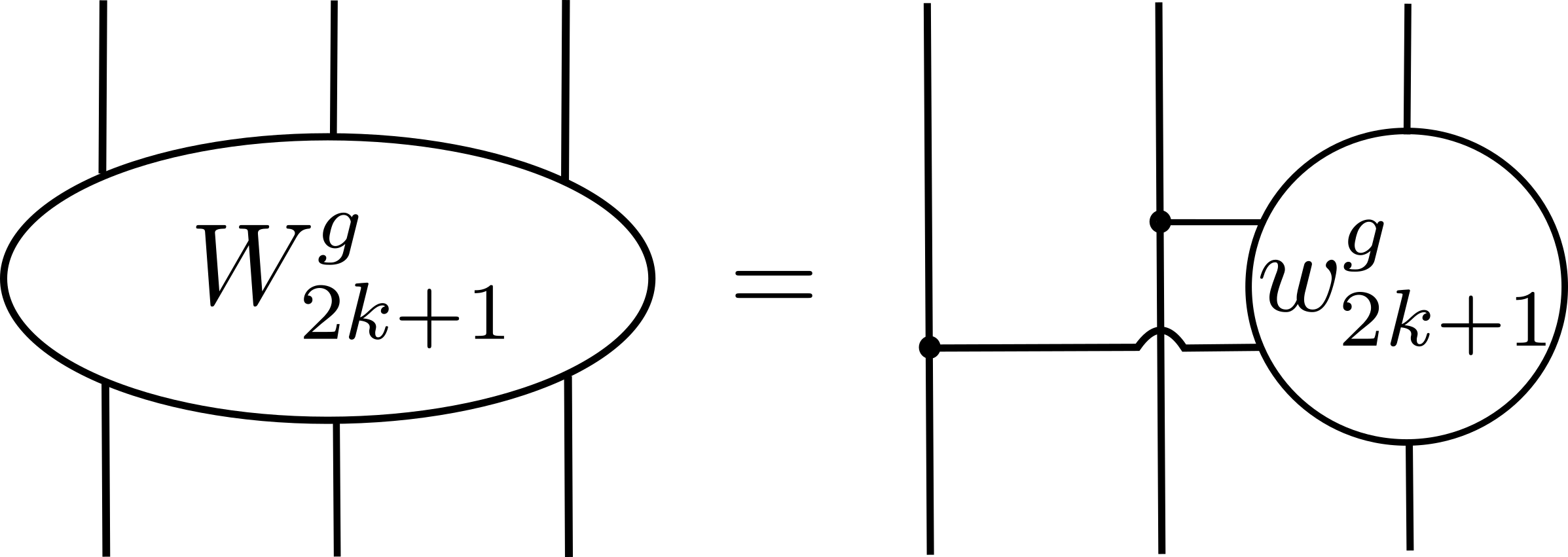}
\end{aligned} \ \  . \label{eq:w_diagonal}
\end{equation}
We denote by $[w^{g}_{2k+1}]_{L, L'}$ the matrix obtained when fixing the indices corresponding to the left two legs to $L$ and $L'$ (each corresponding to $\frac{\ell}{2}$ l-bits). 
By expressing  $W_{2k+1}^g W_{2k+1}^{g \dagger} = \mathbb{1}$ diagrammatically, one can easily check that for all $L, L'$ $[w^{g}_{2k+1}]_{L, L'}$  is also a unitary. Hence, if we fix the ten left lower indices in Eq.~\eqref{eq:combined_trafo} to $L_1, L_2, \ldots, L_{10}$, we obtain a relation similar to the one of matrix product states with the same symmetry (cf. also Eq.~\eqref{eq:fundeq}),
\begin{equation}
\begin{aligned}
\includegraphics[width=0.47\textwidth]{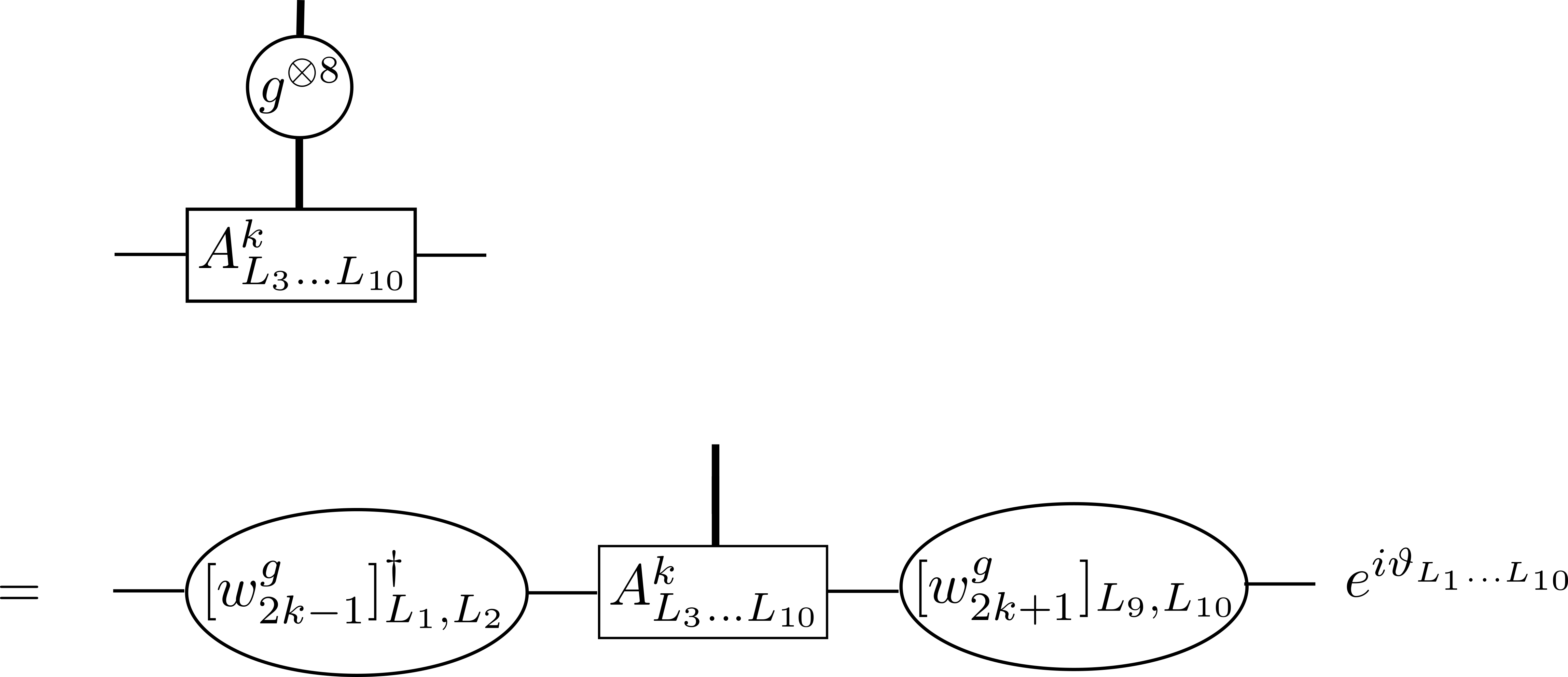},
\end{aligned}   \label{eq:symMPS}
\end{equation}
where the $A^k_{L_3 \ldots L_{10}}$ correspond to the concatenation of the unitaries $u_{4k-2}, v_{4k-2}, \ldots, v_{4k+1}$ and thick lines to eight thin ones, i.e., ${4 \ell}$ original legs. The $A^k_{L_3 \ldots L_{10}}$ are the tensors constituting the matrix product state representation of the eigenstate corresponding to that choice of $l$-bits.

We now consider Eq.~\eqref{eq:combined_trafo} for the group elements $g, h \in G$ and for the element $gh$: If one employs the fact that $\Theta_k^g$ is diagonal, one arrives at (using Eq.~\eqref{eq:w_diagonal}) 
\begin{equation}
\begin{aligned}
\includegraphics[width=0.47\textwidth]{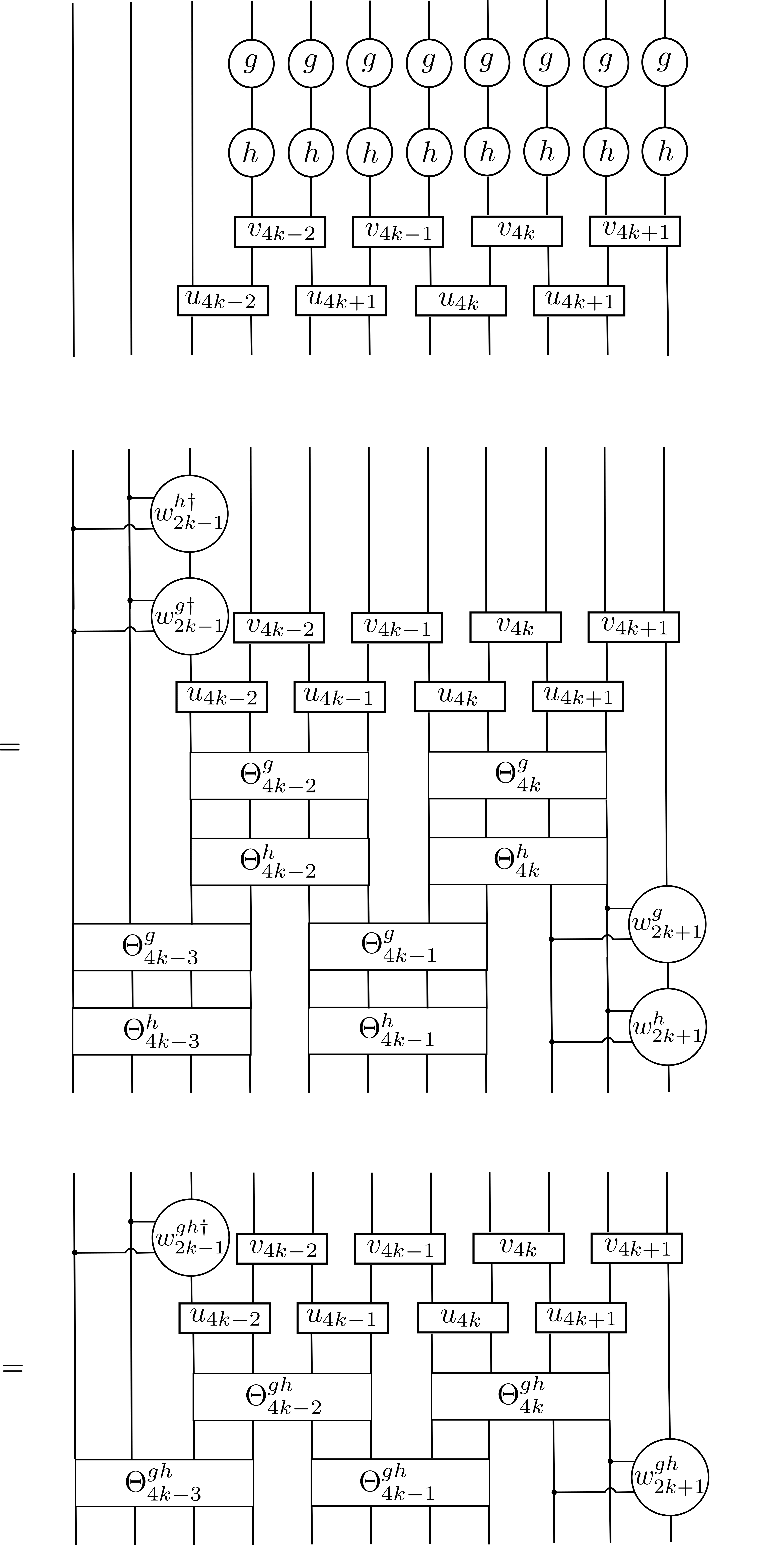} .
 \end{aligned}  \label{eq:gh_trafo}
\end{equation}
In order to unravel this expression, we analyse the relation between $\Theta_j^g$, $\Theta_j^h$ and $\Theta_j^{gh}$: Since $\mathpzc{v}_g$ is a linear representation of the group $G$, Eq.~\eqref{eq:theta} implies $\Theta_{gh} = \Theta_g \Theta_h$. 
If we use the representation of those matrices by two-layer quantum circuits~\eqref{eq:theta_quantum_circuit}, this implies
\begin{equation}
\begin{aligned}
\includegraphics[width=0.46\textwidth]{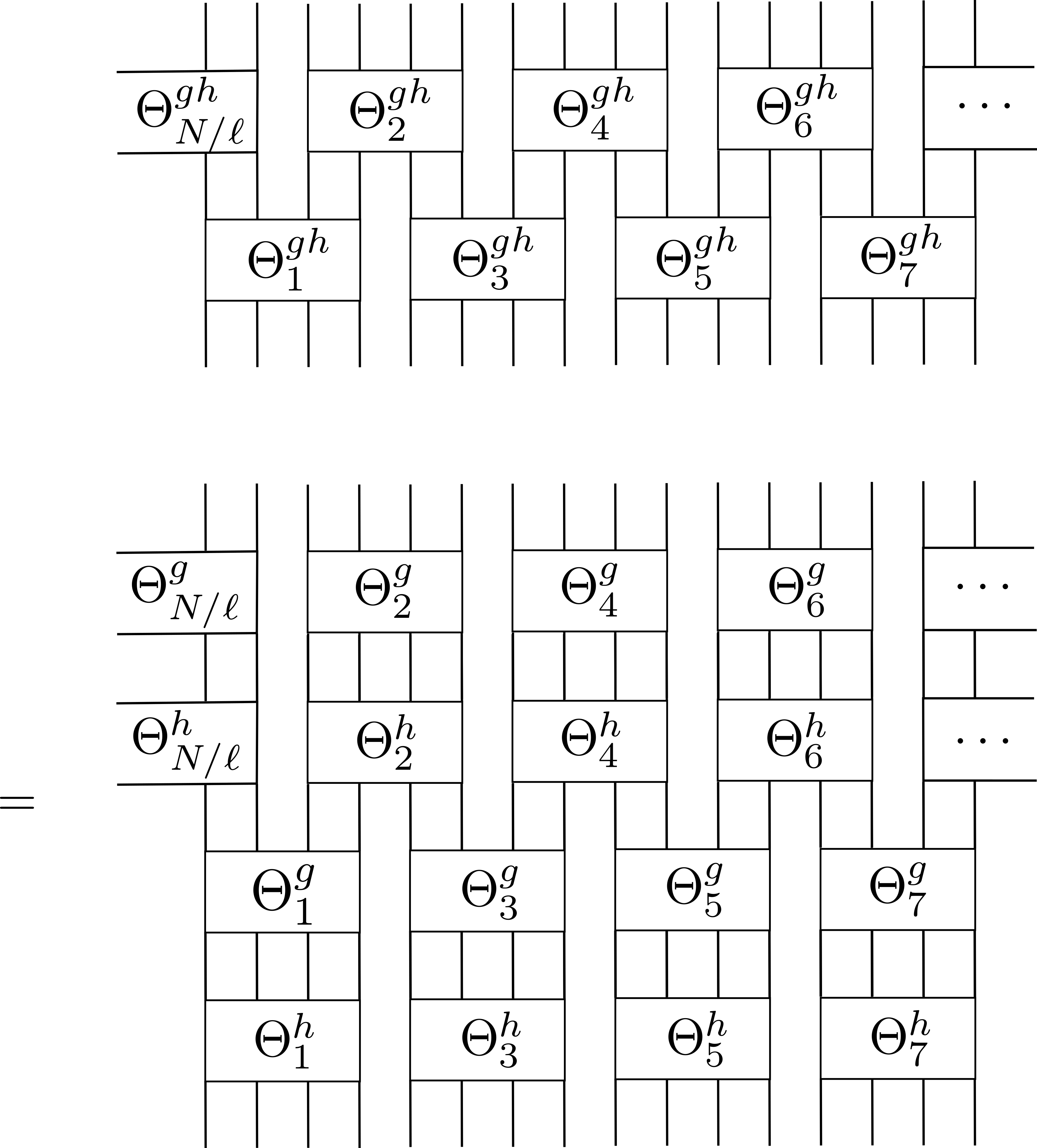} 
 \end{aligned}  \ . \label{eq:theta_gh}
\end{equation}
If one combines all $\Theta_j^g$ and $\Theta_j^h$, this is again an equality of two two-layer quantum circuits, i.e., Eqs.~\eqref{eq:gauge1} and~\eqref{eq:gauge2} apply,
\begin{equation}
\begin{aligned}
\includegraphics[width=0.35\textwidth]{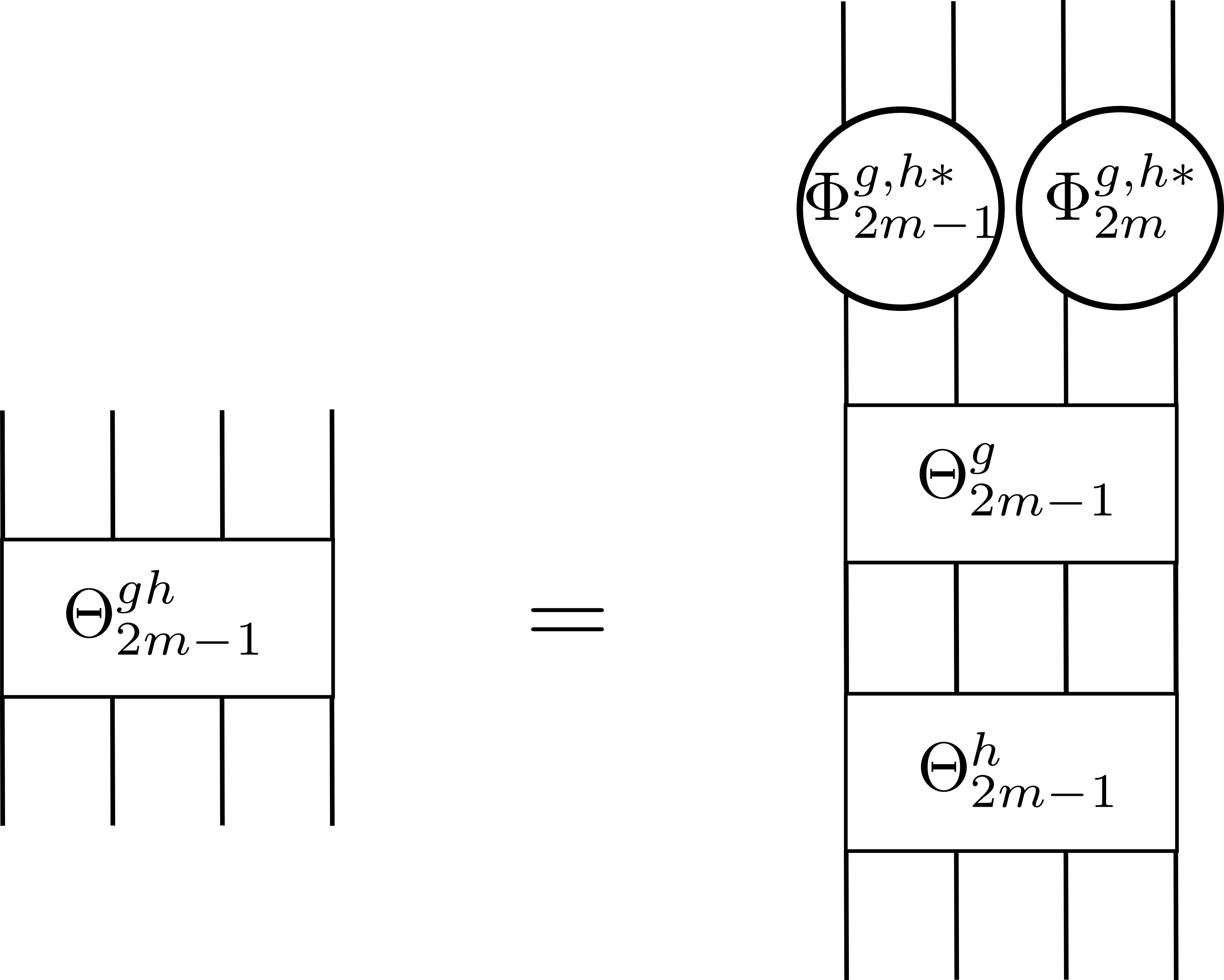} 
 \end{aligned}  \ , \label{eq:gauge_theta1}
\end{equation}
\begin{equation}
\begin{aligned}
\includegraphics[width=0.4\textwidth]{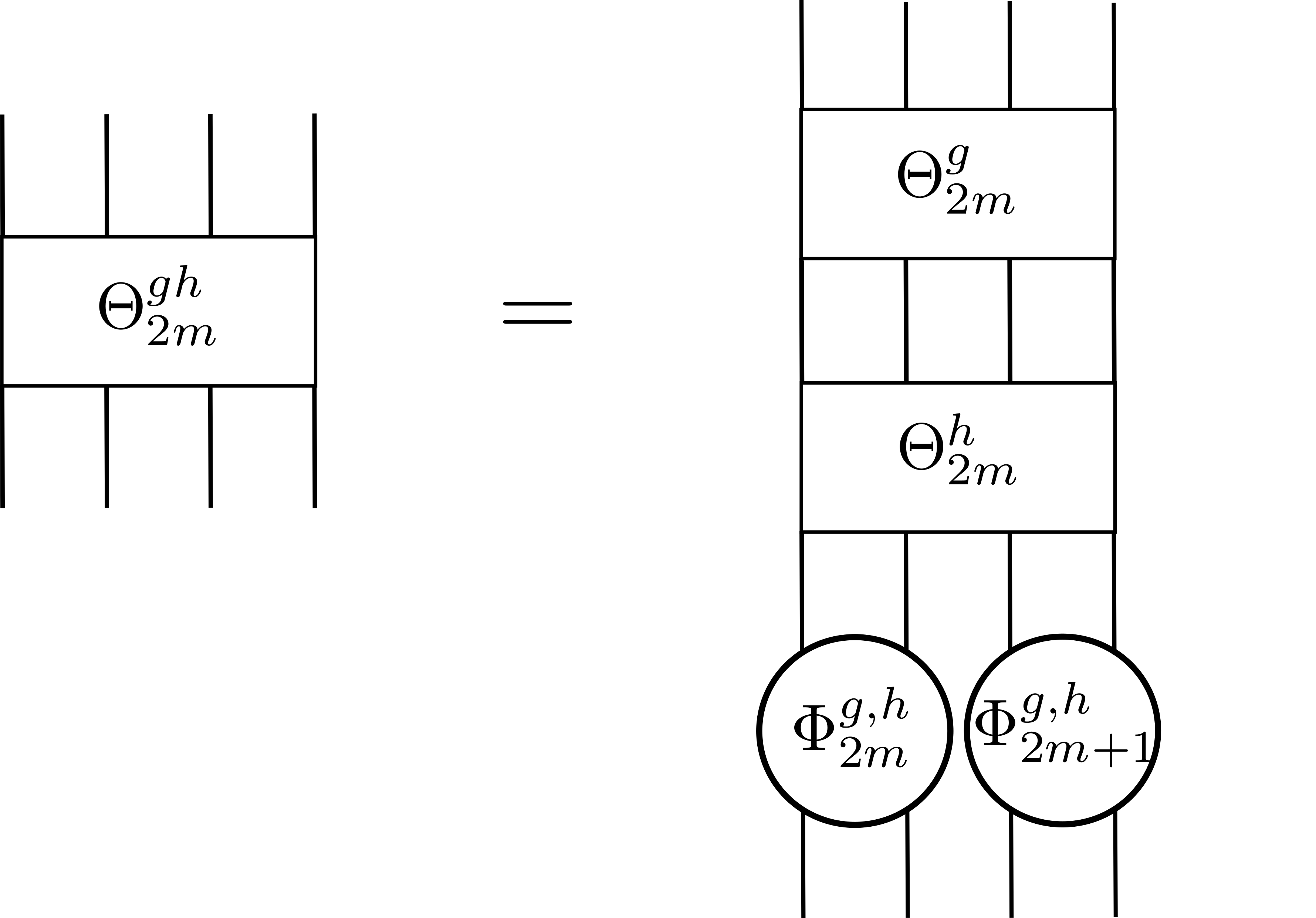} 
 \end{aligned}  \ . \label{eq:gauge_theta2}
\end{equation}
Note that the diagonal unitaries $\Phi_j^{g,h}$ depend on $g$ and $h$ individually. Eqs.~\eqref{eq:gauge_theta1} and~\eqref{eq:gauge_theta2} combined imply
\begin{equation}
\begin{aligned}
\includegraphics[width=0.47\textwidth]{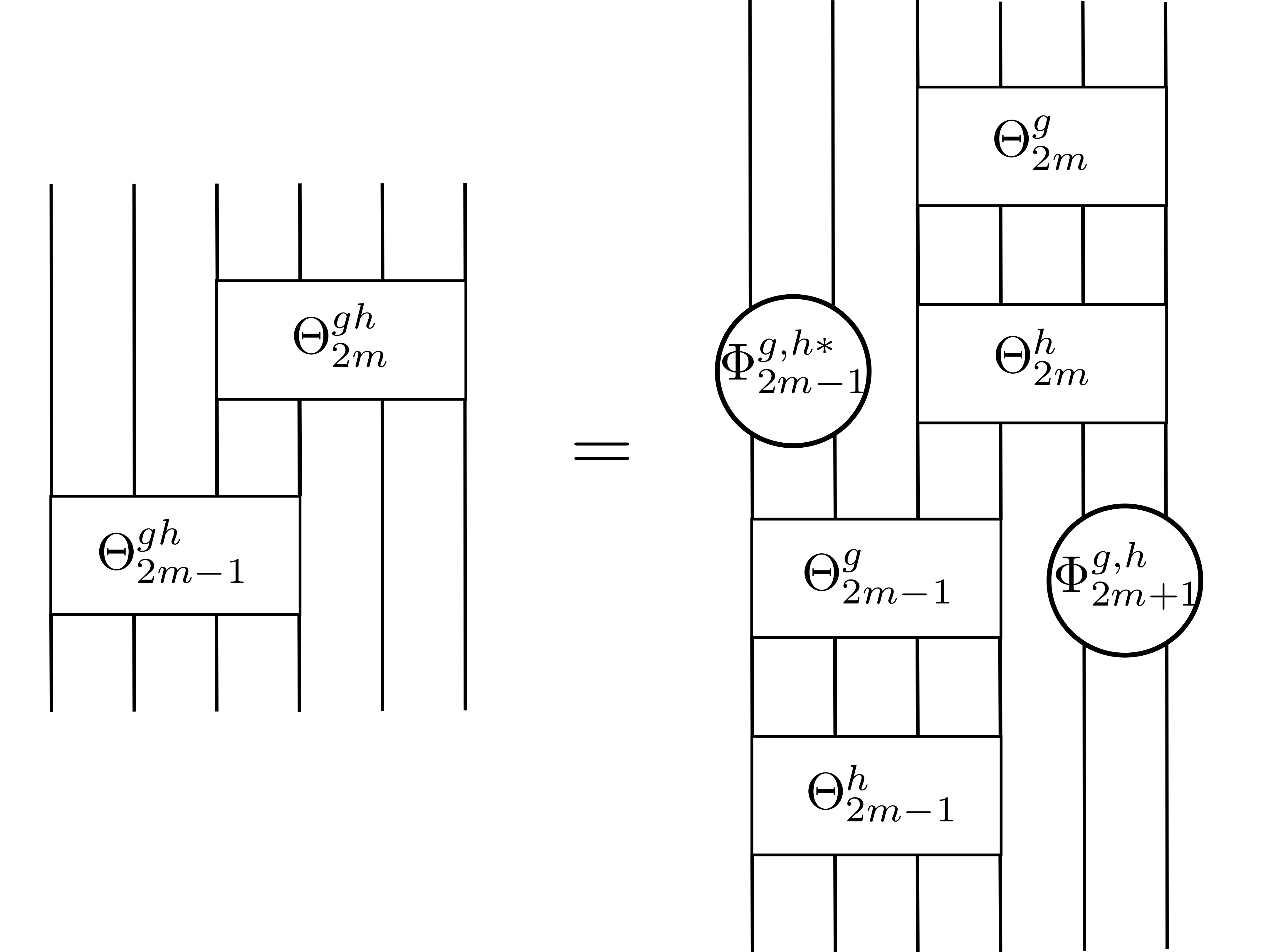} 
 \end{aligned}  . \label{eq:theta_theta}
\end{equation}
We insert this into Eq.~\eqref{eq:gh_trafo} and obtain
\begin{equation}
\begin{aligned}
\includegraphics[width=0.3\textwidth]{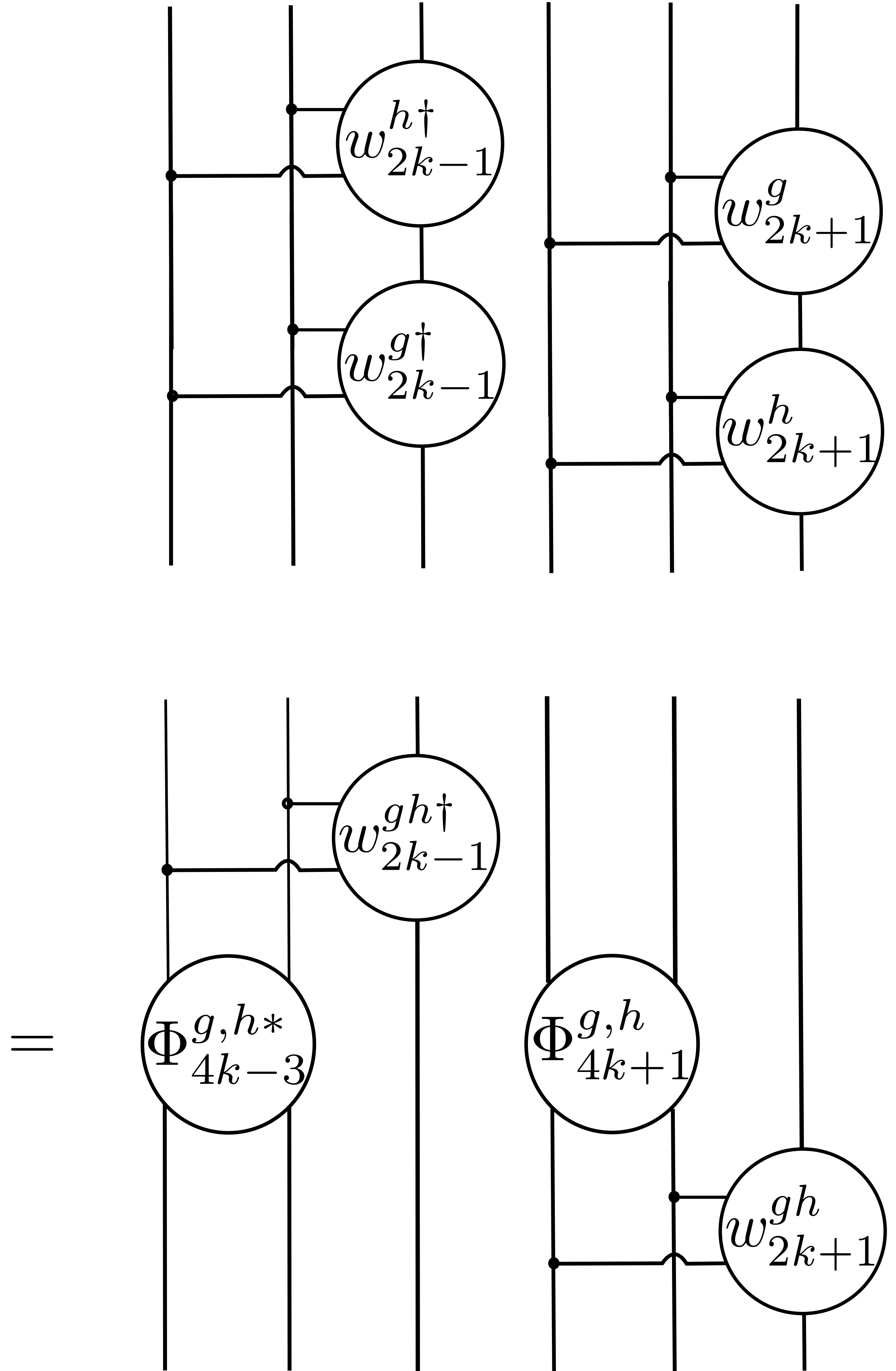} .
 \end{aligned}   \label{eq:w_connection}
\end{equation}
Next, we show that one can choose $\Theta_j^g$ in such a way that the $\Phi_j^{g,h}$ are proportional to the identity, i.e., they give rise only to an overall phase factor: We define $\tilde \Theta_j^g$ such that they also fulfill Eq.~\eqref{eq:theta_quantum_circuit} via
\begin{equation}
\begin{aligned}
\includegraphics[width=0.3\textwidth]{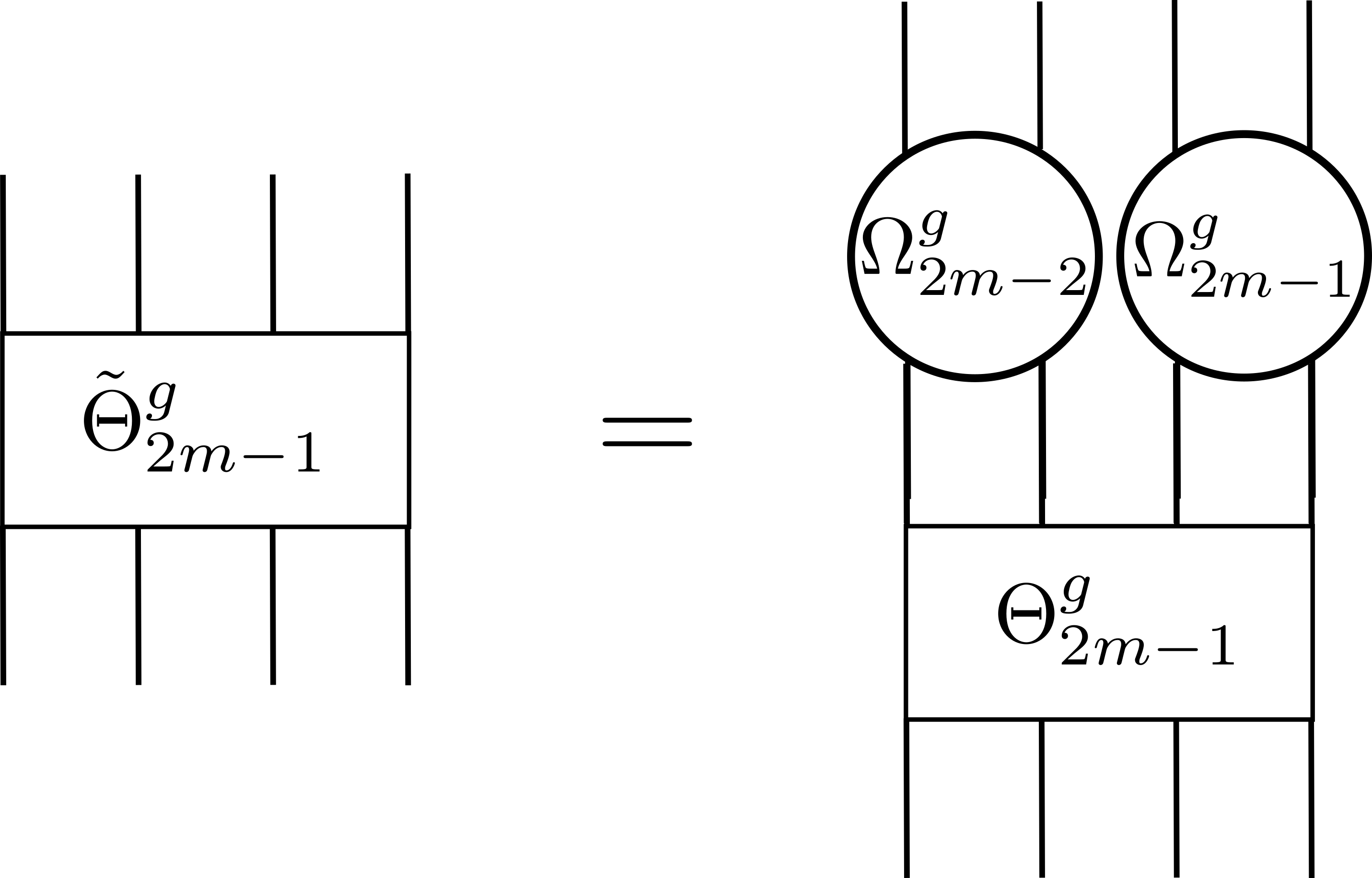} 
 \end{aligned}   \label{eq:redefine_theta1}
\end{equation}
and
\begin{equation}
\begin{aligned}
\includegraphics[width=0.3\textwidth]{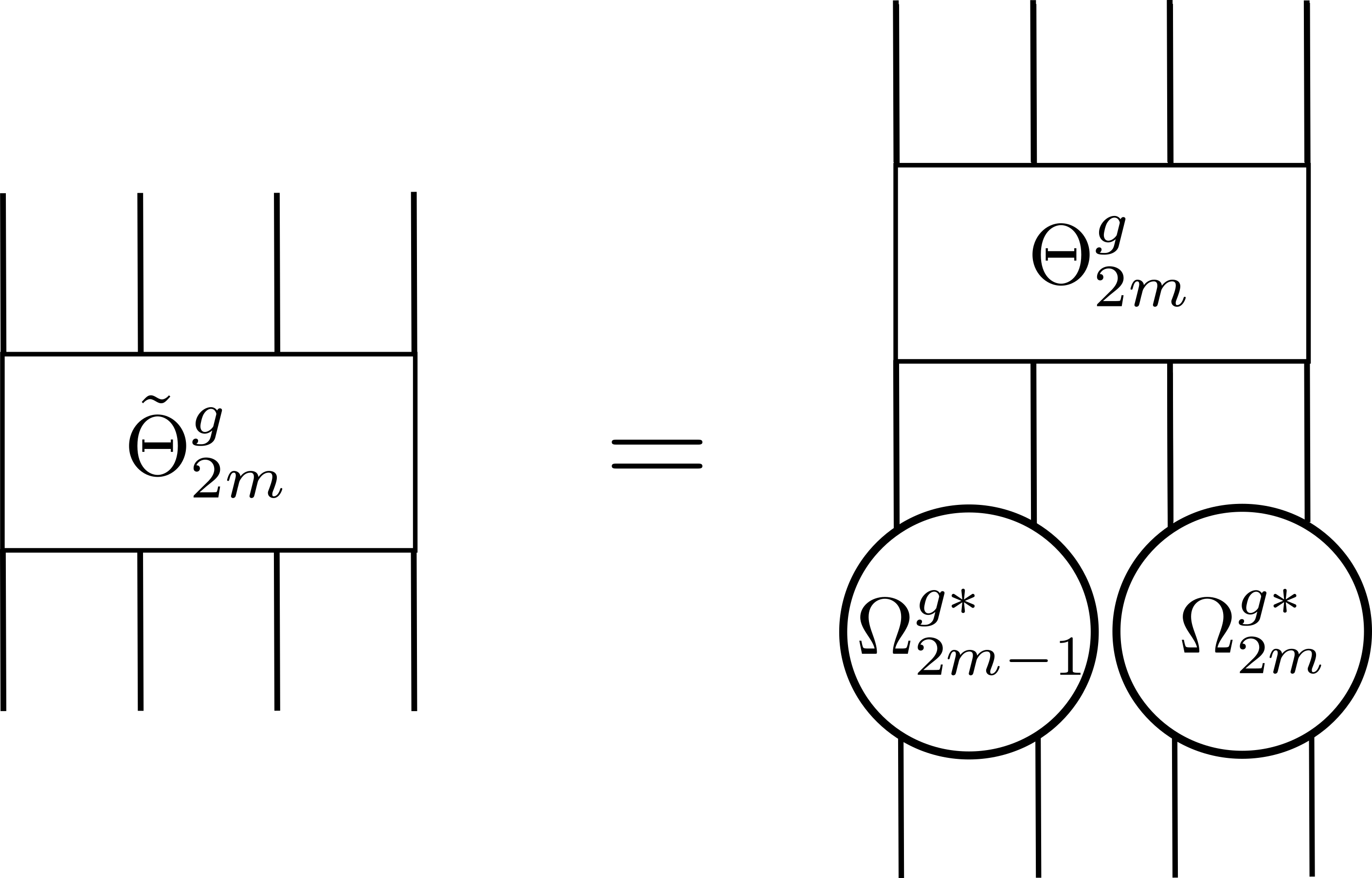} 
 \end{aligned}    \label{eq:redefine_theta2}
\end{equation}
with diagonal matrices $\Omega_j^g$ (whose diagonal elements are also of magnitude 1), which can be chosen arbitrarily. We start by choosing $\Omega^g_{N/\ell}$ and $\Omega_1^g$ as follows
\begin{equation}
\begin{aligned}
\includegraphics[width=0.25\textwidth]{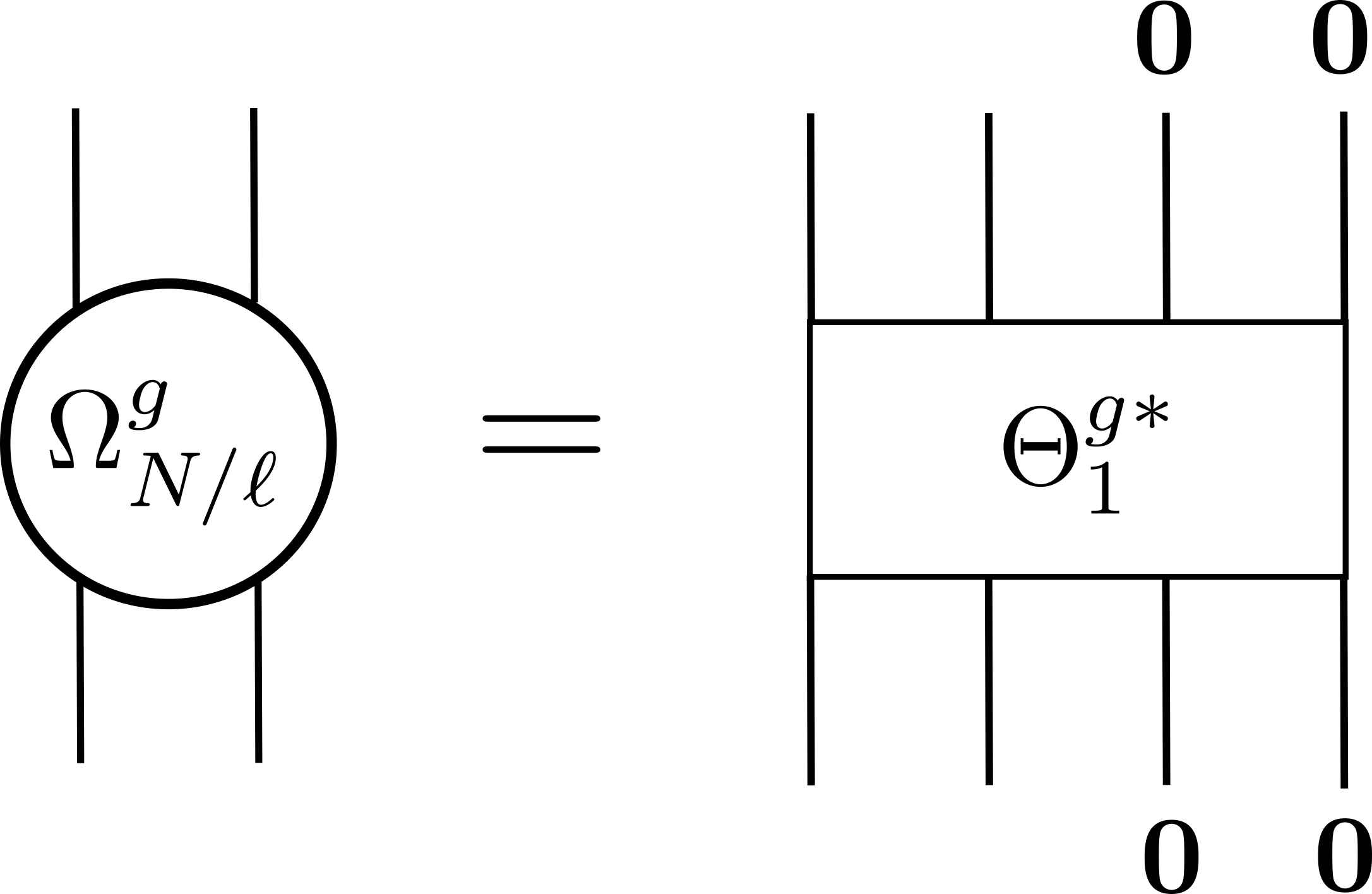} 
 \end{aligned},    \label{eq:OmegaN}
\end{equation}
\begin{equation}
\begin{aligned}
\includegraphics[width=0.34\textwidth]{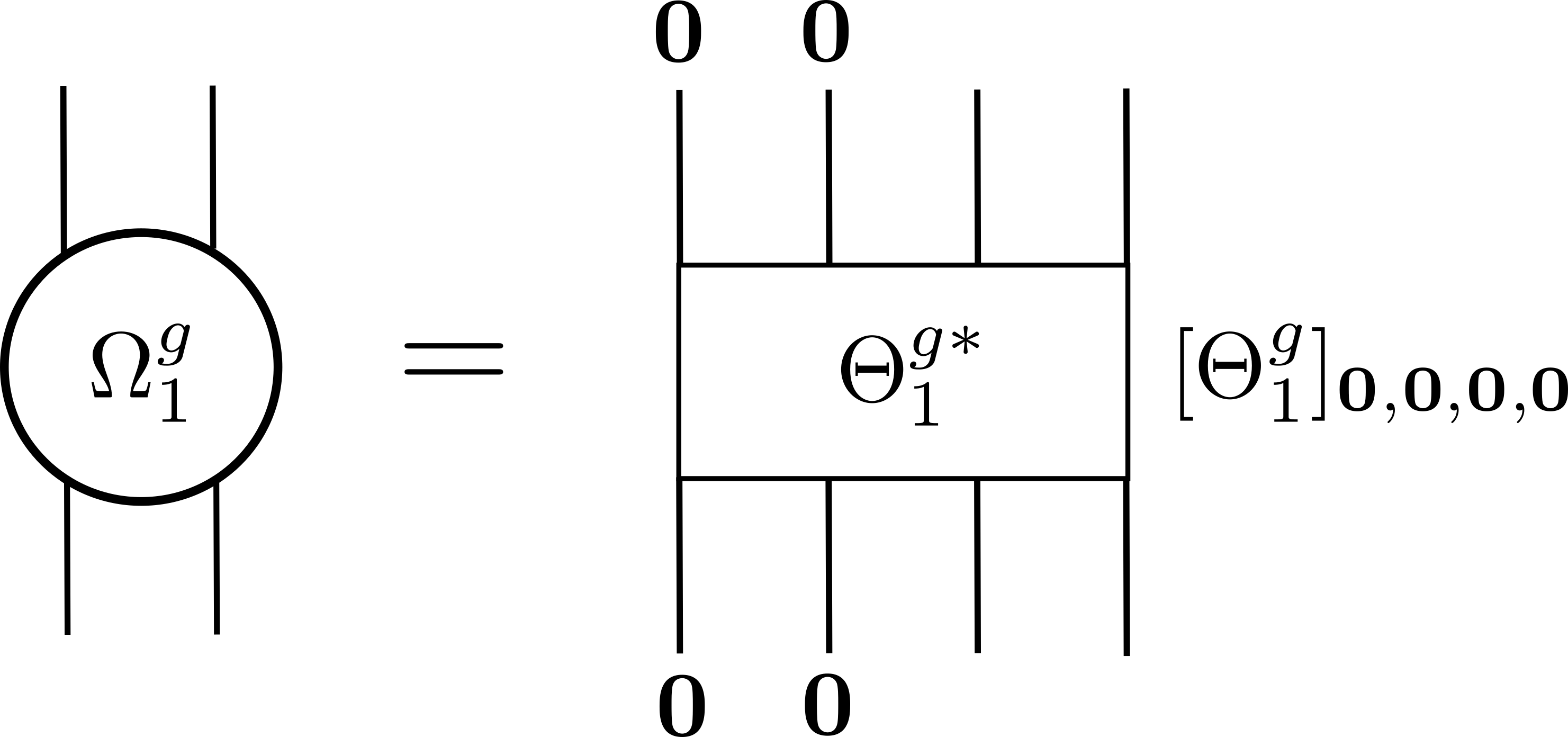} 
 \end{aligned} ,   \label{eq:Omega1}
\end{equation}
where $[\Theta_1^g]_{\mb 0, \mb 0, \mb 0, \mb 0}$ refers to the matrix element for all l-bit indices set to zero. Those two equations imply
\begin{equation}
\begin{aligned}
\includegraphics[width=0.42\textwidth]{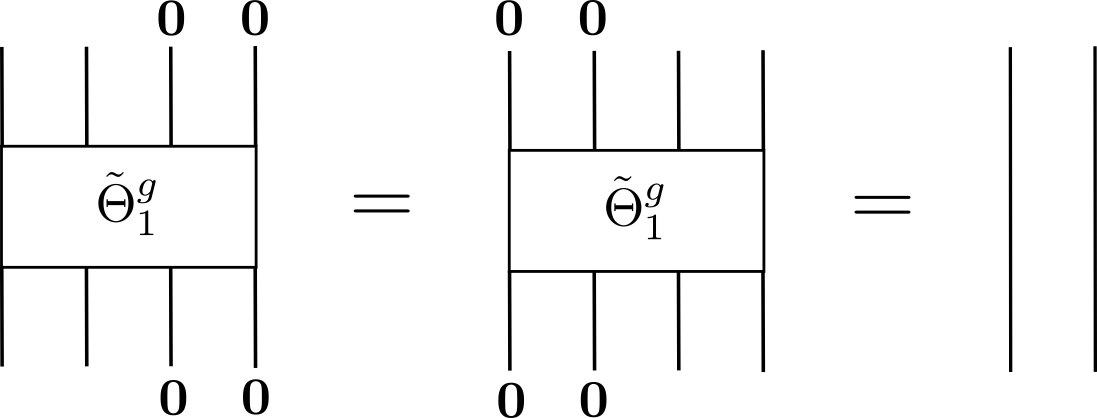} 
 \end{aligned} \ .   \label{eq:tilde_Omega1}
\end{equation}
We now proceed by consecutively fixing $\Omega_2^g$, $\Omega_3^g$, \ldots, $\Omega_{N/\ell-1}^g$ in such a way that
\begin{equation}
\begin{aligned}
\includegraphics[width=0.23\textwidth]{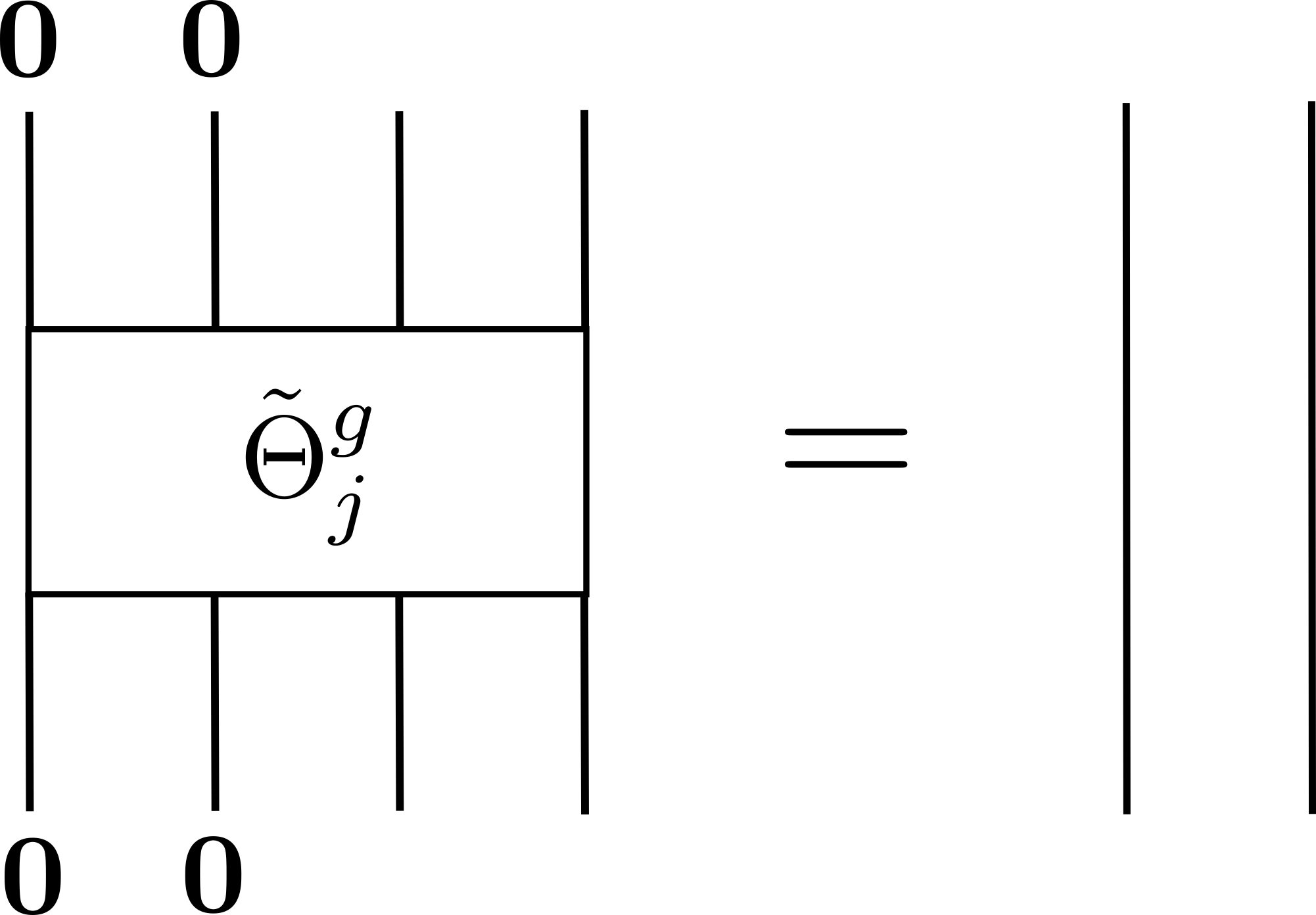} 
 \end{aligned}    \label{eq:tilde_Omegak}
\end{equation}
for all $j = 2, 3, \ldots, \frac{N}{\ell}-1$. According to Eq.~\eqref{eq:tilde_Omega1}, Eq.~\eqref{eq:gauge_theta1} yields by setting the indices of the left two or right two legs to $\mb 0$
\begin{equation}
\begin{aligned}
\includegraphics[width=0.23\textwidth]{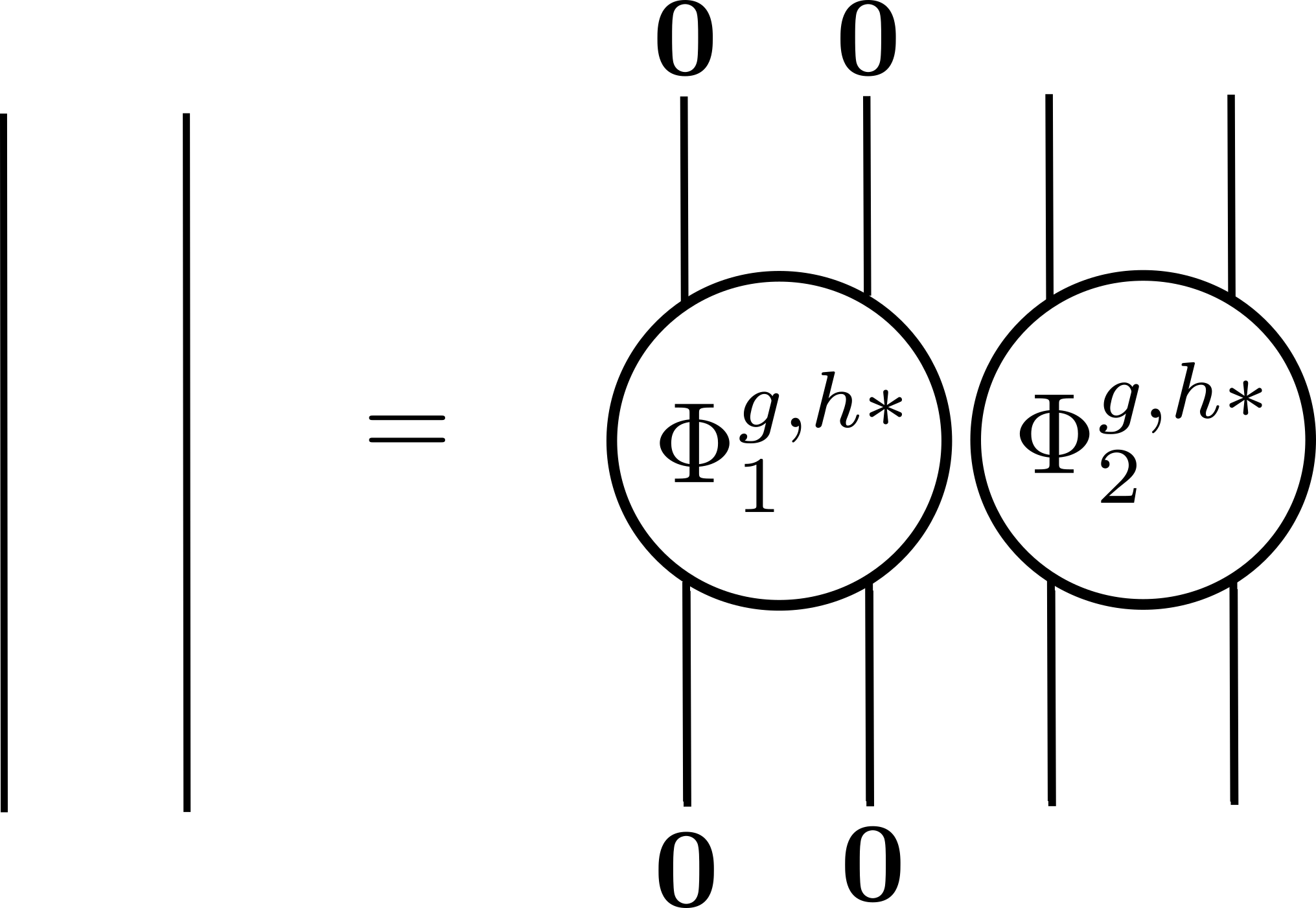} 
 \end{aligned}  \label{eq:Phi1}
\end{equation}
and
\begin{equation}
\begin{aligned}
\includegraphics[width=0.23\textwidth]{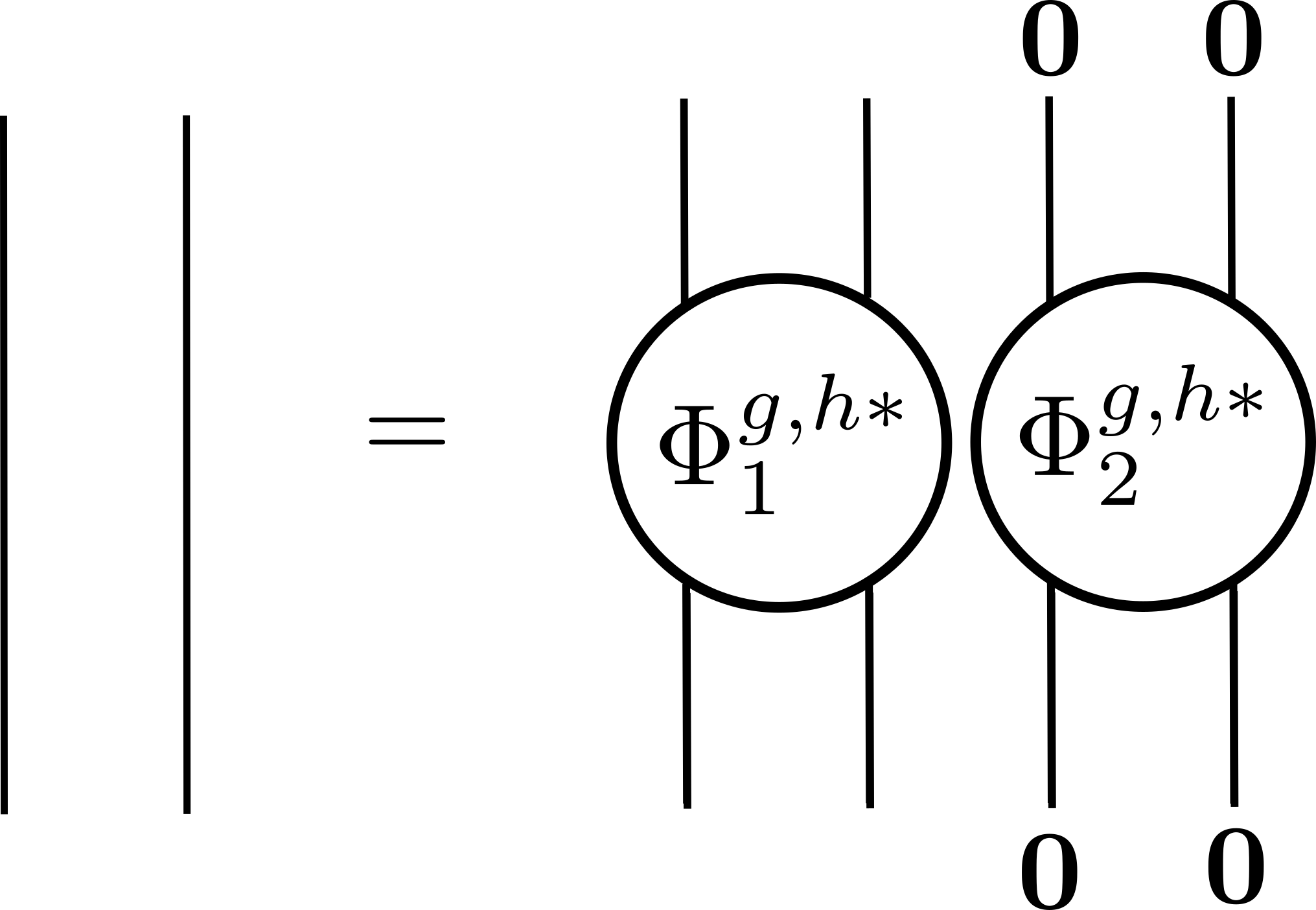}
 \end{aligned} \  .  \label{eq:Phi2}
\end{equation}
(The $\Phi_j^{g,h}$ here refer to $\tilde \Theta_j^g$, but we do not decorate them with tildes for simplicity of notation.) We thus have $\Phi_2^{g,h} = \mathbb{1} [\Phi_{1}^{g,h}]_{\mb 0, \mb 0}^*$ and $\Phi_1^{g,h} = \mathbb{1} [\Phi_2^{g,h}]^*_{\mb 0, \mb 0}$. Hence, 
\begin{align}
\Phi_1^{g,h} &= \mathbb{1}  e^{-i \alpha^{g,h}},\\
\Phi_2^{g,h} &= \mathbb{1}  e^{i \alpha^{g,h}}
\end{align} 
with $\alpha^{g,h} \in [0, 2 \pi)$. Similarly, Eq.~\eqref{eq:tilde_Omegak} implies using Eqs.~\eqref{eq:gauge_theta1} and~\eqref{eq:gauge_theta2} with the indices of the left two legs set to $\mb 0$ that
\begin{align}
\mathbb{1} &= [\Phi^{g,h*}_{2m-1}]_{\mb 0, \mb 0} \, \Phi^{g,h *}_{2m}, \\
\mathbb{1} &= [\Phi^{g,h}_{2m}]_{\mb 0, \mb 0} \, \Phi_{2m+1}^{g,h}
\end{align}
for $m < \frac{N}{2\ell}$, i.e.,
\begin{align}
\Phi_{2m}^{g,h} = \mathbb{1} e^{i \alpha^{g,h}}, \\
\Phi_{2m+1}^{g,h} = \mathbb{1} e^{-i \alpha^{g,h}}.
\end{align}
Hence, $\Phi_{N/\ell}^{g,h}$ is the only such matrix which might not be proportional to the identity. However, in Eq.~\eqref{eq:w_connection} we are only interested in the tensor product
\begin{align}
\Phi_{4k-3}^{g,h *} \otimes \Phi_{4k+1}^{g,h} = \Phi_{N/\ell-3}^{g,h *} \otimes \Phi_1^{g,h} = \mathbb{1} \label{eq:Phi_tensor}
\end{align}
for $1 \leq k < \frac{N}{4 \ell}$. 
Eq.~\eqref{eq:Phi_tensor} inserted into Eq.~\eqref{eq:w_connection} thus implies for all $k$
\begin{equation}
\begin{aligned}
\includegraphics[width=0.3\textwidth]{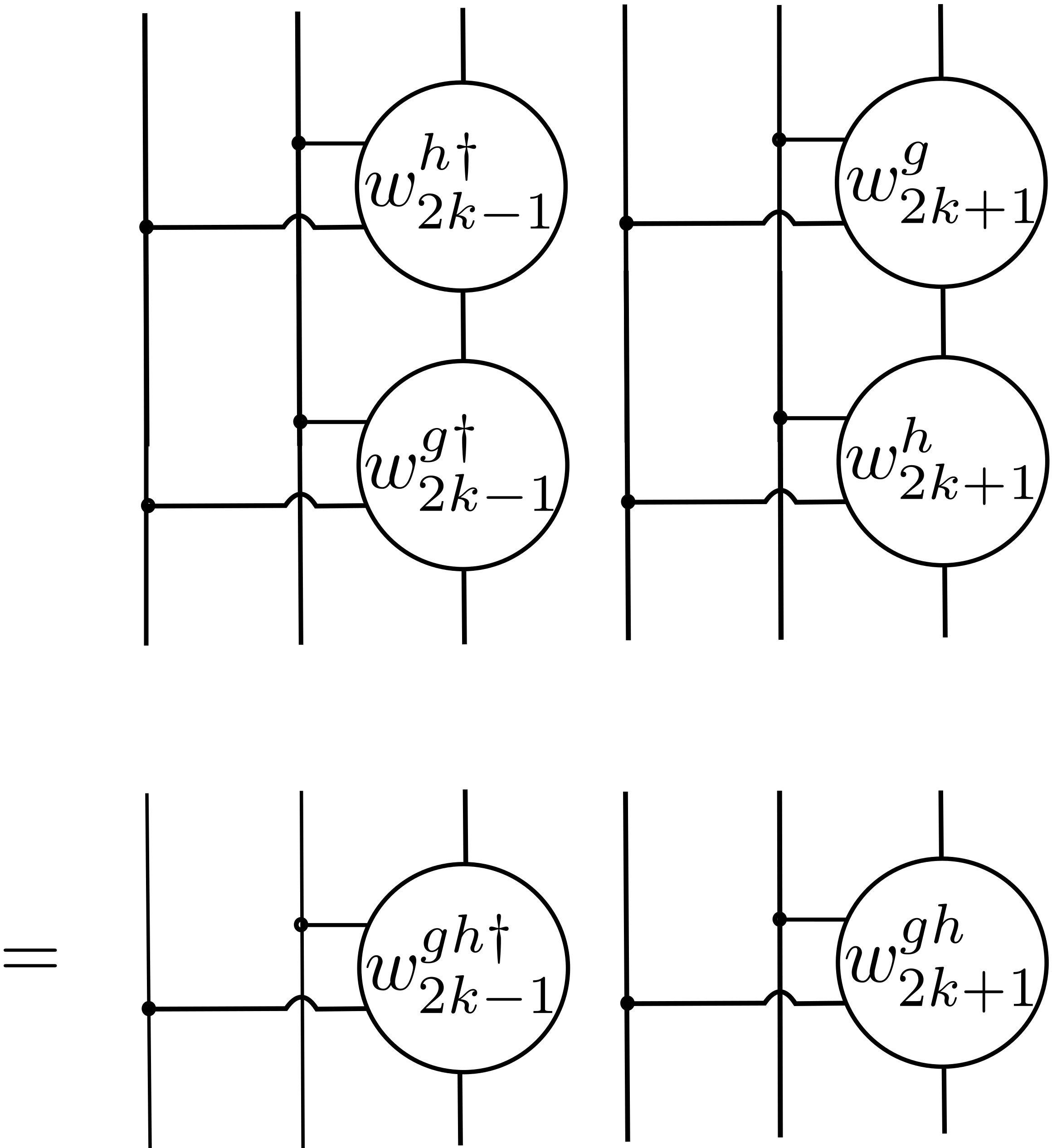}
 \end{aligned} \  .  \label{eq:w_cohomology}
\end{equation}
If we now fix the indices corresponding to the first two legs from the left to $L_1$ and $L_2$ and of the fourth and fifth legs to $L_4$ and $L_5$, this equation reads 
\begin{align}\label{eq:fixlegs}
&\left([w^{h}_{2k-1}]_{L_1, L_2}^\dagger [w^{g}_{2k-1}]_{L_1, L_2}^\dagger\right) \otimes \left([w^{g}_{2k+1}]_{L_4, L_5} [w^{h}_{2k+1}]_{L_4, L_5}\right)  \notag \\
&= [w_{2k-1}^{gh}]_{L_1,L_2}^\dagger \otimes [w_{2k+1}^{gh}]_{L_4,L_5} .
\end{align}
This relation implies (using the fact that $[w_j^g]_{L, L'}$ is also unitary)
\begin{align} 
[w^{g}_{2k-1}]_{L_1, L_2} [w^{h}_{2k-1}]_{L_1, L_2} &= [w_{2k-1}^{gh}]_{L_1,L_2} e^{i \beta_{k,L_1 L_2 L_4 L_5}^{g,h}}, \\
[w^{g}_{2k+1}]_{L_4, L_5} [w^{h}_{2k+1}]_{L_4, L_5}  &= [w_{2k+1}^{gh}]_{L_4,L_5} e^{i \beta_{k,L_1 L_2 L_4 L_5}^{g,h}}
 \end{align}
 with $\beta_{k,L_1 L_2 L_4 L_5}^{g,h} \in [0, 2 \pi)$. Both equations taken together show that $\beta$ must be the same for all $L_1$, $L_2$, $L_4$, $L_5$. 
 Finally, we arrive at
 \begin{align}
[w^{g}_{2k-1}]_{L_1, L_2} [w^{h}_{2k-1}]_{L_1, L_2} 
&= [w_{2k-1}^{gh}]_{L_1,L_2} e^{i \beta^{g,h}}, \label{eq:cohomology1} \\
[w^{g}_{2k+1}]_{L_4, L_5} [w^{h}_{2k+1}]_{L_4, L_5}  
&= [w_{2k+1}^{gh}]_{L_4,L_5} e^{i \beta^{g,h}} \label{eq:cohomology2}
\end{align}
with $\beta^{g,h}$ independent of $k$. Hence $[w^{g}_{2k-1}]_{L_1, L_2}$ and $[w^{g}_{2k+1}]_{L_4, L_5}$ are \textit{projective representations} of the group $G$: Projective representations are matrices $q_g$ which are defined up to a phase factor and represent the group $G$ up to a phase factor,
\begin{align}
q_g q_h = q_{gh} e^{i \omega(g,h)}.
\end{align}
Hence, the equivalent set of matrices defined by $q_g' = q_g e^{i\chi_g}$ obeys
\begin{align}
q_g' q_h' = q_{gh}' e^{i \omega'(g,h)}
\end{align}
with $\omega'(g,h) = \omega(g,h) - \chi_{gh} + \chi_g + \chi_h$. The elements of the second cohomology group of the symmetry group $G$ are the equivalence classes of phases $\omega(g,h)$ under the above transformation, i.e., $\omega(g,h) \rightarrow \omega(g,h) - \chi_{gh} + \chi_g + \chi_h$. Since these are discrete (i.e., the second cohomology group is finite), continuously changing the unitaries $q_g$ cannot change the element of the second cohomology group they correspond to. Hence, continuous changes of the quantum circuit and thus of the unitaries $w^g_{2k-1}$ do not alter the corresponding element of the second cohomology group. 
Thus, these elements correspond to different SPT phases. (For the stability with respect to adiabatic evolutions of the Hamiltonian, see Sec.~\ref{sec:robustness}.)  
 Eq.~\eqref{eq:cohomology1} thus implies that the projective representations $[w^g_{2k-1}]_{L_1,L_2}$ all correspond to the same element of the second cohomology group. 
 Therefore, according to Eq.~\eqref{eq:cohomology2}, $[w_{2k+1}^{g}]_{L_4,L_5}$ all correspond to the same element of the second cohomology group as $[w^g_{2k-1}]_{L_1,L_2}$. Hence, an FMBL system with a symmetry possesses one topological label for all eigenstates. We demonstrate in Section~\ref{sec:robustness} that the topological label does not change under symmetry-preserving perturbations to the Hamiltonian unless they violate the FMBL condition. 


\subsection{MBL systems with an anti-unitary on-site symmetry}\label{sec:anti-unitary}

An anti-unitary on-site symmetry corresponds to the presence of both time-reversal symmetry and an on-site symmetry (with symmetry group $G$). In that case, the Hamiltonian is invariant under a local unitary $\mathpzc{v}_g$, up to complex conjugation, that is, for given group element $g$, either Eq.~\eqref{eq:Ham_sym} or 
\begin{align}
H = \mathpzc{v}_g^{\otimes N} H^* (\mathpzc{v}_g^\dg)^{\otimes N}. \label{eq:Ham_antisym}
\end{align}
holds. In the latter case, Eq.~\eqref{eq:theta} reads
\begin{align}
\Theta_g = U^\dagger \mathpzc{v}_g^{\otimes N} U^*.
\label{eq:antitheta}
\end{align}
Let us define~\cite{Bultnick2017}  $\gamma(g) = 0$ if the corresponding operation does not involve complex conjugation,  $\gamma(g) = 1$ if it does and
\begin{align}
\lfloor X \rceil^{\gamma(g)} = \begin{cases}
X   \ &\mathrm{if} \ \gamma(g) = 0, \\
X^* \ &\mathrm{if} \ \gamma(g) = 1.
\end{cases}
\end{align}
One can now repeat the derivation of Eqs.~\eqref{eq:theta_equality} to~\eqref{eq:combined_trafo} replacing $u_j$ by $\lfloor u_j  \rceil^{\gamma(g)}$ and  $v_j$ by $\lfloor v_j  \rceil^{\gamma(g)}$ on the sides of the equations containing $g = \mathpzc{v}_g^{\otimes \frac{\ell}{2}}$. That is, Eq.~\eqref{eq:combined_trafo} now reads
\begin{widetext}
\begin{equation}
\begin{aligned}
\includegraphics[width=0.6\textwidth]{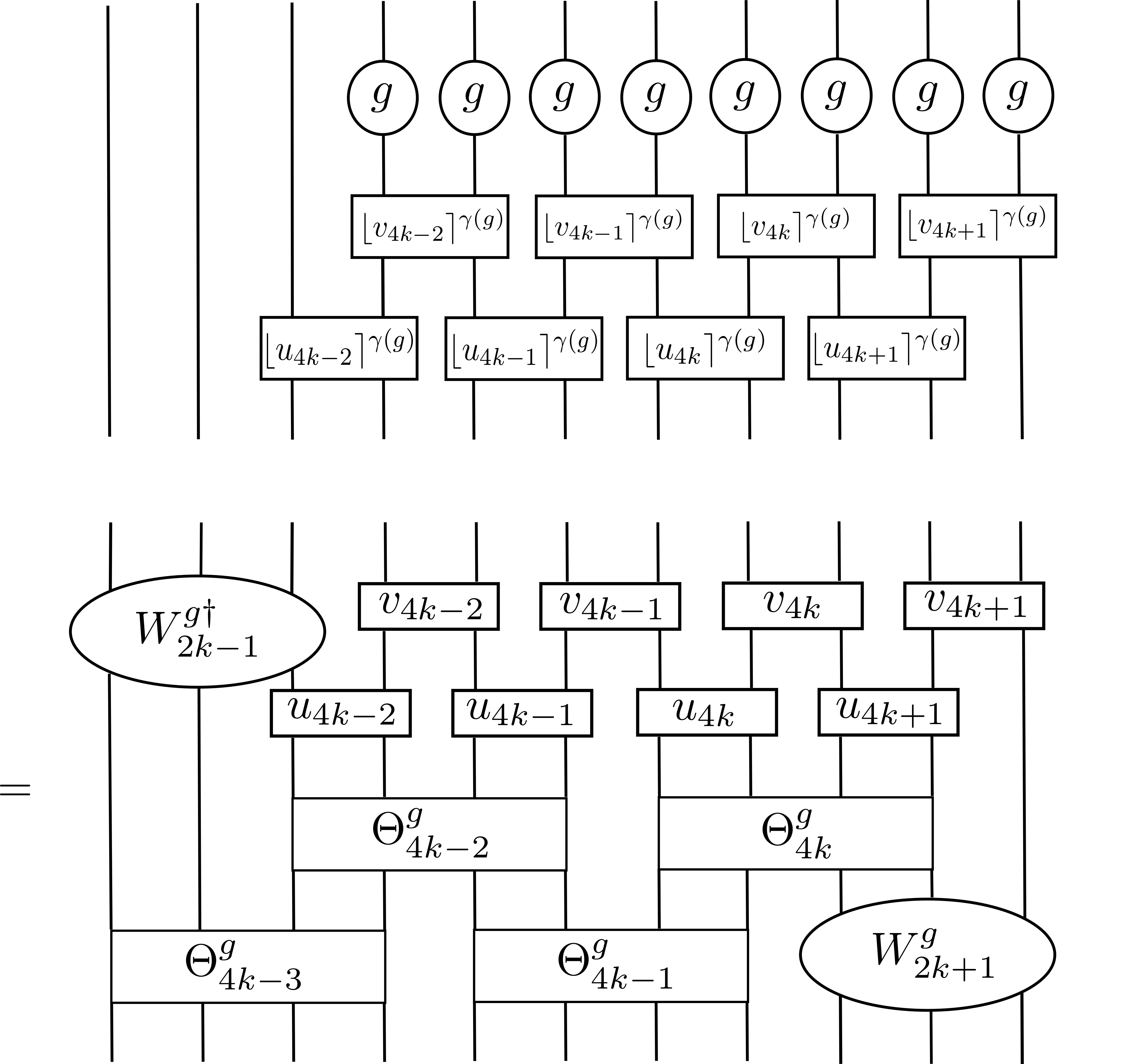}.
 \end{aligned}  \label{eq:anticombined_trafo}
\end{equation}
Eq.~\eqref{eq:gh_trafo} now takes the form
\begin{equation}
\begin{aligned}
\includegraphics[width=0.8\textwidth]{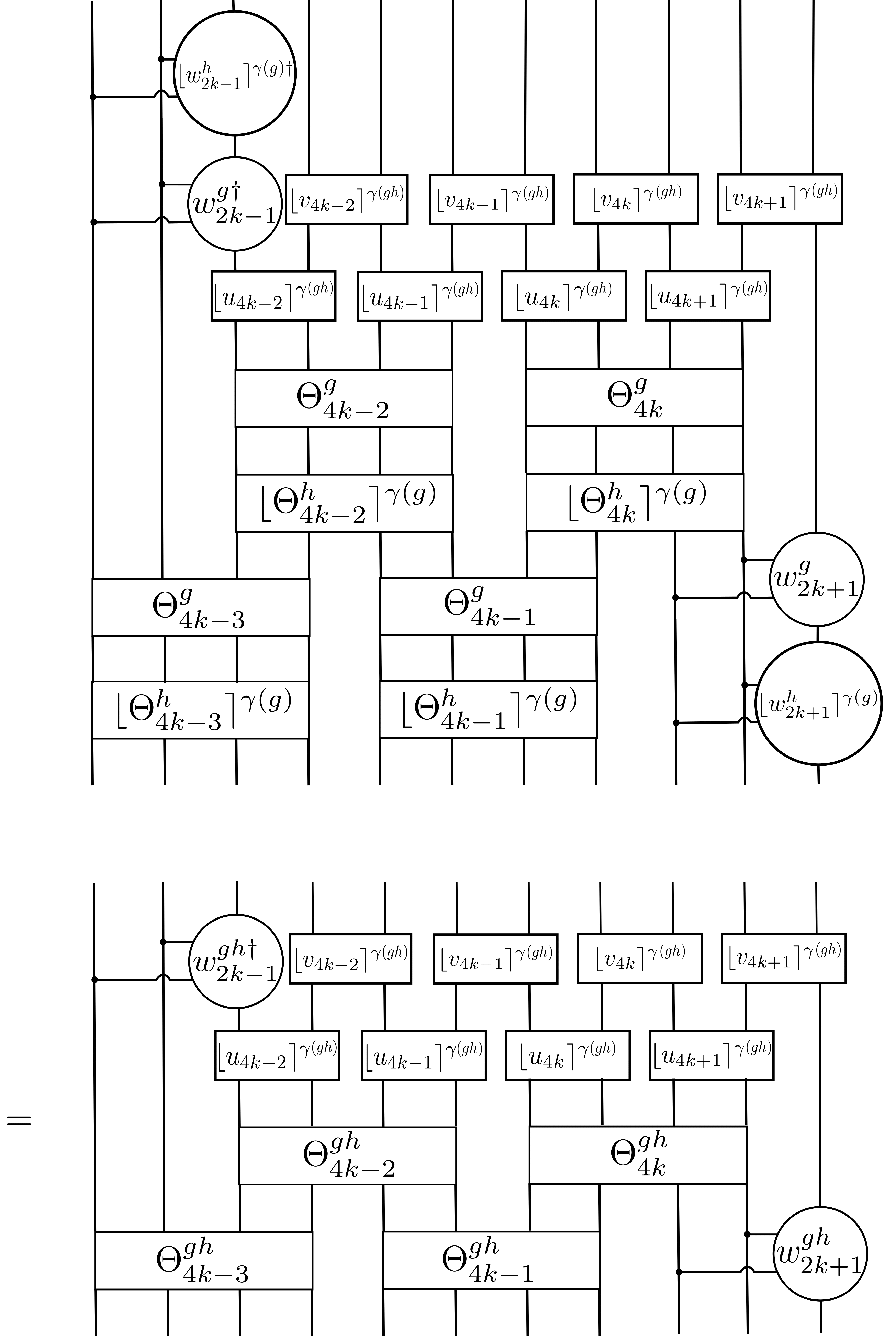}
 \end{aligned} . \label{eq:antigh_trafo}
\end{equation}
\end{widetext}
 Now, one can similarly derive Eqs.~\eqref{eq:theta_gh} to~\eqref{eq:cohomology2} if one replaces $\Theta_j^h$ with $\lfloor \Theta_j^h \rceil^{\gamma(g)}$ and  $w_j^h$ with $\lfloor w_j^h \rceil^{\gamma(g)}$. Hence, we have (see Eqs.~\eqref{eq:cohomology1} and~\eqref{eq:cohomology2})
 \begin{align}
[w^{g}_{2k-1}]_{L_1, L_2} \lfloor w^{h}_{2k-1}\rceil^{\gamma(g)}_{L_1, L_2} 
&= [w_{2k-1}^{gh}]_{L_1,L_2} e^{i \beta_k^{g,h}}, \label{eq:anticohomology1} 
\\
[w^{g}_{2k+1}]_{L_4, L_5} \lfloor w^{h}_{2k+1}\rceil^{\gamma(g)}_{L_4, L_5}  
&= [w_{2k+1}^{gh}]_{L_4,L_5} e^{i \beta_{k+1}^{g,h}} \label{eq:anticohomology2}
\end{align}
with $\beta^{g,h}_{k+1} = \beta^{g,h}_k$. The phase factors on the right hand sides are again independent of the $l$-bit configuration. Now, the topological label is given by the equivalence class these phase factors belong to under the equivalence relation~\cite{Bultnick2017} $\beta_k^{g,h} \rightarrow \beta_k^{g,h} - \chi_{gh} + \chi_g + (-1)^{\gamma(g)} \chi_h$, which corresponds to a generalization of the second cohomology group. 
All eigenstates are again in the same topological phase.


\subsection{Time-reversal symmetry}\label{sec:TRS}

Time-reversal symmetry is a special case of the anti-unitary symmetry considered in the previous section if one chooses $G = \{e,z\}$. Then Eq.~\eqref{eq:anticohomology1} reads (for $g = h = z$)
\begin{align}
[w^{z}_{2k-1}]_{L_1, L_2} [w^{z}_{2k-1}]^*_{L_1, L_2} &= [w_{2k-1}^{e}]_{L_1,L_2} e^{i \beta_k^{e}} \notag \\
&= \mathbb{1} e^{i \beta_k^{z,z}}.
\end{align}
Hence, $[w^{z}_{2k-1}]_{L_1, L_2} = [w^{z}_{2k-1}]_{L_1, L_2}^\top e^{i \beta_k^{z,z}}$, which implies inserted into itself~\cite{Pollmann2010} that $e^{i \beta_k^{z,z}} = \pm 1$, i.e., we have a $\mathbb{Z}_2$ classification for the full spectrum of eigenstates, as shown in Ref.~\onlinecite{Thorsten}. For the sake of completeness, we explicitly rederive this result using the formalism introduced above:  Time-reversal invariant systems fulfill (setting $\mathpzc{v} = \mathpzc{v}_z$)
\begin{align}
H = \mathpzc{v}^{\otimes N} H^* (\mathpzc{v}^\dagger)^{\otimes N}
\end{align} 
with $\mathpzc{v} \mathpzc{v}^* = \pm \mathbb{1}$. For the unitary $U$ diagonalizing the Hamiltonian this implies
\begin{align}
\mathpzc{v}^{\otimes N} U^* = U \Theta.
\end{align}
The corresponding condition on the quantum circuit $\tilde U$ is the same as Eq.~\eqref{eq:theta_equality} if on the right hand side the unitaries are replaced by their complex conjugates and $g$ by $\mathcal{V} = \mathpzc{v}^{\otimes \ell/2}$.  The changes in the equations directly thereafter are similar; note in particular that Eq.~\eqref{eq:combined_trafo} now reads 
\begin{equation}
\begin{aligned}
\includegraphics[width=0.46\textwidth]{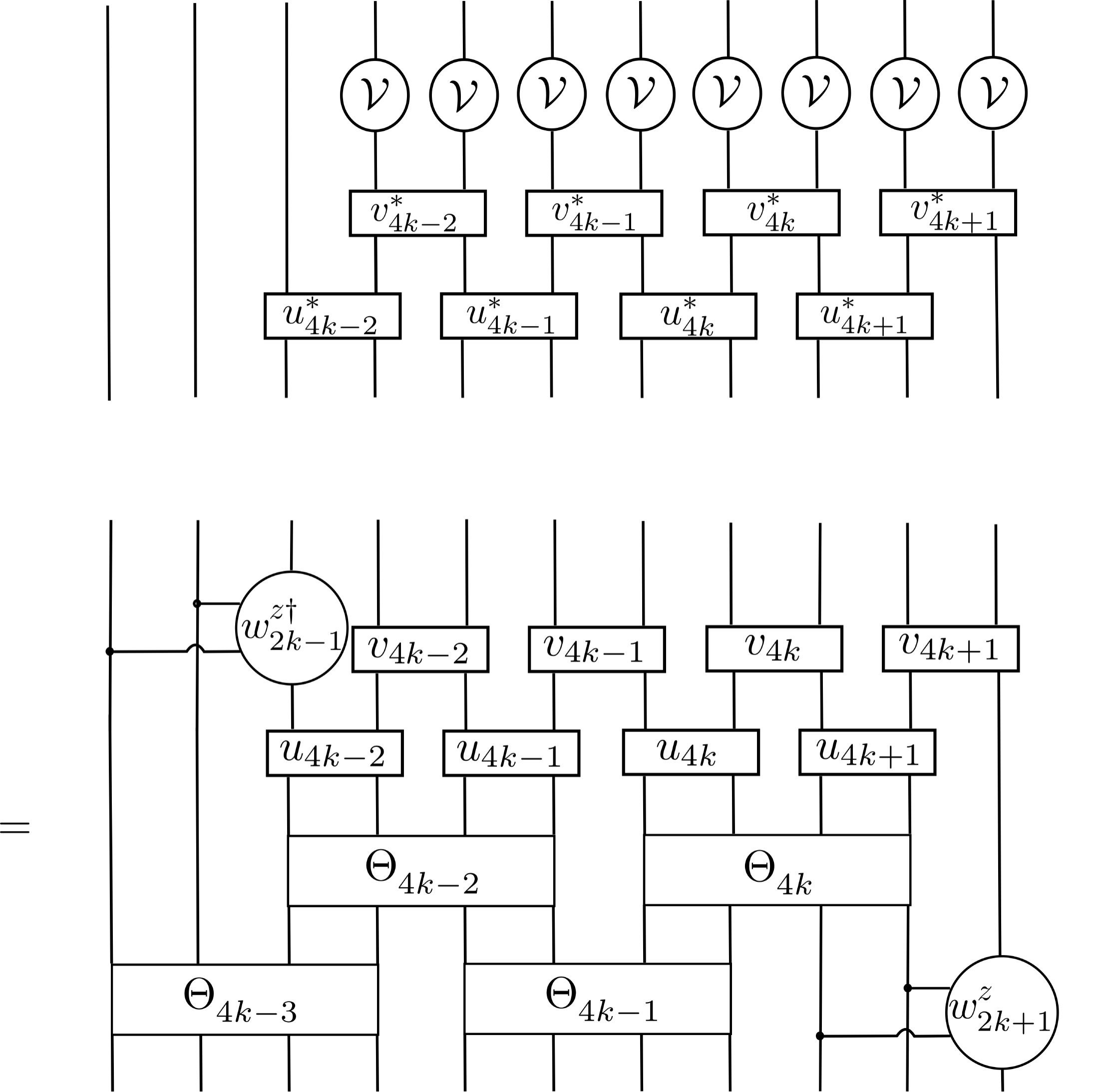} .
 \end{aligned} \label{eq:combined_TRS}
\end{equation}
If we insert this equation into its complex conjugate, we arrive at
\begin{equation}
\begin{aligned}
\includegraphics[width=0.46\textwidth]{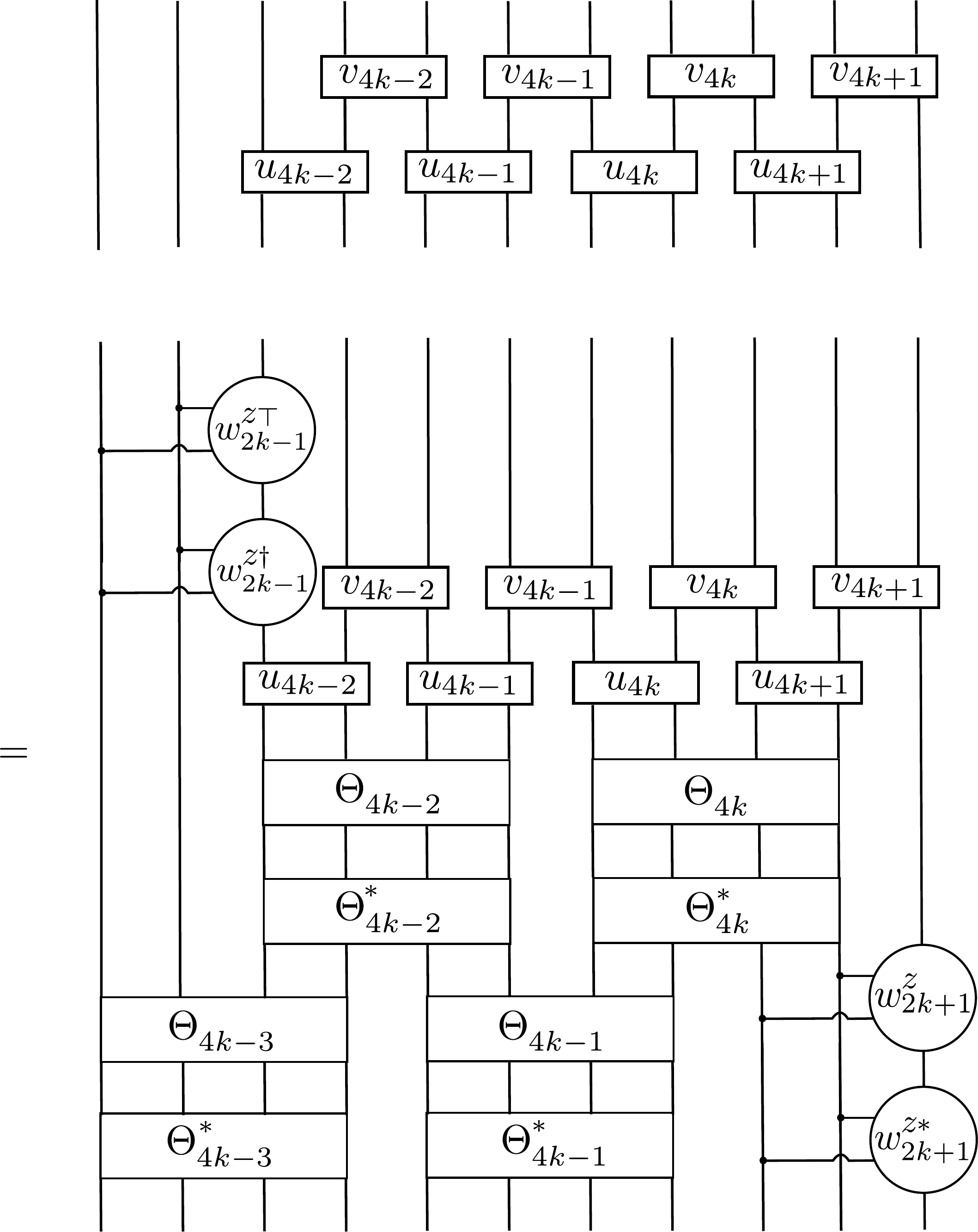},
 \end{aligned} \label{eq:TRS_TRS}
\end{equation}
which implies
\begin{equation}
\begin{aligned}
\includegraphics[width=0.34\textwidth]{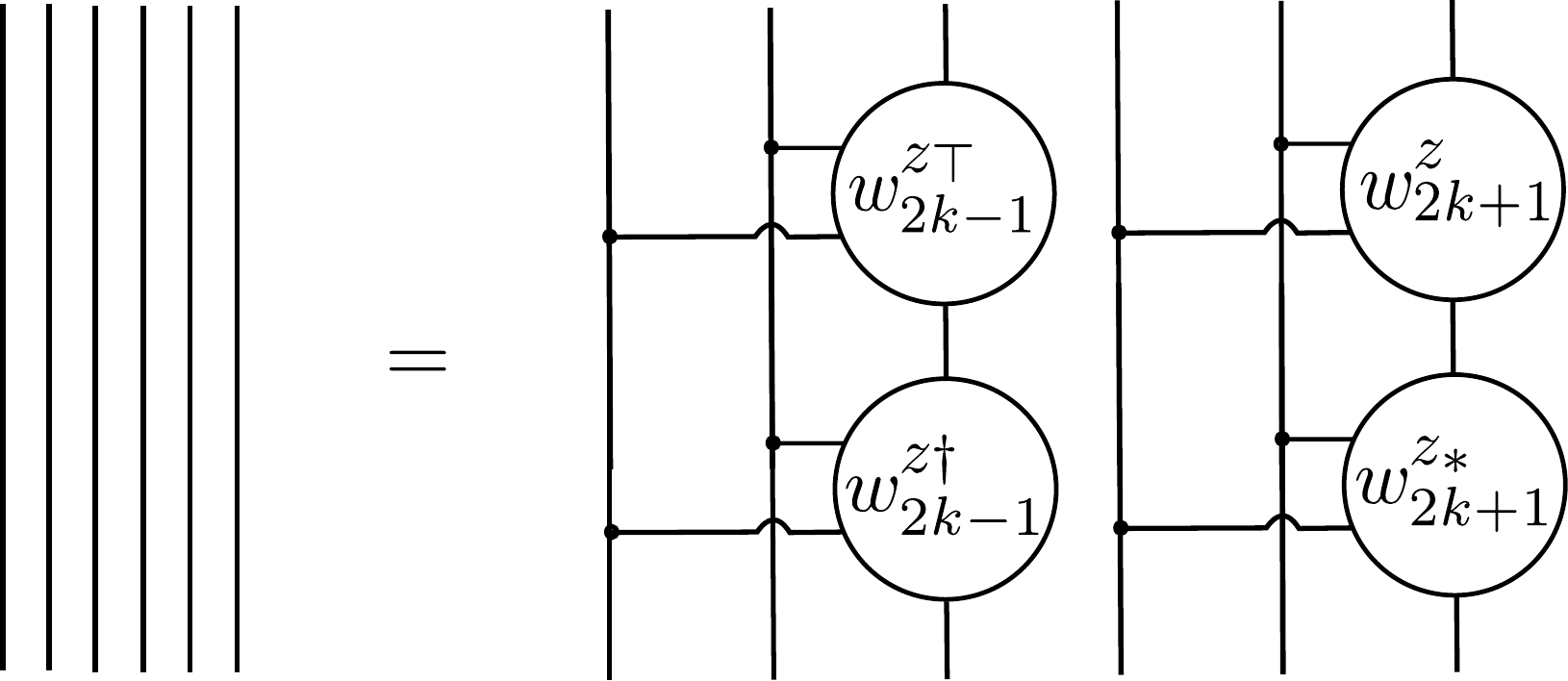}
 \end{aligned} \label{eq:ww} .
\end{equation}
Fixing the indices of the first two legs from the left again to $L_1$ and $L_2$ and of the fourth and fifth ones to $L_4$ and $L_5$ results in
\begin{align}
[w^z_{2k-1}]_{L_1,L_2} [w^z_{2k-1}]_{L_1,L_2}^* &= \mathbb{1} e^{i \beta_{k,L_1 L_2 L_4 L_5}^{z,z}}, \label{eq:ww1} \\ 
[w^z_{2k+1}]_{L_4,L_5} [w^z_{2k+1}]_{L_4,L_5}^* &= \mathbb{1} e^{i \beta_{k,L_1 L_2 L_4 L_5}^{z,z}}.
\end{align} 
This shows again that $\beta_{k,L_1 L_2 L_4 L_5}^{z,z}$ must be the same for all $L_1, L_2, L_4, L_5$ and $k$. Using the fact that $[w_{2k-1}^{z}]_{L_1,L_2}$ and $[w^z_{2k+1}]_{L_4,L_5}$ are unitaries, we multiply Eq.~\eqref{eq:ww1} from the right by $[w^z_{2k-1}]_{L_1,L_2}^\top$ and insert the obtained relation into itself~\cite{Pollmann2010}, arriving at $e^{2 i \beta^{z,z}} = 1$, i.e., $\beta^{z,z} = 0, \pi$. Since this index is the same for all positions $k$ and $l$-bit indices, we again obtain one topological index, which has to be the same for all eigenstates.



\section{Classification of fermionic SPT MBL phases} \label{sec:fermclass}

In this section, we classify one-dimensional fermionic SPT MBL phases by extending the spinful  treatment using a diagrammatic representation of fermionic tensor networks. 
We obtain that a classification in terms of the elements of the (generalized) second cohomology group of the extended symmetry group $G' = G \times \mathbb{Z}_2$, where $G$ is the original on-site (anti-)unitary symmetry group and the additional $\mathbb{Z}_2$ stems from parity conservation. 
We explicitly derive two out of the three $\mathbb{Z}_2$ topological invariants (shared by all eigenstates)  associated with the $\mathbb{Z}_8$ classification for systems with time-reversal symmetry. 
	Recall that the $\mathbb{Z}_8$ classification of fermionic SPT ground states is given by three topological invariants defined as follows\cite{vers1610}: (i) 
	An index $\kappa = 0,1$ which arises from the fact that the time-reversal operator on the virtual level squares to positive or negative identity; 
	(ii) An index $\mu = 0,1$ 
	which indicates whether the time reversal operator on the virtual level commutes or anti-commutes with the parity operator;
	and 
(iii) the fractionalization of the parity of the ground state~\cite{pollmann2011ferm}, i.e., whether decoupled Majorana modes appear at the edges after tracing out part of the system or for open boundary conditions.  
The derivation  below shows that all eigenstates share the same $\kappa$ and $\mu$. However, the current approach cannot be used to detect the topological invariant associated with (iii), because the classification uses two-layer quantum circuits that locally preserve fermionic parity.
Only a long-range parity-preserving quantum circuit can exactly represent the topological phase associated with fermionic parity fractionalization~\cite{pollmann2011ferm}, and we illustrate this with the example of the Kitaev chain in Sec.~\ref{sec:kit}.


We will use fermionic tensor networks as defined in Refs. \onlinecite{guif200910, guif201001, gu201004, ign201005, orus2010, vers1610}. We will review fermionic tensor networks in the language of super vector spaces (Secs.~\ref{sec:supervec} and~\ref{sec:fermtn}), following closely Ref.~\onlinecite{vers1610}, propose a diagrammatic representation (Sec.~\ref{sec:diagrep}), and sketch the extension of the above treatment in this diagrammatic representation (Sec.~\ref{sec:extproof}). 
Lastly, we explicitly derive a classification of fermionic FMBL systems in the presence of time reversal symmetry (Sec.~\ref{fermtrs}).



\subsection{Formalism} \label{sec:fermform}

\subsubsection{Super vector spaces} \label{sec:supervec}
A super vector space $V = V^0 \oplus V^1$ is a direct sum of the vector spaces $V^0$ and $V^{1}$ containing even and odd parity vectors. The parity of a (super-)vector $\ket{i} \in V$ is denoted by $|i| \in {0,1}$ (0 for even and 1 for odd). The graded tensor product of two vectors $\ket{i}$ and $\ket{j}$ is $\ket{i} \otimes_{\mathfrak{g}} \ket{j} \in V \otimes_{\mathfrak{g}}  V$, and its parity is  $|i| + |j| \mod 2$. The reordering of vectors $\mathcal{F}$ within a graded tensor product is the isomorphism
\begin{align}
\mathcal{F}: \quad &  V \otimesg W  \to W \otimes_{\mathfrak{g}}  V \nonumber
\\
& \ket{i }  \otimesg \ket{j} \to (-1)^{|i| |j|} \ket{j}  \otimes_{\mathfrak{g}} \ket{i } \; .  \label{eq:fermreord}
\end{align}

The reordering of graded tensor products in $V^* \otimesg W$, $V \otimesg W^*$ and $V^* \otimesg W^*$ is similarly defined. The contraction $\mathcal{C}$ is the homomorphism 
\begin{align}
\mathcal{C}: \quad &  V^* \otimesg V  \to \mathbb{C} \nonumber
\\
& \bra{\psi  }  \otimesg \ket{\phi } \to \braket{\psi | \phi}
\end{align}
An operator acting on the super vector space $V$ is 
\begin{equation}
\mathrm{M} = \sum_{i,j} M_{i,j} \ket{i} \otimesg \bra{j} \quad \in V \otimesg V^* , 
\end{equation}
which has parity $|\mathrm{M}| := |i| + |j| \mod 2$. Higher rank operators are similarly defined.

\subsubsection{Fermionic tensor networks} \label{sec:fermtn}
Consider a rank three tensor in $ V_j \otimesg \mathcal{H}_j \otimesg (V_{j+1})^*$
\begin{align}
\mathrm{A}[j] = \sum_{i,\alpha ,\beta} & A[j]_{\alpha, \beta}^{i}
{\vk{\alpha}}_{j-1} \otimesg \ket{i}_j \otimesg {\vb{\beta}}_{j} 
\end{align}
where the round bras and kets are bases of virtual spaces $V_j $ and $V_{j}^*$. A fermionic matrix product state (fMPS) with periodic boundary conditions is obtained by 
\begin{align}
\ket{\psi} = \mathcal{C}_{v} (\mathrm{A} [1] \otimesg \mathrm{A}[2] \otimesg \dots \otimesg \mathrm{A}[N])
\end{align}
where $\mathcal{C}_{v} $ is the contraction over all virtual indices. For even parity fMPS, one can always choose every $A[j]$ to have even parity. For odd parity fMPS, it is convenient to choose only a single tensor to have odd parity, so that the fermionic reordering of the tensors in the fMPS is trivial.

\subsubsection{Diagrammatic representation} \label{sec:diagrep}

The diagrammatic rules of our fermionic tensor network approach are as follows:
\begin{itemize}
	\item \textit{Fermionic ordering of $\mathbb{Z}_2$  graded tensor products} is represented by a single directed line in red passing through all elements of super vector space (represented as \textit{open} legs in black below).
	
	\item \textit{Kets (Bras) of the (dual) super vector space $V = V^0 \oplus V^1$} are represented as open legs in black that point along (against) the direction of the arrow. 	
	
	\item \textit{Fermionic reordering} of $\ket{i}$ and $\ket{j}$ gives rise to a parity-dependent sign $(-1)^{|i||j|}$ which is represented as a crossing between two open legs, denoted as a black dot. 
\end{itemize}

We use two examples to demonstrate these rules:
\begin{itemize}
	\item The \textit{supertrace of a rank-2 operator} is written algebraically as
	\begin{align}
	\mathcal{C} \big( \sum_{i, j}  M_{ij} \ket{i} \otimesg \bra{j} \big) 
	&= \mathcal{C} \big( \sum_{i, j} M_{ij} (-1)^{|i| |j|} \bra{j} \otimesg \ket{i} \big) 
	\nonumber 
	\\
	&= \sum_{i} (-1)^{|i|} \,  M_{ii}  \label{eq:supertrace}
	\end{align}	
	Diagrammatically, we have,
	\includegraphics[width=0.88\columnwidth]{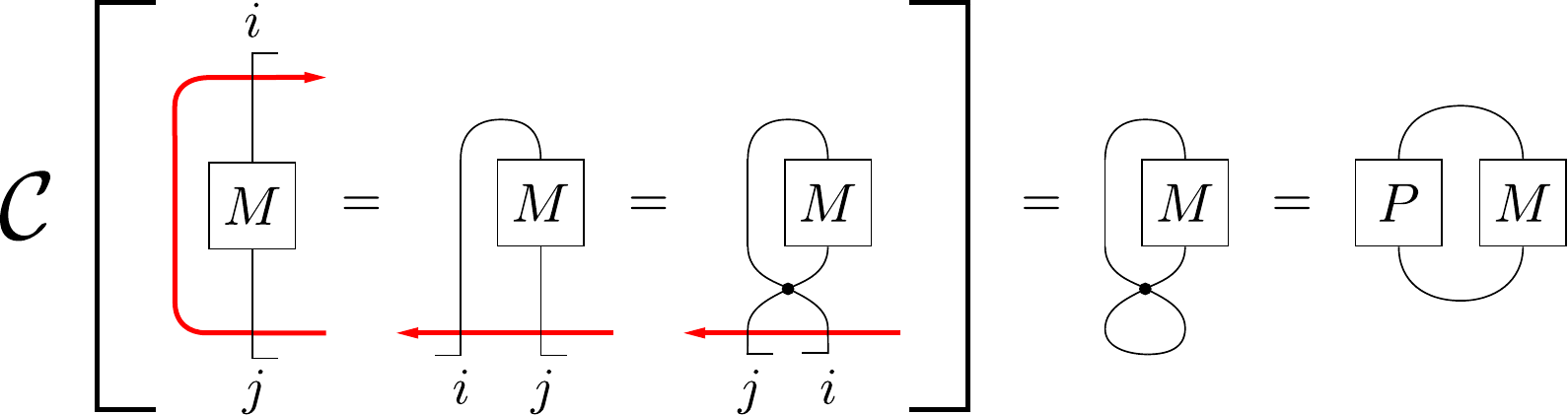} 
	where  the red directed line represents the fermionic ordering of graded tensor products. The open leg pointing along (against) the fermionic ordering represents ket (bra). The black dot represents the parity-dependent sign upon reordering the vector in the graded tensor product. Lastly,  $P_{ij} =(-1)^{|i|} \delta_{ij}$ is the parity operator.
	

	\item \textit{fMPS with even parity.} Suppose we have a translation invariant system which can be represented by an fMPS with $A[1] = A[2] = \dots = A[N]$. 
	
	\begin{center}
		\includegraphics[width=0.7\columnwidth]{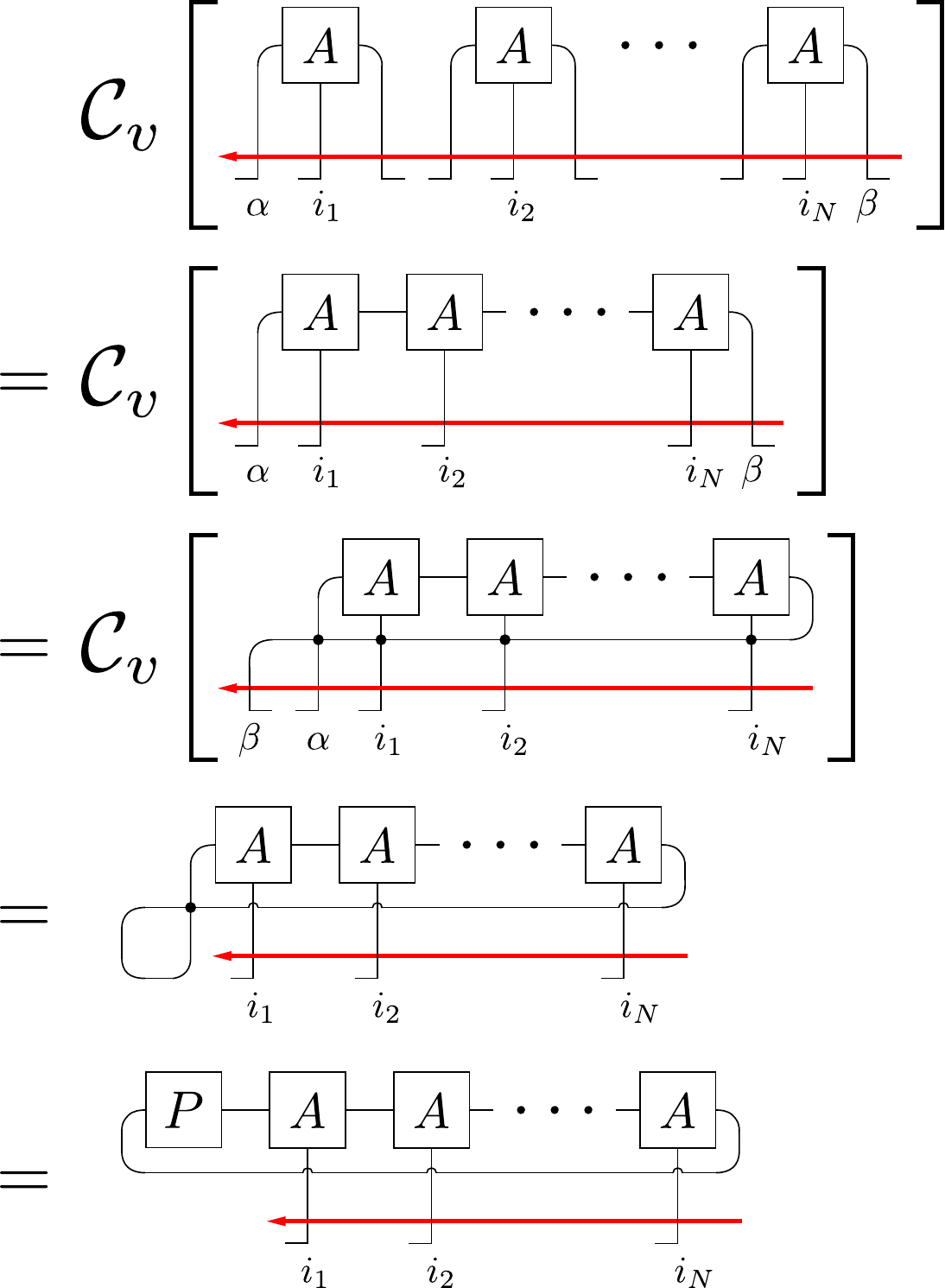}
	\end{center}
	where $\mathcal{C}_v$ represents the contraction of all virtual legs, we took advantage of all non-vanishing tensor contractions fulfilling $|i_1| + |i_2| + \ldots + |i_N| = 0 \mod 2$ and $P$ is defined as before. We have used diagrammatics to recover the even parity fMPS as described in Ref.~\onlinecite{vers1610}.

\end{itemize}

\subsection{Classification of fermionic MBL systems with an (anti-)unitary on-site symmetry} \label{sec:extproof}
After the incorporation of the above diagrammatic formulation, the derivation of the classification of fermionic  SPT MBL phases is very similar to the spinful one. For clarity's sake, we will demonstrate two key steps of the derivation with the fermionic diagrams, (i) the derivation of Eq.~\eqref{eq:thetaqc}, and (ii) the derivation of Eqs.~\eqref{eq:gauge1} and \eqref{eq:gauge2}.

Firstly, as mentioned previously and shown in the Appendix, non-abelian symmetries are incompatible with FMBL. Additionally, Kramers degeneracies  arising from time-reversal symmetry would also ruin the stability of FMBL~\cite{vass201612}. Therefore, we assume the absence of symmetry-enforced degeneracies. To derive Eq.~\eqref{eq:thetaqc}, we begin by writing the fermionic analogue of Eq.~\eqref{eq:tildeU}
\begin{equation}
\begin{aligned}
\includegraphics[width=0.95\columnwidth]{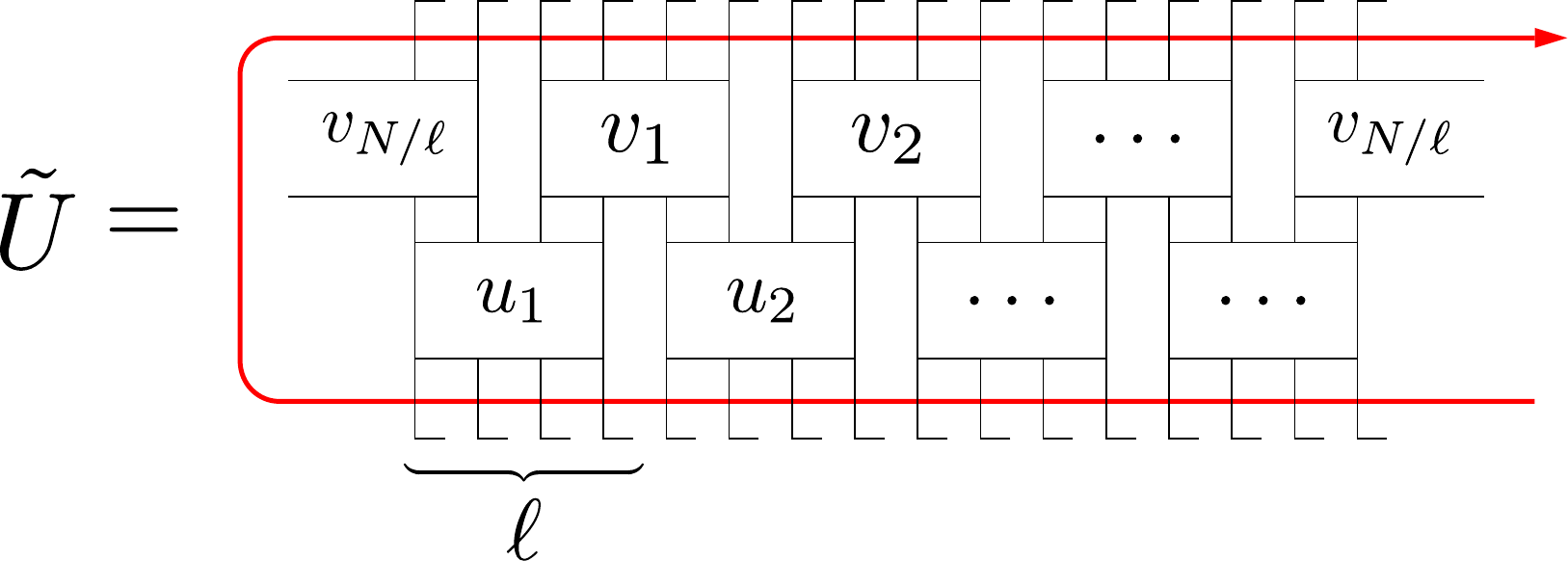}
\end{aligned}, 
\end{equation}
where again, the lower legs represent l-bits indices, and by fixing them, one obtains a matrix product state representation of the eigenstate corresponding to those l-bits. Each tensor ($u$ or $v$) in the two-layer quantum circuit is required to have even parity. Diagrammatically, this means
\begin{equation}
\begin{aligned}	
\includegraphics[width=0.6\columnwidth]{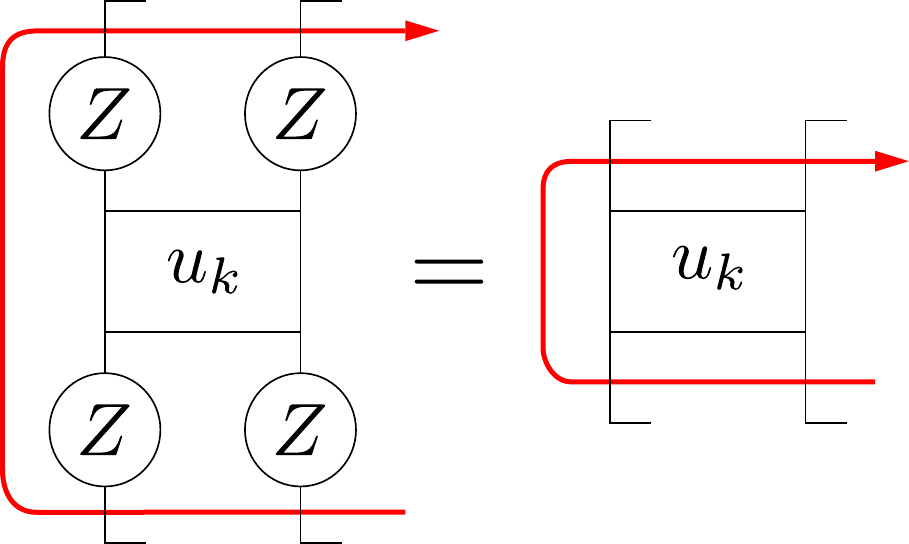}
\end{aligned}, \label{paritycon}
\end{equation}
where $Z = \sigma_z^{\otimes \ell/2}$ and similarly for $v_k$. 
Consider the extended symmetry group $G' = G \times \mathbb{Z}_2$, where $\mathbb{Z}_2$ corresponds to parity symmetry. $\mathpzc{v}_g$ is now the linear representation of $g \in G'$, and again we denote $g  \equiv \mathpzc{v}_g^{\otimes \ell/2 }$ in the diagrams below. So, the fermionic analogue of Eq.~\eqref{eq:Ham_sym} is
\begin{equation}
\begin{aligned}
	\includegraphics[width=0.95\columnwidth]{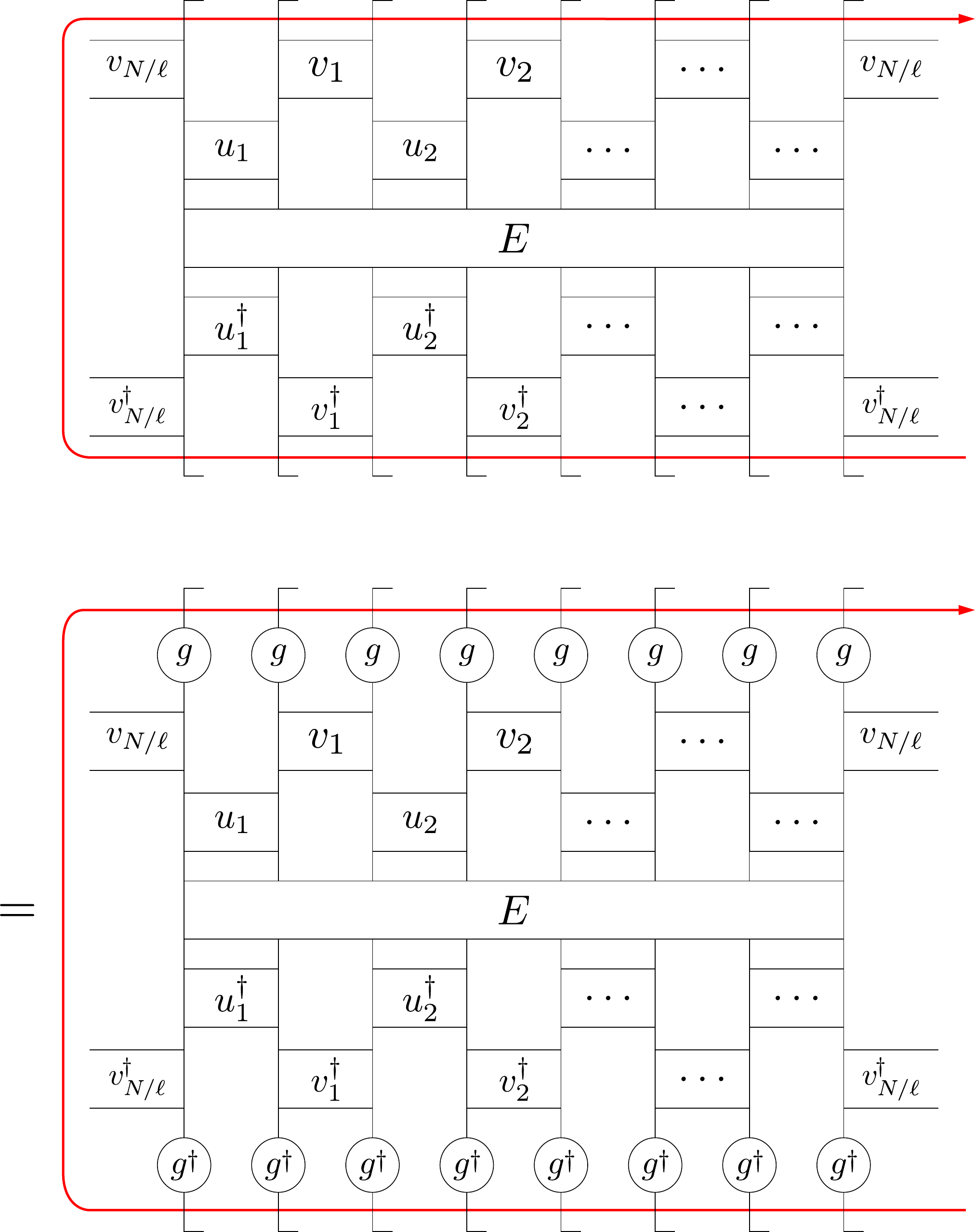}
\end{aligned} \label{eq:fermionic_diag}
\end{equation}
Now contract the left hand side with conjugates of $u$'s and $v$'s from the top and bottom. 
\begin{center}
	\includegraphics[width=0.8\columnwidth]{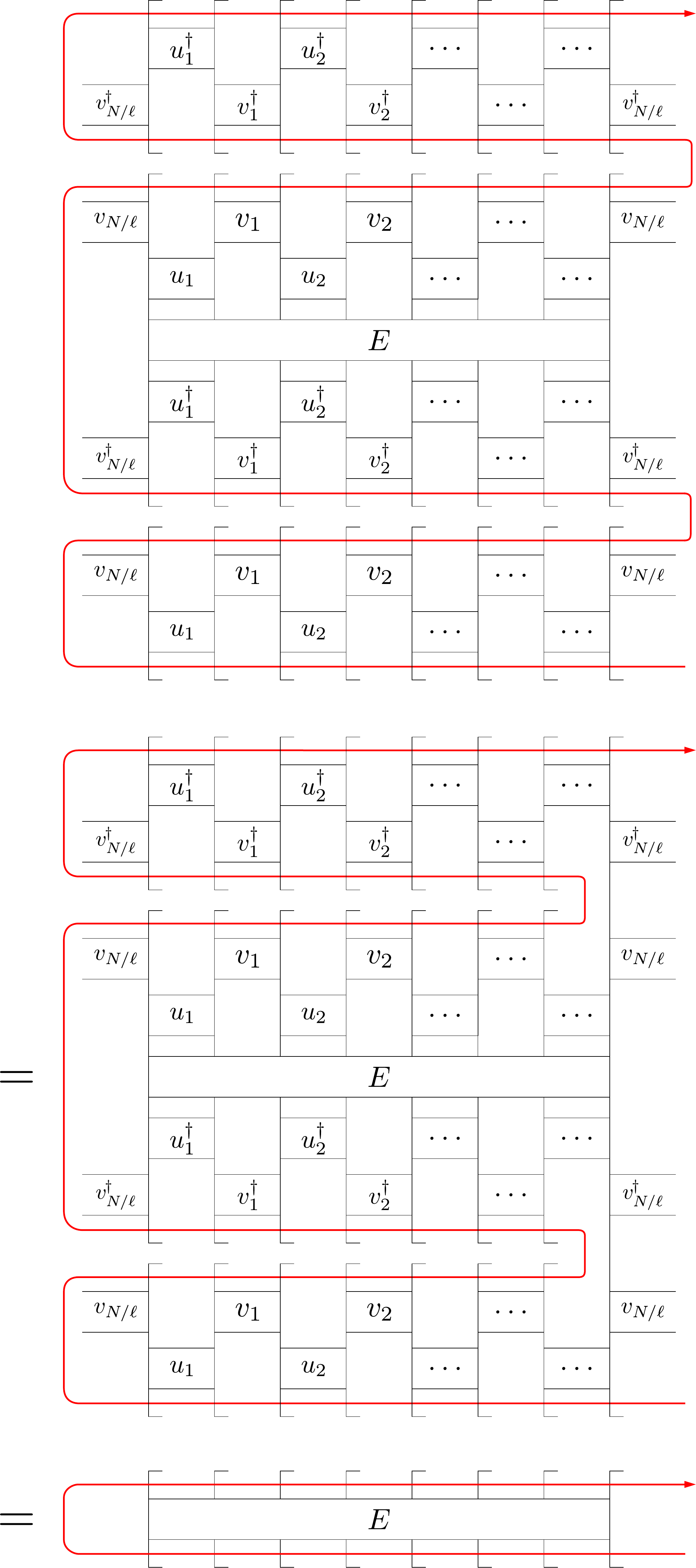}
\end{center}
where, in the first equality, we have iteratively contracted pairs of open legs in the middle of the diagram from the right. Do the same to the right hand side of Eq.~\eqref{eq:fermionic_diag}, and this gives us the fermionic analogue of Eq.~\eqref{eq:theta}.
\begin{equation}
\begin{aligned}
\includegraphics[width=0.8\columnwidth]{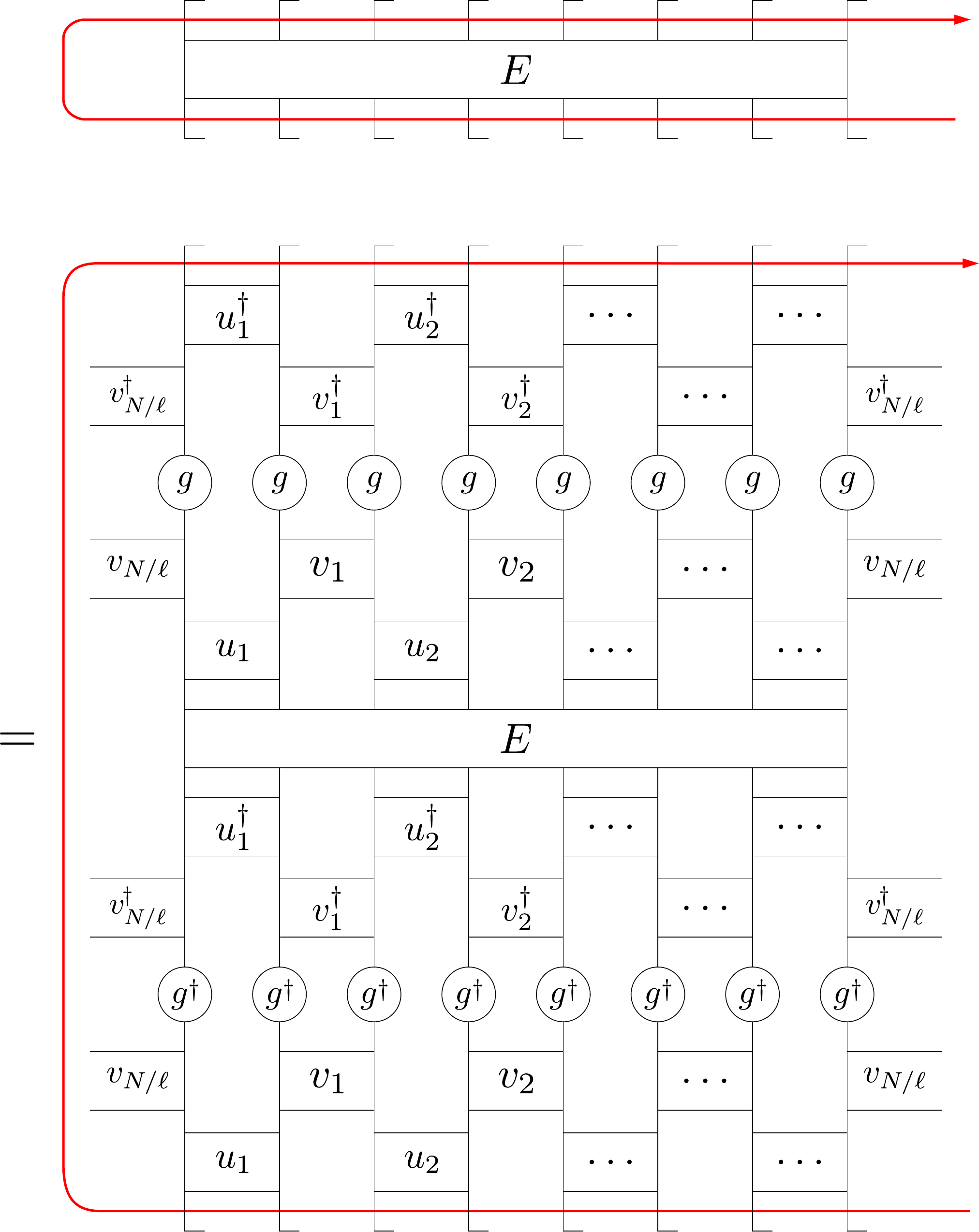} 
\end{aligned} \label{fermE}
\end{equation}
Since $E$ is non-degenerate, Eq.~\eqref{fermE} implies 
\begin{center}
	\includegraphics[width=\columnwidth]{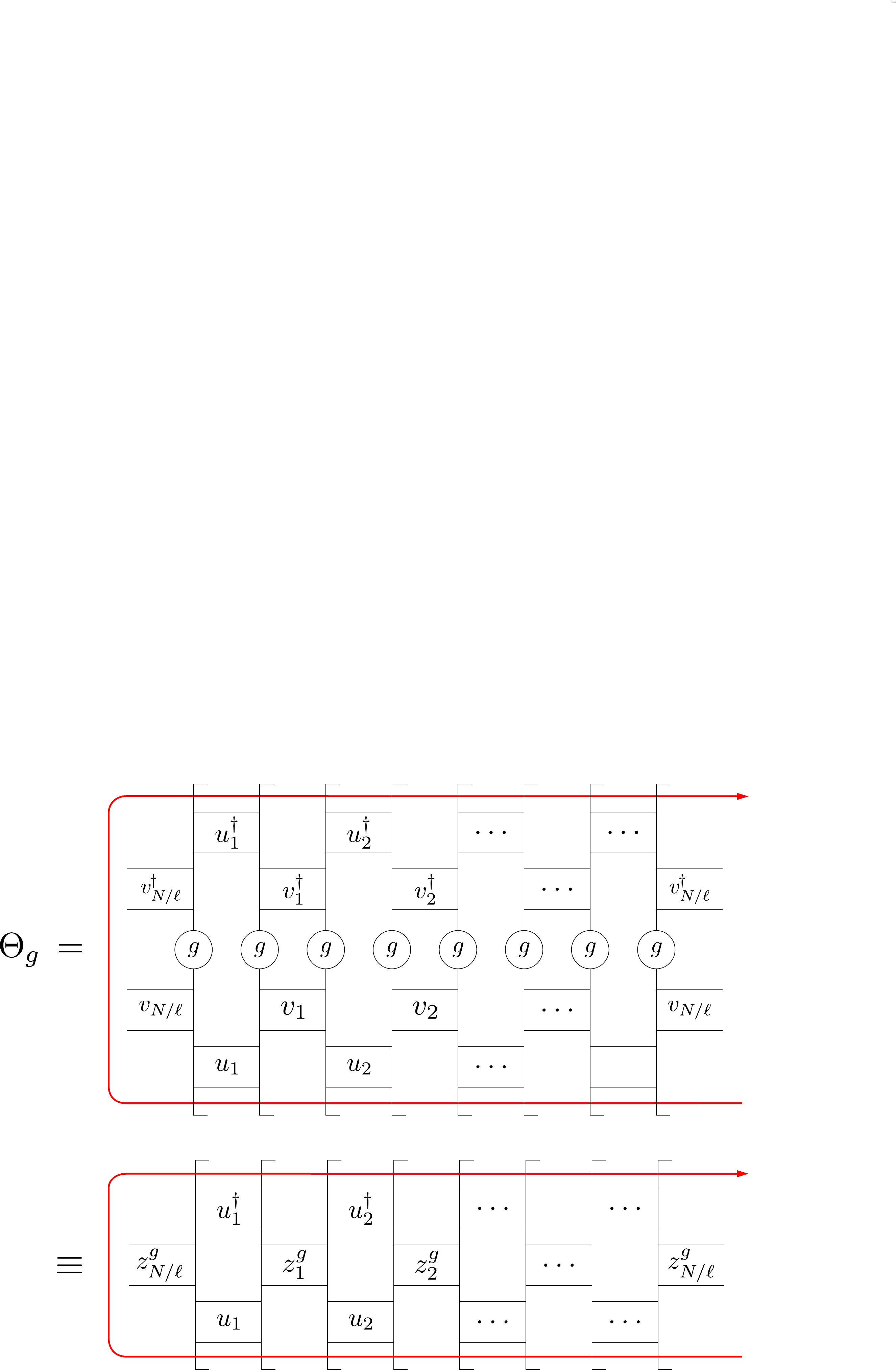}
\end{center}
We define the basis vectors of (the $k$-th set of) $\ell$ consecutive sites with the following ket and bra labelled by $\mathbf{l}_k$:
\begin{center}
	\includegraphics[width=0.8\columnwidth]{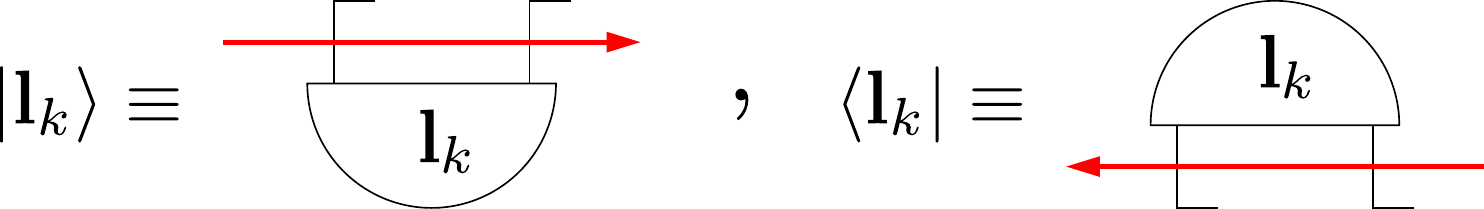}
\end{center}
Then the diagonal matrix element of $\Theta_g$ with respect to this basis is
\begin{center}
	\includegraphics[width=0.8\columnwidth]{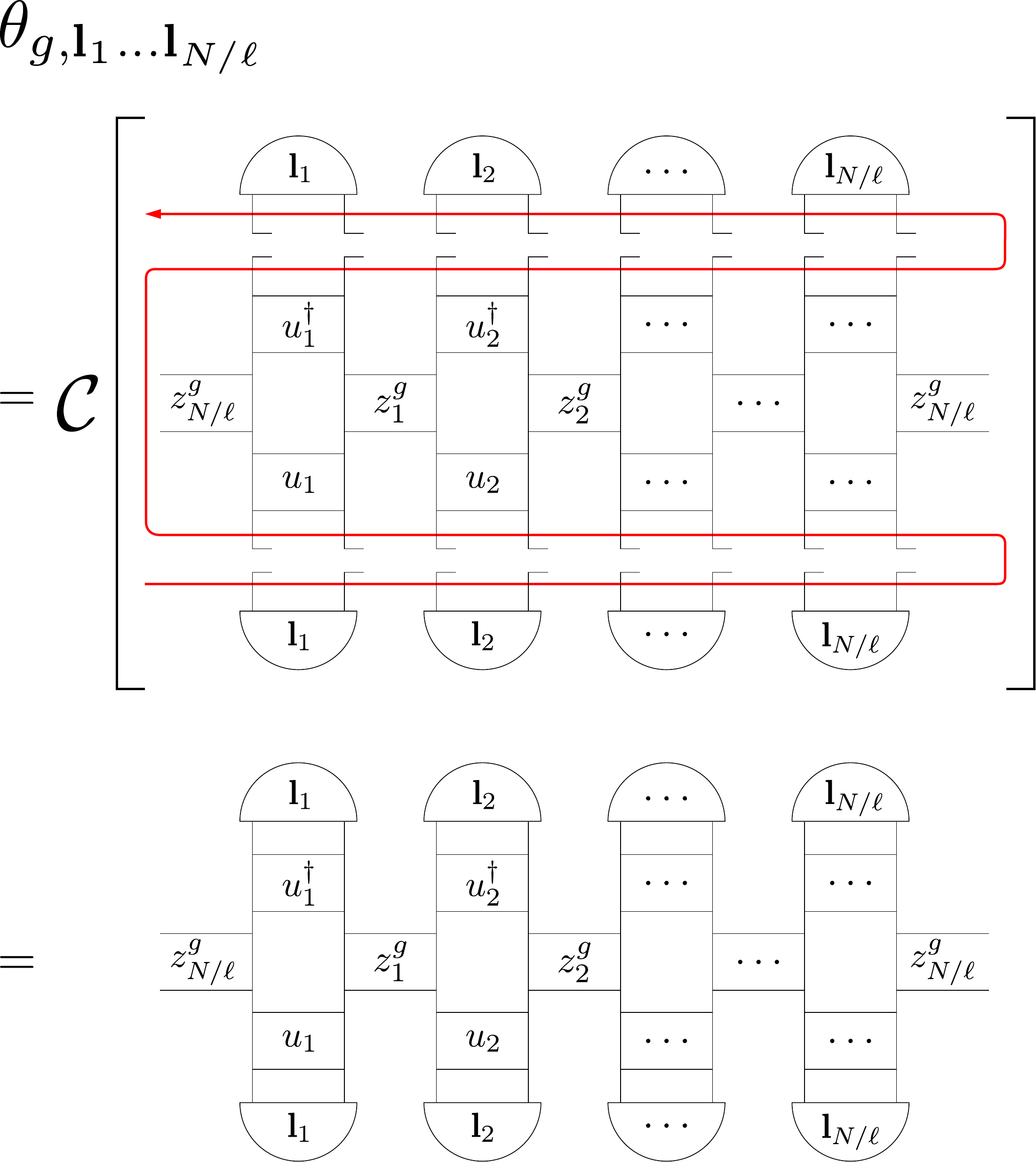}
\end{center}
where, again, we have contracted the right-most pairs of open legs iteratively. Now consider two matrix elements whose all but one indices coincide. 
\begin{center}
	\includegraphics[width=0.95\columnwidth]{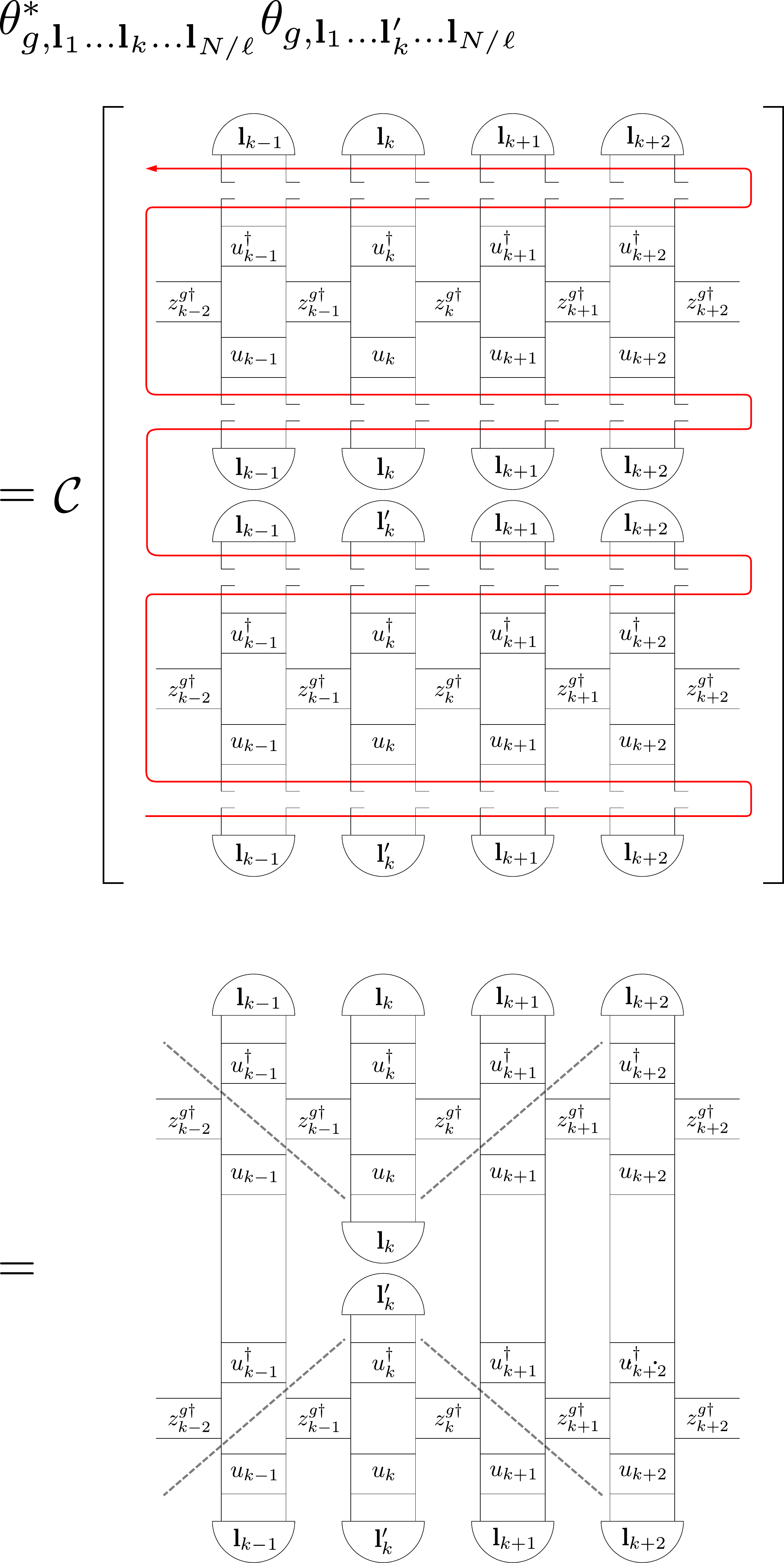}
\end{center} 
This establishes Eqs.~\eqref{eq:lc1} and Eq.~\eqref{eq:thetaqc} as before.


Secondly, we derive the fermionic analogue of Eqs.~\eqref{eq:gauge1} and \eqref{eq:gauge2}.

\begin{center}
	\includegraphics[width=0.8\columnwidth]{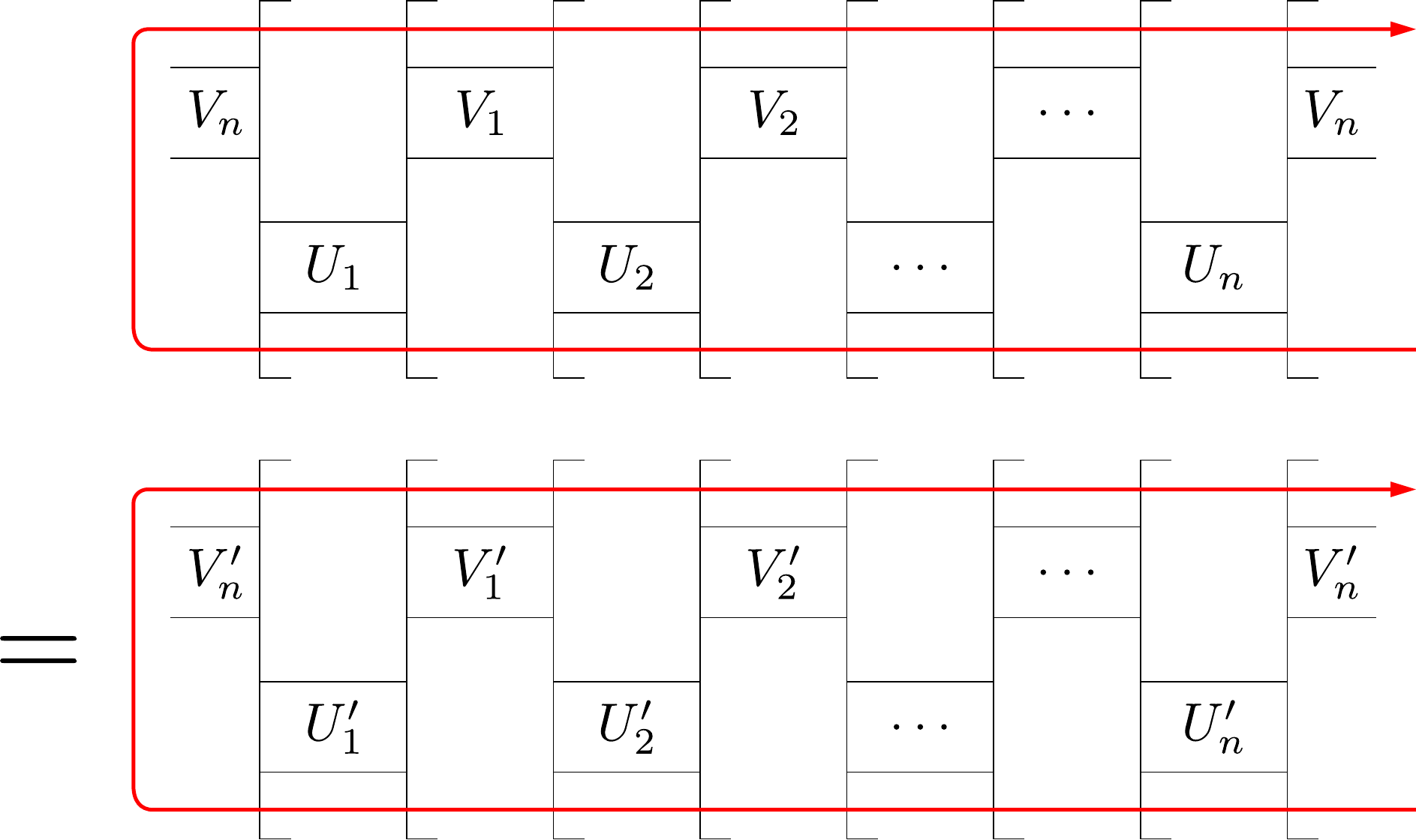}
\end{center}

We contract with the adjoint of the layer of $V$ and $U'$ from the top and bottom, respectively. 
\begin{center}
	\includegraphics[width=0.8\columnwidth]{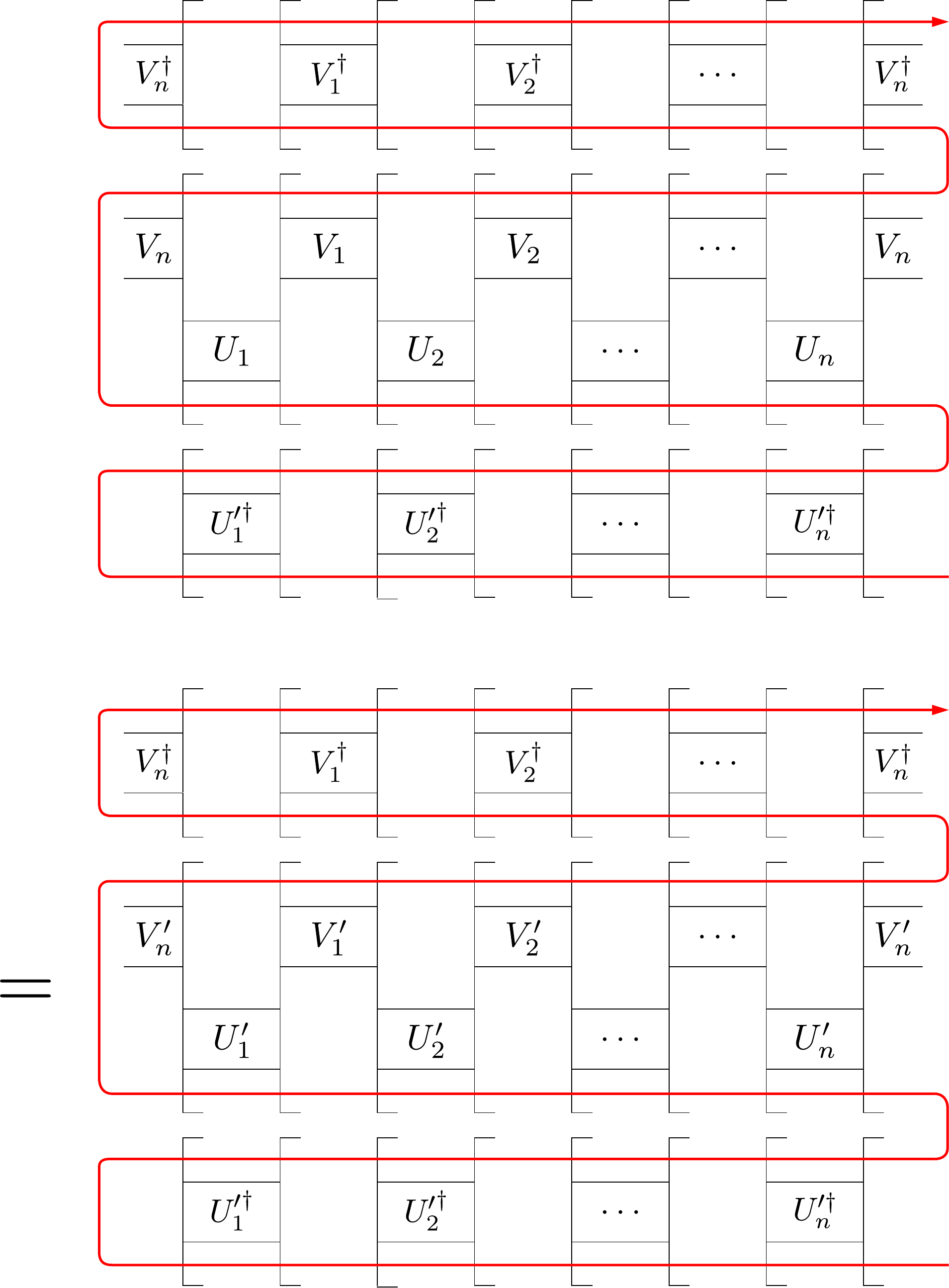}
\end{center}
After annihilation of conjugate pairs of unitaries, we have

\begin{center}
	\includegraphics[width=0.8\columnwidth]{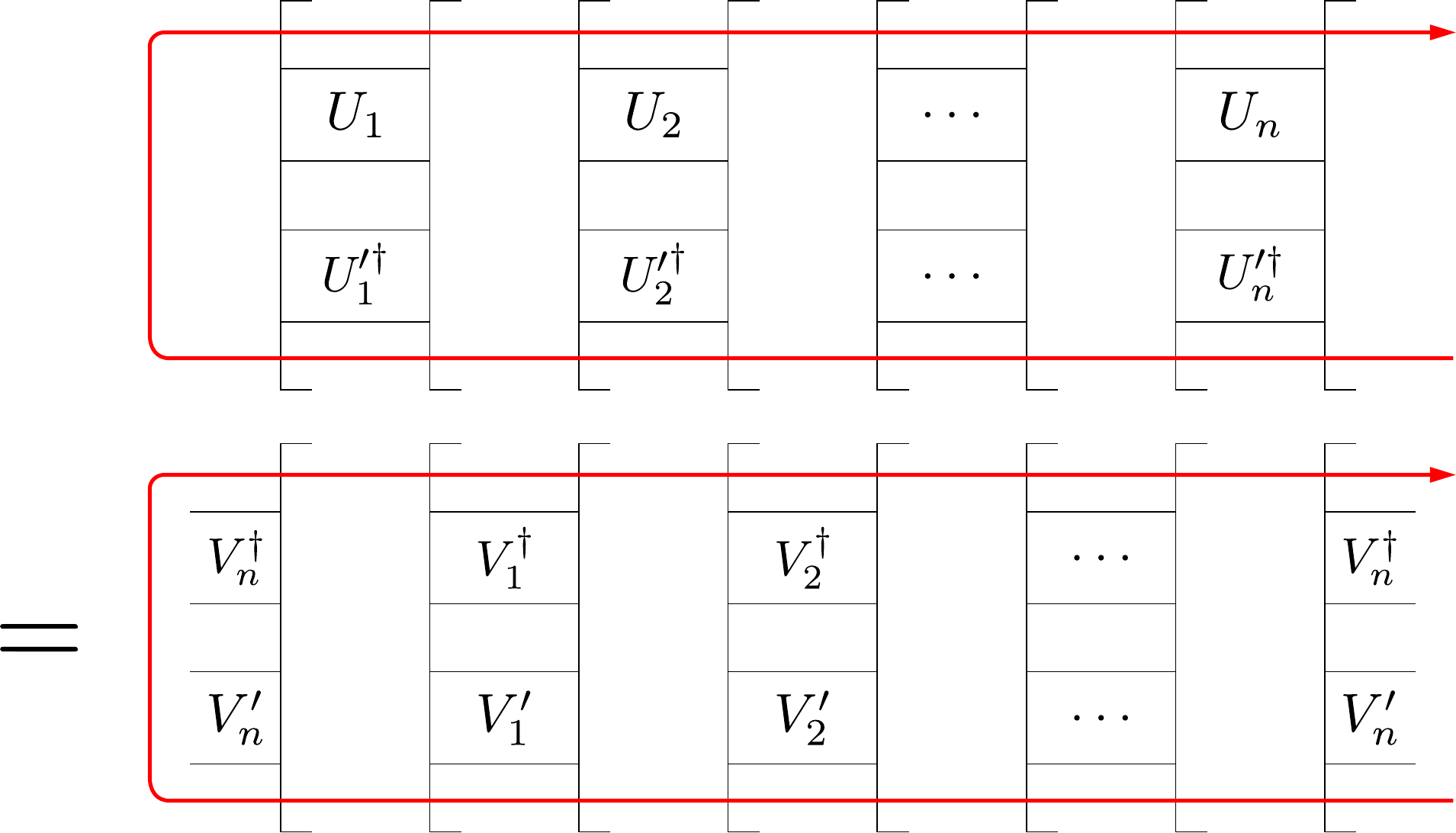}
\end{center}
Since all involved operators $U_j {U_j'}^\dagger$, $V_j^\dagger V_j'$ have even parity, this equation implies that each block of $U$'s and $V$'s must subdivide into tensor products of tensors acting on the corresponding sites.
\begin{center}
	\includegraphics[width=0.6\columnwidth]{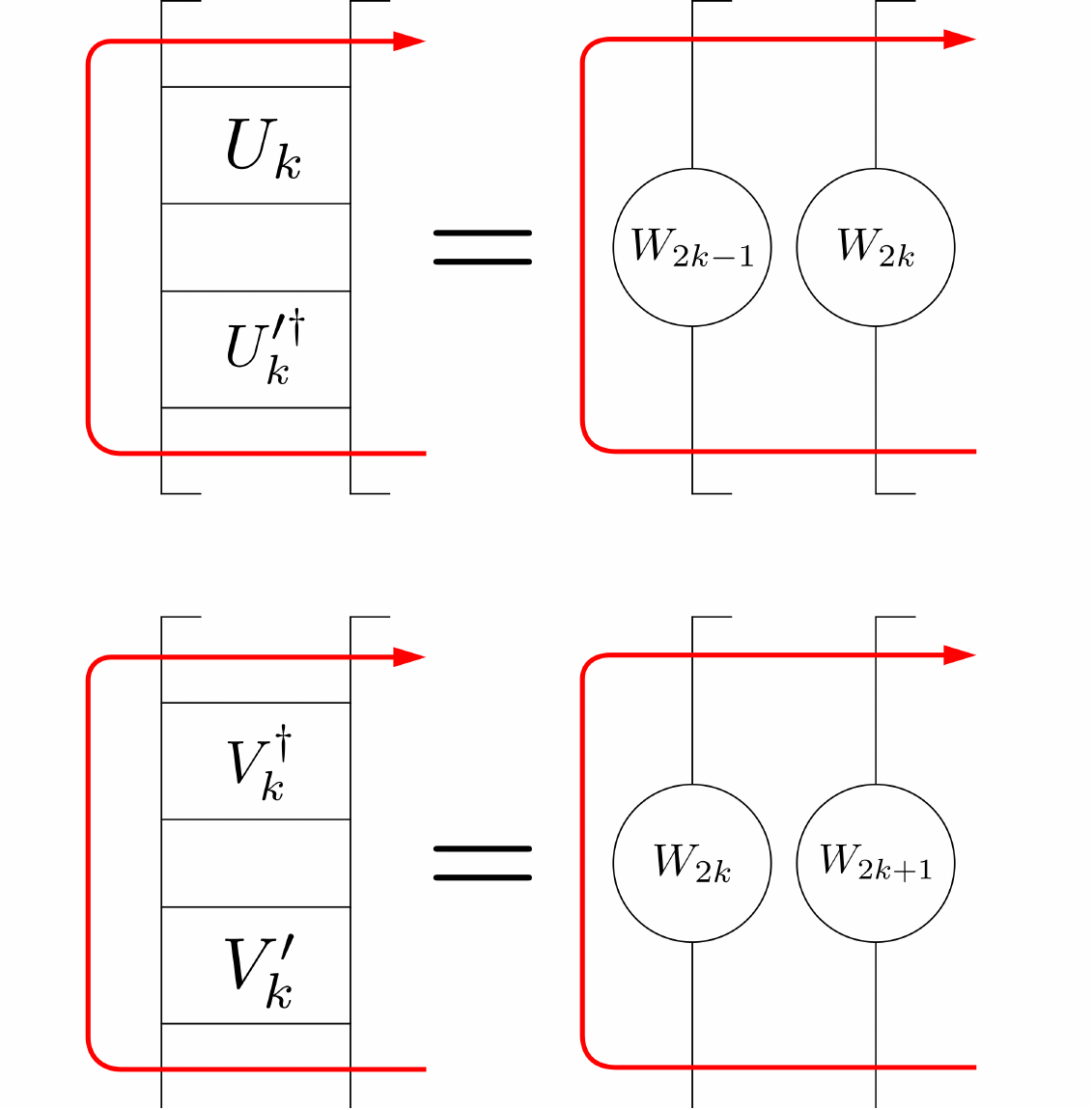}
\end{center}
After removing the fermionic bra's and ket's, we recover Eqs.~\eqref{eq:gauge1} and \eqref{eq:gauge2}.
This concludes the demonstration of two key steps in the fermionic classification. Using this diagrammatic approach, the derivations in Sec.~\ref{sec:clbos} can be readily repeated for fermionic systems. We thus arrive at Eqs.~\eqref{eq:cohomology1} and~\eqref{eq:cohomology2} with the symmetry group elements $g$ taken from the extended group $G' = G \times \mathbb{Z}_2$. This results in a classification by the elements of the second cohomology group of $G'$. If some of the symmetry operations are anti-unitary, the classification is given by a generalization of the second cohomology group of $G'$ allowing for complex conjugations, similarly to the result of Sec.~\ref{sec:anti-unitary}. However, we miss a $\mathbb{Z}_2$ topological index related to the fractionalization of fermionic parity symmetry compared to the classification of fermionic SPT ground states~\cite{pollmann2011ferm,Bultnick2017}. The origin of the failure of our method to capture this index will become clear in the following two Sections. 

\subsection{An example: Time-reversal symmetry} \label{fermtrs}

Here we discuss how the special case of the classification of fermionic FMBL systems with time-reversal symmetry arises from the above classification. The topological index $\kappa$ can be derived in an analogous way as in the spin case in Sec.~\ref{sec:TRS}, where $\kappa$ is related to $\beta^{z,z}$ via $\beta^{z,z} = \kappa \pi$ and can take values of $0$ and $1$. In the following, we will demonstrate the existence of a second topological index $\mu = 0, 1$. We define the symmetry group corresponding to parity conservation as $\mathbb{Z}_2 = \{e,p \}$ with $p^2 = e$. 
 Observe that $\Theta_z = \sigma_z^{\otimes N}$. This follows because each two-gate $u$ or $v$ preserves parity as in Eq.~\eqref{paritycon}, so from Eq.~\eqref{eq:theta1}, we have \footnote{We take a consistent a fermionic ordering such that no additional signs arise.} $\Theta_p = \tilde{U}^\dagger \sigma_z^{\otimes N} \tilde{U} = \tilde{U}^\dagger \tilde{U} \sigma_z^{\otimes N}  = \sigma_z^{\otimes N}$. Furthermore, if we cast $\Theta_p$ as the fermionic analogue of Eq.~\eqref{eq:theta_quantum_circuit}, we can set $\Theta^p_{i} = \mathbb{1}$ for odd $i$, while for even $i$, we have 
\begin{center}
	\includegraphics[width=0.9\columnwidth]{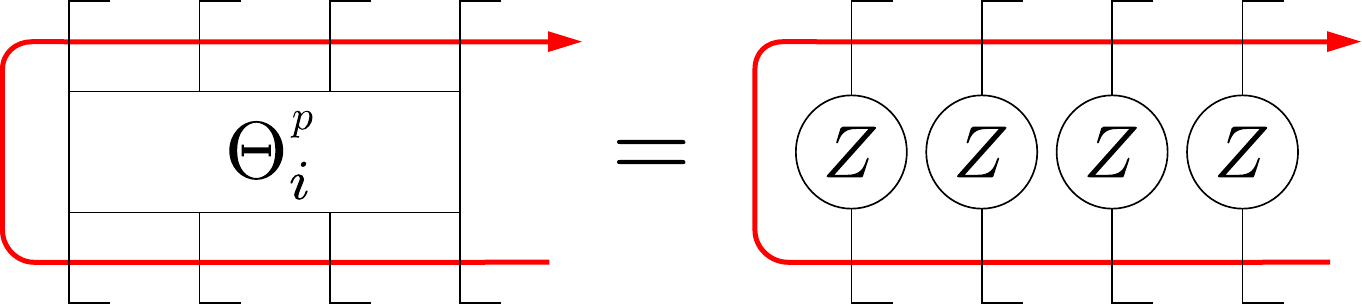}
\end{center}
Substitute the above into the fermionic analogue of Eq.~\eqref{eq:combined_trafo}, 
\begin{center}
	\includegraphics[width=0.9\columnwidth]{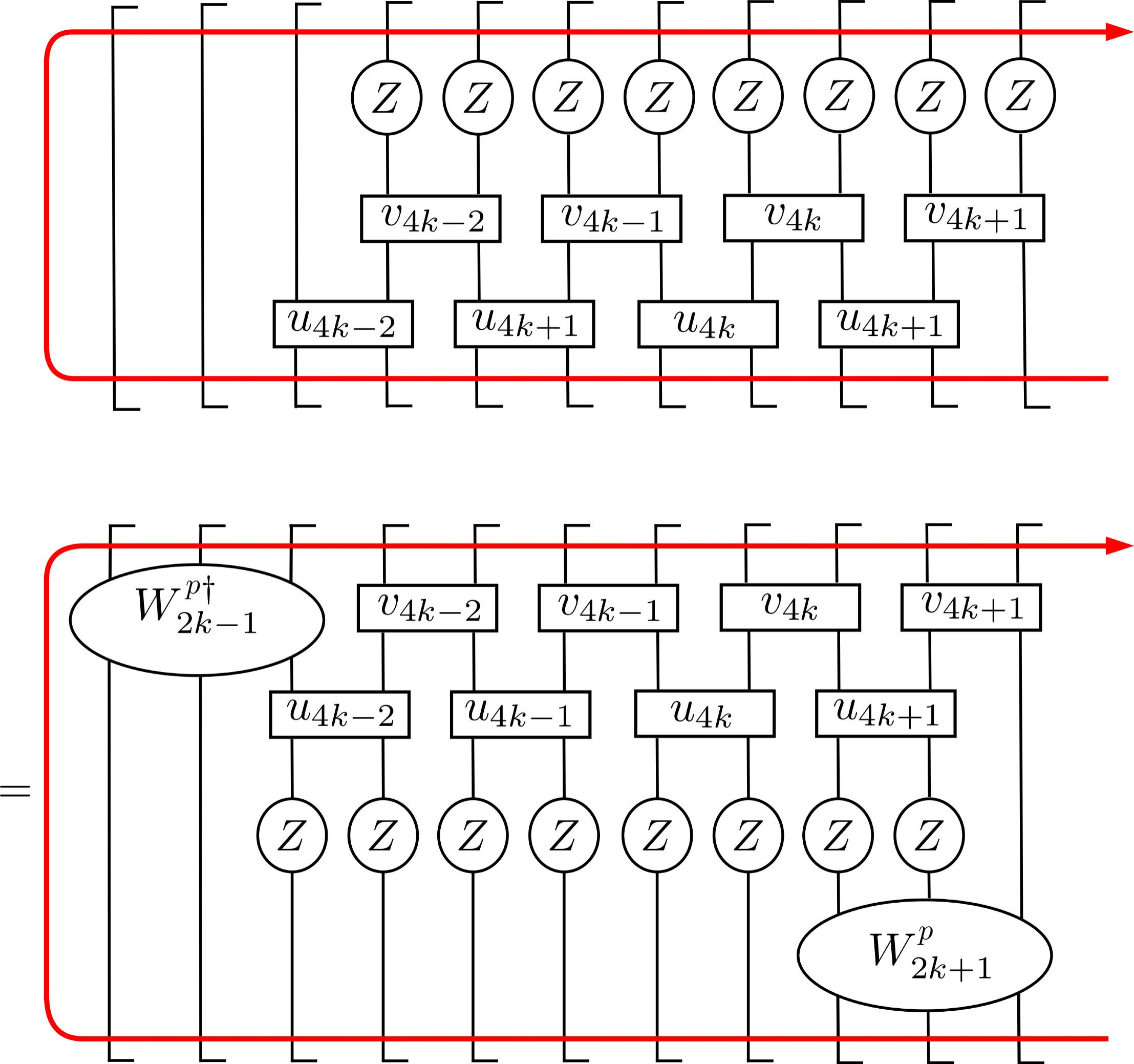}
\end{center}
We then see that the following equation holds for all $k$,
\begin{equation}
\begin{aligned}	
\includegraphics[width=\columnwidth]{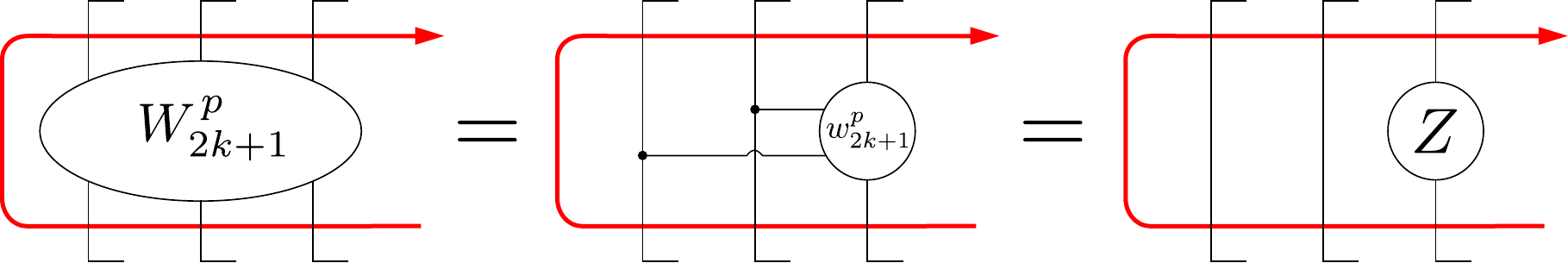}
\end{aligned} . \label{eq:paritywz}
\end{equation}
We note in particular that $w^p_{2k+1}$ is real. Now we consider the fermionic analogue of Eq.~\eqref{eq:gh_trafo} obtained by consecutive actions of the time reversal operator $\mathpzc{v}^{\otimes N}$ (combined with complex conjugation) and parity operator $\sigma_z^{\otimes N}$. Since these operators have to commute, we obtain
\begin{center}
	\includegraphics[width=\columnwidth]{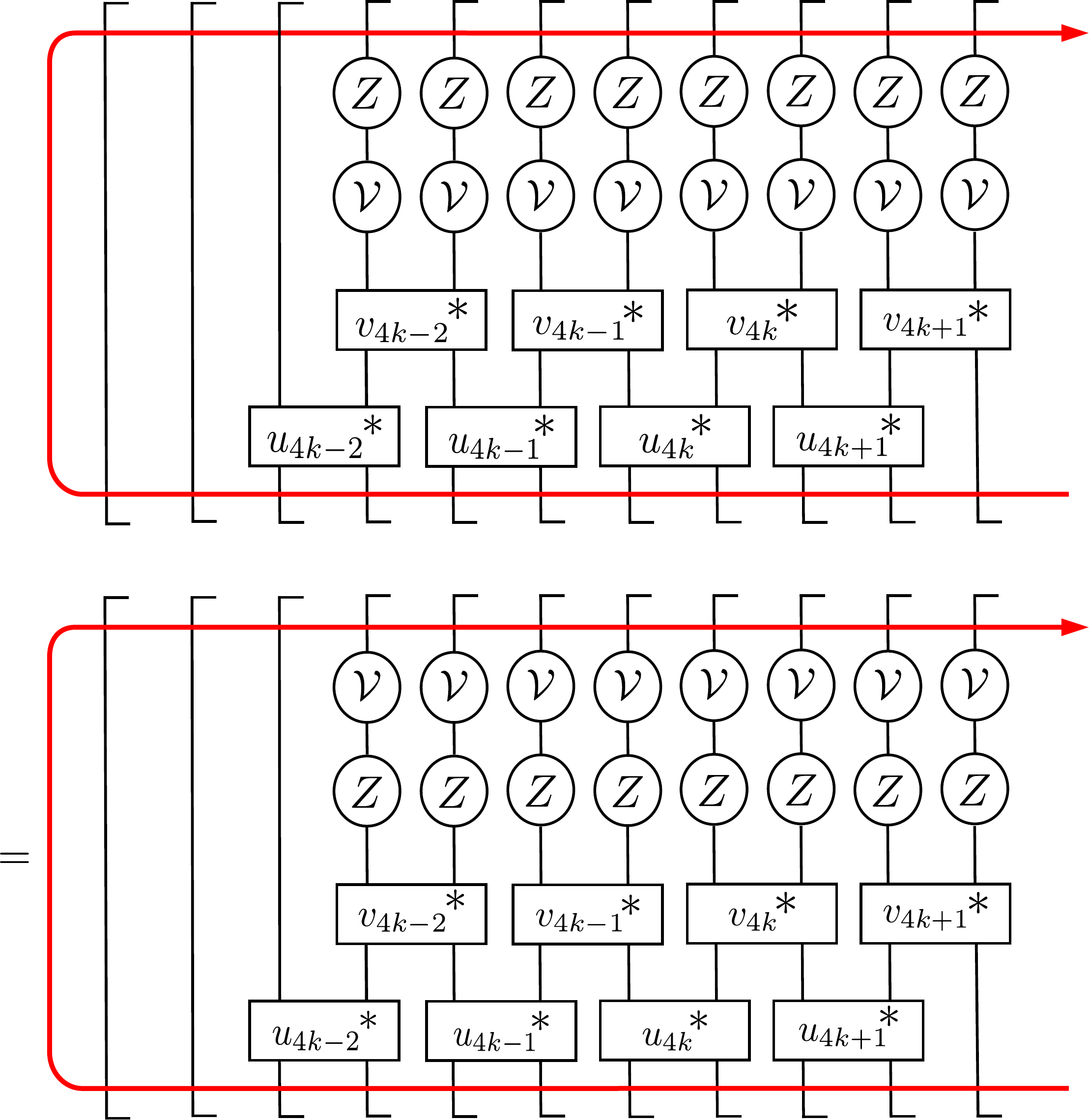}
\end{center}
i.e., analogously to Eqs.~\eqref{eq:gh_trafo} and \eqref{eq:antigh_trafo}
\begin{equation}
\begin{aligned}	
\includegraphics[width=\columnwidth]{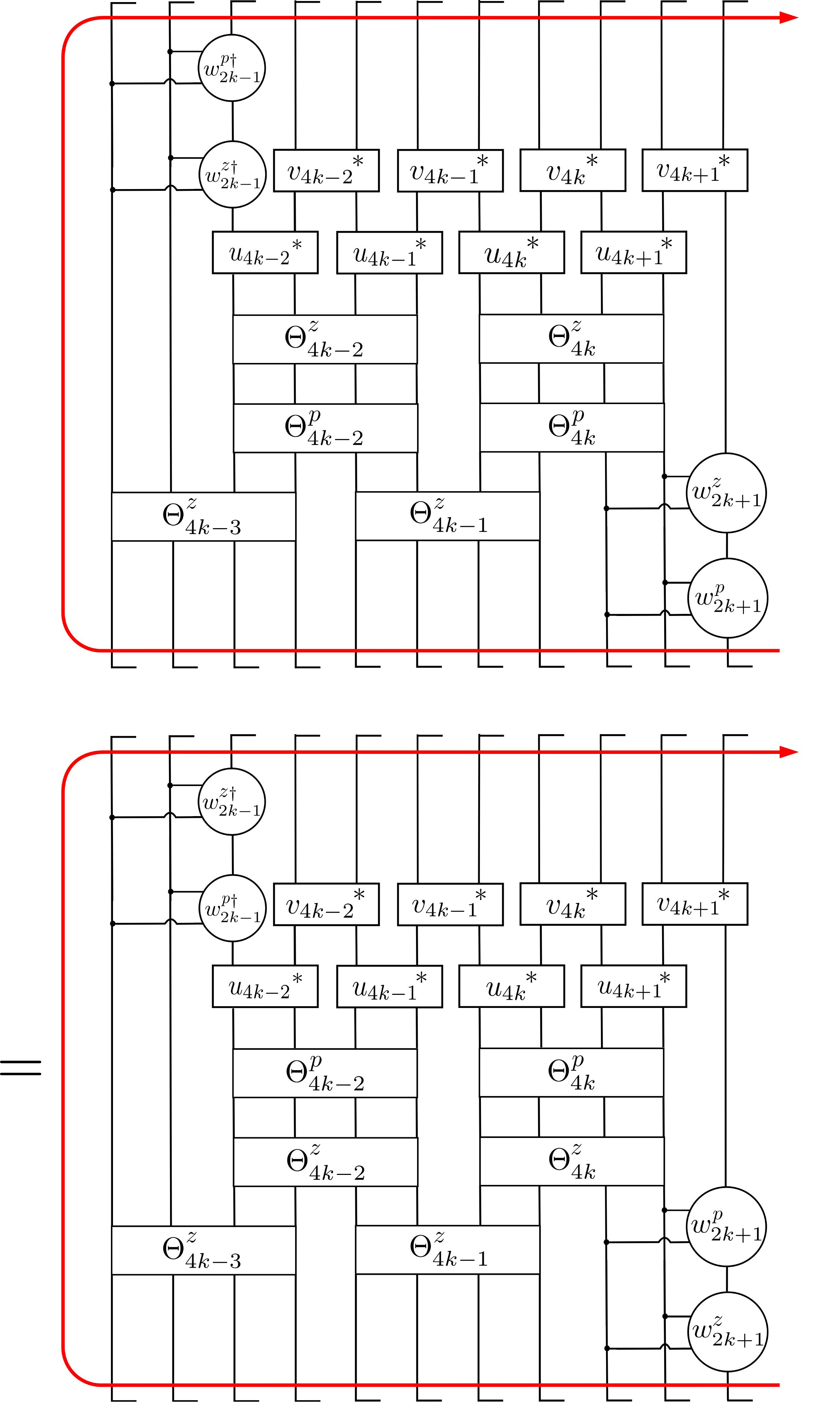}
\end{aligned}. \label{eq:paritystep2}
\end{equation}
 Note that since $w_{2k+1}^p$ is real, the additional complex conjugation appearing in Eqs.~\eqref{eq:anticohomology1} and \eqref{eq:anticohomology2} due to anti-unitarity can be neglected. 
Hence, we arrive the following equation
\begin{equation}
\begin{aligned}	
\includegraphics[width=1\columnwidth]{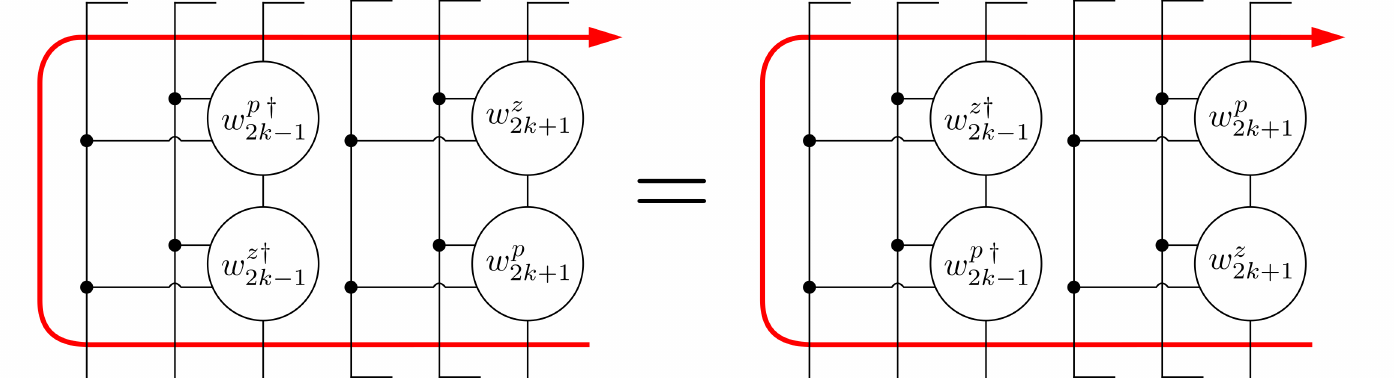}
\end{aligned} \label{eq:paritycom}
\end{equation}
Similarly to  Eq.~\eqref{eq:fixlegs}, we fix the indices corresponding to the first two legs from the left as $L_1$ and $L_2$, and to the fourth and fifth legs as $L_4$ and $L_5$. We thus have for each $k$  
\begin{align}	
\nonumber
[w^{z}_{2k-1}& ]^\dagger_{L_1, L_2}   [w^{p }_{2k-1}]^\dagger_{L_1, L_2} =
\\ 
& [w^{p }_{2k-1}]^\dagger_{L_1, L_2} [w^{z }_{2k-1}] ]^\dagger_{L_1, L_2} e^{-i \pi \mu_{k,L_1,L_2, L_4, L_5}} 
\label{eq:trspar1}
\\
\nonumber
[w^{p}_{2k+1}& ]_{L_4, L_5}[w^{z}_{2k+1}]_{L_4, L_5} =
\\
& [w^{z}_{2k+1}]_{L_4, L_5} [w^{p}_{2k+1}]_{L_4, L_5} e^{+i \pi \mu_{k,L_1,L_2, L_4, L_5}} 
\label{eq:trspar2}
\end{align} 
where $\mu_{k ,L_1,L_2, L_4, L_5} = 0,1$. 
 Comparing both equations, we see that $\mu$ is independent of $L_1, L_2, L_4, L_5$, i.e. $\mu_{k ,L_1,L_2, L_4, L_5} = \mu_k$. 
Now, if we compare Eq.~\eqref{eq:trspar2} for $k$ and Eq.~\eqref{eq:trspar1} for $k+1$, we conclude that 
	$\mu_k = \mu_{k+1} $,
	i.e. $\mu_k = \mu$ does not depend on $k$.

 Using $w_{2k+1}^z w_{2k+1}^{z *} = (-1)^\kappa \mathbb{1}$, Eq.~\eqref{eq:trspar2} yields $w_{2k+1}^p w_{2k+1}^{z \top} = w_{2k+1}^{z \top} w_{2k+1}^p e^{i \pi \mu}$. Due to Eq.~\eqref{eq:paritywz}, this implies $w_{2k+1}^{p \dg} w_{2k+1}^{z \dg} = w_{2k+1}^{z \dg} w_{2k+1}^{p \dg} e^{-i \pi \mu}$. Comparison with Eq.~\eqref{eq:trspar2} gives $1 = e^{-2i \pi \mu}$. Hence, $\mu = 0, 1$ is the second topological index. In other words, 
\begin{align}	
&
w^{z}_{i} w^{p}_{i} = (-1)^\mu w^{p}_{i} w^{z}_{i}.
\end{align}  
Since the  index $\mu$ is the same for all $k$ and for all $l$-bit indices, this index is a topological index shared by all eigenstates. 



\subsection{Missing topological index}	\label{sec:kit}
	\begin{figure}[htb]
		\includegraphics[width=0.4\textwidth]{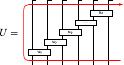}
		\caption{A quantum  circuit with local quantum gates in a ladder geometry that diagonalizes the Kitaev chain at $\mu=0, t= \Delta$.}
		\label{fig:ladder}
	\end{figure}
	
We illustrate the nature of the missing topological index with the example of the Kitaev chain~\cite{Kitaev_2001}. For open boundary conditions it is governed by the Hamiltonian 	
	\begin{equation}\label{eq:kit}
	H = - \mu \sum_{i=1}^N n_i 
	- \sum_{i=1}^{N-1} \big(t c_i^\dagger c_{i+1} 
	+ \Delta c_i c_{i+1}
	+ \mr{h.c.}\big) 
	\;,
	\end{equation}
	where $\mu$ is the chemical potential, $c_i$ the fermion annihilation operator at site $i$, $n_i= c^\dagger_i c_i$ the particle number operator at site $i$, $t$ the hopping parameter, $\Delta$ the superconducting gap, and $\mr{h.c.}$ denotes the Hermitian conjugate. For $|\mu| <2t$ and $t=\Delta$, this model is known to have decoupled Majorana edge modes (with exponential tails extended into the bulk), and is in a topologically non-trivial phase, which is only protected by fermionic parity conservation and characterized by a two-fold degenerate ground state for open boundary conditions and $N \rightarrow \infty$. For $\mu=0, t=\Delta$, the Hamiltonian of Eq.~\eqref{eq:kit} can be diagonalized exactly by a quantum  circuit  $U = U_{L-1,L} \otimes \ldots \otimes U_{2,3} \otimes U_{1,2}$ with local gates in a ladder geometry (Fig.~\ref{fig:ladder}), where $U_{i,i+1}= \mathbb{1}_{2^{i-1} \times 2^{i-1}} \otimes u_0 \otimes \mathbb{1}_{2^{N-i-1} \times 2^{N-i-1}}$ acts non-trivially only on sites $i$ and $i+1$, where  
	\begin{equation}
	u_0 =
	\frac{1}{\sqrt{2}}
	\begin{pmatrix}
	0 & -1 &-1&0
	\\
	-1& 0 &0 & 1
	\\
	-1& 0 &0 & -1
	\\
	0 & -1 &1&0
	\end{pmatrix} \; .
	\end{equation}
The same quantum circuit also diagonalizes the Anderson localized case of spatially fluctuating $t_i = \Delta_i$ (with $\mu_i = 0$), where all eigenstates have two-fold degenerate single-particle entanglement spectra~\cite{2015Slagle}). 	
	This example illustrates the expectation that the topological phase associated with fermionic parity fractionalization can only be exactly represented by a long-range parity-preserving quantum circuit, and consequently, the two-layer quantum circuit approach is unable to detect the corresponding topological index.

Lastly, we note that $\mathbb{Z}_2$ spontaneous symmetry breaking phases in spin chains are Jordan-Wigner dual to the fermionic topologically non-trivial phase just described. One may think that the classification of fermionic SPT MBL phases can be obtained from the spin case using the Jordan-Wigner transformation. As our method is not able to classify spontaneous symmetry-breaking phases (but still applies in the presence of spontaneous symmetry breaking), we cannot take advantage of the  Jordan-Wigner transformation.


\section{Robustness to perturbations}\label{sec:robustness}


In Ref.~\onlinecite{Thorsten} it was pointed out that if the Hamiltonian $H(\lambda)$ is changed adiabatically such that $H(0)$ corresponds to the original Hamiltonian and $H(1)$ to the final one, one can always define a unitary $U_\mr{cont}(\lambda)$ which changes continuously as a function of $\lambda$ and diagonalizes the Hamiltonian for all $\lambda \in [0,1]$. We assume that the Hamiltonian stays FMBL along the path and does not break the symmetry. Hence, there exists a quantum circuit $\tilde U(\lambda)$ which efficiently diagonalizes the Hamiltonian for all $\lambda \in [0,1]$. 
However, even within small approximation error~\cite{Thorsten} $U_\mr{cont}(\lambda)$ might not be the same unitary (at least for some $\lambda$) as the one given by the quantum circuit: For almost all $\lambda$ (those without degeneracies for finite $N$), the unitary $U_\mr{cont}(\lambda)$ is related to the quantum circuit by a permutation matrix $P(\lambda)$ whose non-vanishing matrix elements may have phases,
\begin{align}
\tilde U(\lambda) = U_\mr{cont}(\lambda) P(\lambda) \label{eq:cont_permutation}
\end{align}
up to an error that vanishes in the thermodynamic limit~\cite{Thorsten}. We want to use this property to show that the topological index of $\tilde U(\lambda)$ for $\lambda_1$ and $\lambda_2 = \lambda_1 + \epsilon$ is the same in the limit $\epsilon \rightarrow 0$. First, note that Eq.~\eqref{eq:cont_permutation} implies up to the above error that
\begin{align} 
\tilde U(\lambda_1) P^\dagger(\lambda_1) = U_\mr{cont}(\lambda_1) U_\mr{cont}^\dagger(\lambda_2) \tilde U(\lambda_2) P^\dagger(\lambda_2).
\end{align}
The product $U_\mr{cont}(\lambda_1) U_\mr{cont}^\dagger(\lambda_2)$ can be brought arbitrarily close to $\mathbb{1}$ by taking $\epsilon$ sufficiently small. Hence, we have up to small error
\begin{align}
\tilde U(\lambda_1)  P^\dagger (\lambda_1) P(\lambda_2) = \tilde U(\lambda_2) .  \label{eq:permutation}
\end{align}
The eigenstates encoded in $ \tilde U(\lambda_1)$ and $ \tilde U(\lambda_2)$ are thus up to phase factors the same just relabeled. Since the topological index of all eigenstates is the same (and determines the overall topological index derived above), the unitaries $\tilde U(\lambda_1)$ and $\tilde U(\lambda_2)$ have the same overall topological index. (Note the element of the second cohomology group of $[w_{2k+1}^g]_{L_4, L_5}$ can be determined from a single eigenstate using for instance  Eq.~\eqref{eq:gh_trafo}.) Therefore, the topological index of the SPT MBL phase cannot change along the adiabatic evolution in the thermodynamic limit unless the symmetry or FMBL condition is broken. 

For a rigorous treatment of error bounds, follow the approach of Ref.~\onlinecite{Thorsten}.


\section{Completeness of classification for individual eigenstates}\label{sec:completeness}

Here we demonstrate that the classification derived in Sec.~\ref{sec:clbos} is complete in the sense that  there cannot be any additional topological indices which affect the properties of individual eigenstates (such as degeneracies in the entanglement spectra). This does not rule out the possibility that there are topological obstructions to connecting different Hamiltonians with the same topological index as defined above (i.e., that the overall unitary $U$ has additional topological indices). However, we show that if there are Hamiltonians disconnected by such a topological obstruction, their topological distinctness cannot be visible on their individual eigenstates. The main idea is that the topological indices derived above are the same as the ones for (non-translationally invariant) ground states of local gapped Hamiltonians~\cite{Pollmann2010,2011Schuch,2011Chen,Bultnick2017}. 

Concretely, we use the result of Ref.~\onlinecite{2011Chen} that for two states in the same SPT MBL phase, there must exist a finite time evolution by a local Hamiltonian $H_\mr{loc}(t)$ preserving the symmetry, which transforms the two states into each other. That is, the unitary
\begin{align}
U_\mr{loc} = \mathcal{P}\left( e^{-i \int_0^1 \mr{d} t H_\mr{loc}(t)}\right) \label{eq:unitary_evolution}
\end{align}
applied on one state gives the other. ($\mathcal{P}$ denotes path ordering of the integral.) Suppose there was at least one  topological SPT MBL index that has been missed so far, i.e., different SPT MBL phases $A$ and $B$ with the same eigenstate topological index as determined above, but which are separated from each other by an FMBL-breaking transition. Since the topological indices found above are complete when restricting to only one eigenstate, any eigenstate from phase $A$ can be connected to an arbitrary eigenstate from phase $B$ via a unitary transformation of the type~\eqref{eq:unitary_evolution}. 
Let us consider a Hamiltonian which continuously implements that
\begin{align}
H(\lambda) = \mathcal{P}\left( e^{i \int_0^\lambda \mr{d} t H_\mr{loc}(t)}\right) H_A \mathcal{P}\left( e^{-i \int_0^\lambda \mr{d} t H_\mr{loc}(t)}\right). \label{eq:Ham_evolution}
\end{align}
For $\lambda = 1$ it shares at least one eigenstate with Hamiltonian $H_B$, even though it is not in phase $B$ itself. Consequently, a single eigenstate cannot be employed to distinguish the two phases $A$ and $B$. 
Note that along the path FMBL is preserved, as 
there exist exponentially localized operators 
\begin{align}
\tau_z^i(\lambda) = \mathcal{P}\left( e^{i \int_0^\lambda \mr{d} t H_\mr{loc}(t)}\right) U_A \sigma_z^i U_A^\dg \mathcal{P} \left( e^{-i \int_0^\lambda \mr{d} t H_\mr{loc}(t)}\right)
\end{align}
for all $\lambda \in [0,1]$, i.e., $H(\lambda)$ is in phase $A$ for all $\lambda \in [0,1]$. 

\section{Conclusions}\label{sec:conclusions}

We used two-layer quantum circuits with long gates in order to classify spinful and fermionic one-dimensional MBL phases with an (anti-)unitary on-site symmetry. For spin systems with unitary symmetries, we demonstrated that all eigenstates correspond to the same element of the second cohomology group. For anti-unitary on-site symmetries, a similar classification is obtained in terms of a generalization of the second cohomology group. This leads to a $\mathbb{Z}_2$ classification for time-reversal invariant systems~\cite{Thorsten}. 
Hence, spinful MBL phases in one dimension are characterized by a topological index which is the same for all eigenstates.  
We showed that all those SPT MBL phases are stable with respect to arbitrary symmetry-preserving local perturbations.  As a result, the four-fold degeneracy of the entanglement spectra of the eigenstates of the disordered cluster model are protected by both $\mathbb{Z}_2 \times \mathbb{Z}_2$ symmetry and time-reversal symmetry.  
Furthermore, we demonstrated that the classification is complete in terms of eigenstate topological indices, i.e., while there might be topological obstructions to connecting FMBL Hamiltonians with the same topological index as identified above, their topological distinctness cannot be visible on individual eigenstates. Note that we only classified symmetry-protected topological MBL systems. We did not classify local orders, but our classification also applies in the presence of spontaneous symmetry breaking.  

For fermionic systems,
	we extended the above classification by proposing a fermionic tensor network diagrammatic formulation. We obtained a classification given by the (generalized) second cohomology group of the overall (anti-)unitary symmetry group $G' = G \times \mathbb{Z}_2$.  
This classification lacks the topological label corresponding to the fractionalization of fermionic parity as found for ground state fermionic SPT phases~\cite{pollmann2011ferm}. 
We explicitly derived two out of three $\mathbb{Z}_2$ topological invariants (shared by all eigenstates)  associated with the $\mathbb{Z}_8$ classification of one-dimensional fermionic ground states with time-reversal symmetry~\cite{kitaevferm2010, kitaevferm2011, pollmann2011ferm}. 
Note that our demonstration of the robustness of the found topological indices to symmetry-preserving perturbations also applies to the fermionic case. 

	Our results give rise to important directions for future research: One is the possibility of the mentioned topological obstructions, which would correspond to a topological index that is defined only for the diagonalizing unitary $U$ as a whole, but cannot be defined for individual eigenstates.

Finally, the approach presented here can be extended to two dimensions~\cite{2DMBL}. While MBL might not strictly exist in two dimensions~\cite{deRoeck2017Stability,Altman2018stability}, the relaxation times are likely so long that strongly disordered systems in two dimensions can be viewed as MBL for all experimental and technological purposes. Hence, topological properties such as the protection of quantum information against local noise would be present on all practically relevant time scales. Our procedure  enables the classification of such SPT MBL-like phases in two dimensions. However, the extension of our classification to topologically ordered MBL phases, which do not allow for an exact representation by short-depth quantum circuits, is not obvious. This case would be particularly interesting, as it would include topological MBL phases allowing for fault-tolerant quantum computations at finite energy density.

\section*{Acknowledgments} 

TBW is grateful to Frank Pollmann for pointing out an error in a previous manuscript. We would also like to thank Christoph S{\"u}nderhauf, Andrea De Luca, Norbert Schuch, David P{\'e}rez-Garc{\'i}a and Frank Verstraete for helpful discussions.  TBW was supported by the European Commission under the Marie Curie Programme. AC was supported by the EPSRC Grant No. EP/N01930X/1. The contents of this article reflect only the authors' views and not the views of the European Commission.

\appendix

\section{Instability of FMBL with a non-abelian symmetry}\label{app:A}

We provide an argument using our formalism that FMBL is inconsistent with symmetry-preserving eigenstates for non-abelian symmetry groups. Note that even in the case of a logarithmic violation of the area law of entanglement (as suggested in Ref.~\onlinecite{Protopopov2019}), our formalism is expected to still apply: $\ell \propto N^\beta$ ($0 < \beta < 1$) as assumed below is sufficient to represent states whose entanglement may grow like $N^\beta$. 
 As a specific example, we take a spin-1/2 chain with $G =$ SU(2), but the line of reasoning is similar for other non-abelian symmetry groups. In that case, the Hamiltonian $H$ fulfills
\begin{align}
[H, S_\alpha] = [H, \mb S^2] = [\mb S^2, S_\alpha] = 0 \label{eq:spin_commutation}
\end{align}
with $\mb S$ the overall spin of the system and $\alpha = x,y,z$. Thus, there exists a complete set of exact eigenstates $|\psi_n\rangle$ which are also eigenstates of $\mb S^2$ and $S_z$, $H |\psi_n\rangle = E_n |\psi_n\rangle$. After defining the lowering and raising operators $S_\pm = S_x \pm i S_y$, Eq.~\eqref{eq:spin_commutation} implies that $S_\pm |\psi_n \rangle$ is a different eigenstate with the same energy $E_n$ unless $S_\pm |\psi_n \rangle = 0$. If we restrict ourselves to the subset of eigenstates $|\psi_n\rangle$ with maximal overall spin $S = \frac{N}{2}$, it follows that $S_+^N$ applied on all of them must at least in one instance create the eigenstate $|\uparrow \uparrow \ldots \uparrow\rangle$. Applying $S_-^N$ on $|\uparrow \uparrow \ldots \uparrow\rangle$, we see that $|\downarrow \downarrow \ldots \downarrow \rangle$ is an eigenstate with the same energy. Thus, there are symmetry-protected degeneracies, i.e., $\Theta_g$ no longer has to be diagonal, cf. Eq.~\eqref{eq:insert_diagonalisation}. If without loss of generality we assume $U |00 \ldots 0\rangle = |\downarrow \downarrow \ldots \downarrow \rangle$ and $U |11 \ldots 1\rangle = |\uparrow \uparrow \ldots \uparrow \rangle$, we obtain for $\mathpzc{v}_g = \sigma_x$
\begin{align}
\Theta_{x, 1 1 \ldots 1}^{0 0 \ldots 0} &= \langle 0 0 \ldots 0 | U^\dg \sigma_x^{\otimes N} U | 1 1 \ldots 1 \rangle \notag \\
&= \langle \downarrow \downarrow \ldots \downarrow | \sigma_x^{\otimes N} | \uparrow \uparrow \ldots \uparrow \rangle = 1
\end{align} 
The approximation of the actual value $\Theta_{x, 1 1 \ldots 1}^{0 0 \ldots 0}$ by a quantum circuit with gate length $\ell$ has an error upper bounded by $2 \delta(N) = 8 \sqrt{\frac{\sqrt{2}c}{3} N} e^{-\frac{\ell}{2 c' N^{1-\mu}}}$, cf. Eqs.~(32) and~(38) in Ref.~\onlinecite{Thorsten} with $c, c' >0$, $0 < \mu < 1$. Hence, we have
\begin{equation}
\begin{aligned}
\includegraphics[width=0.45\textwidth]{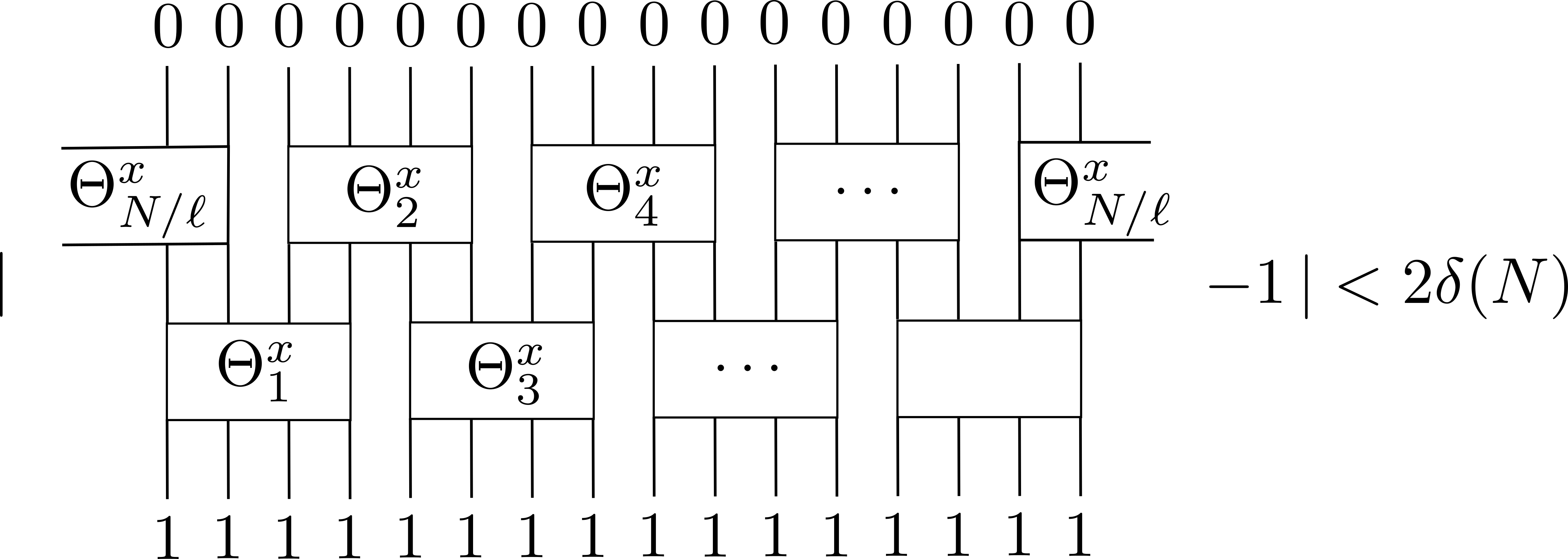}.
 \end{aligned} \label{eq:Theta_element}  
\end{equation}
Note that the individual $\Theta_j^x$ no longer have to be diagonal, but nevertheless $\Theta_x$ can be written as a two-layer quantum circuit with unitaries of range $2 \ell$, as can be seen from blocking unitaries in triples in Eq.~\eqref{eq:theta_elements}. Eq.~\eqref{eq:Theta_element} implies
\begin{equation}
\begin{aligned}
\includegraphics[width=0.3\textwidth]{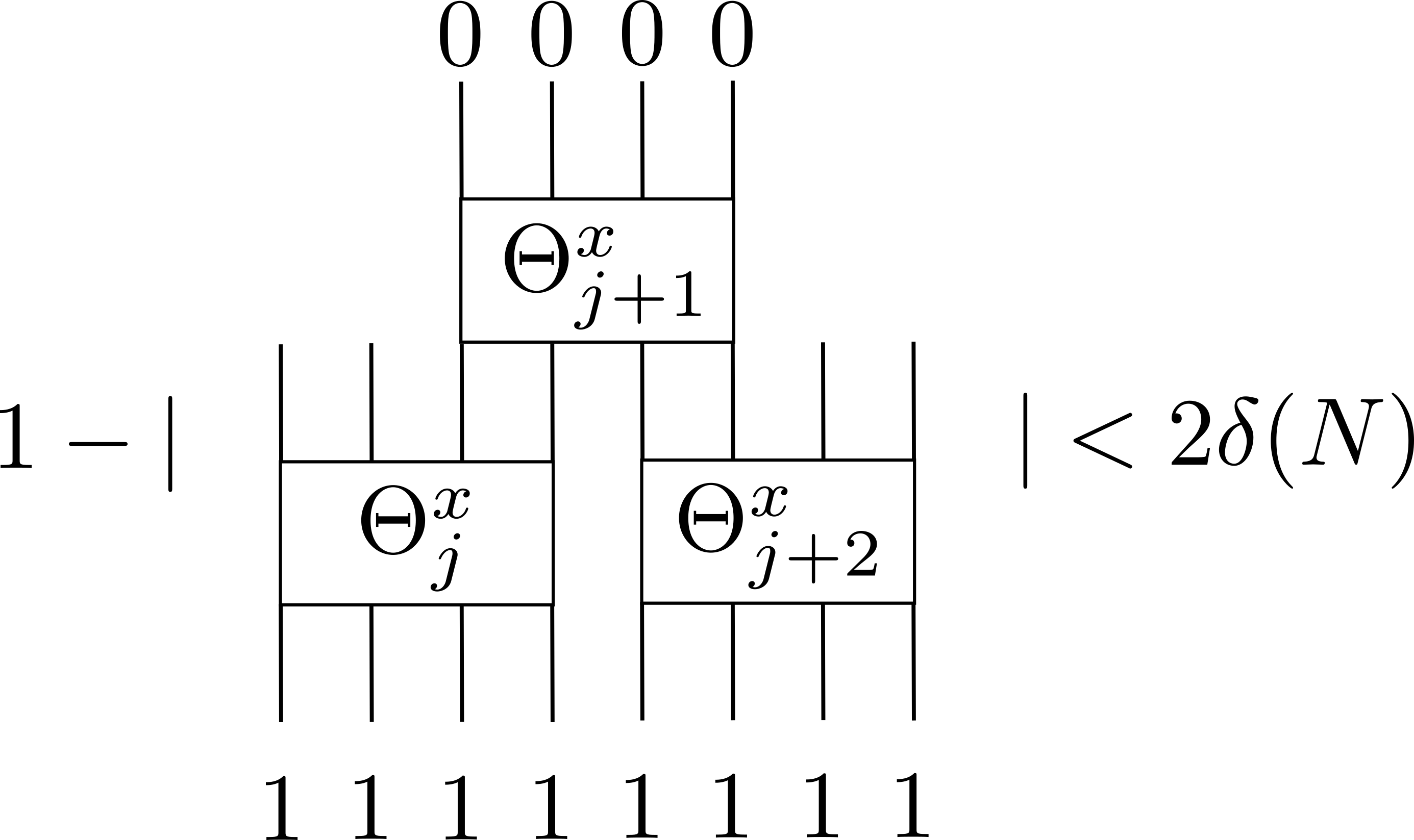}
 \end{aligned} , \label{eq:Theta01}  
\end{equation}
and since Eq.~\eqref{eq:Theta_element} also has to hold after swapping 0's and 1's, we also have
\begin{equation}
\begin{aligned}
\includegraphics[width=0.3\textwidth]{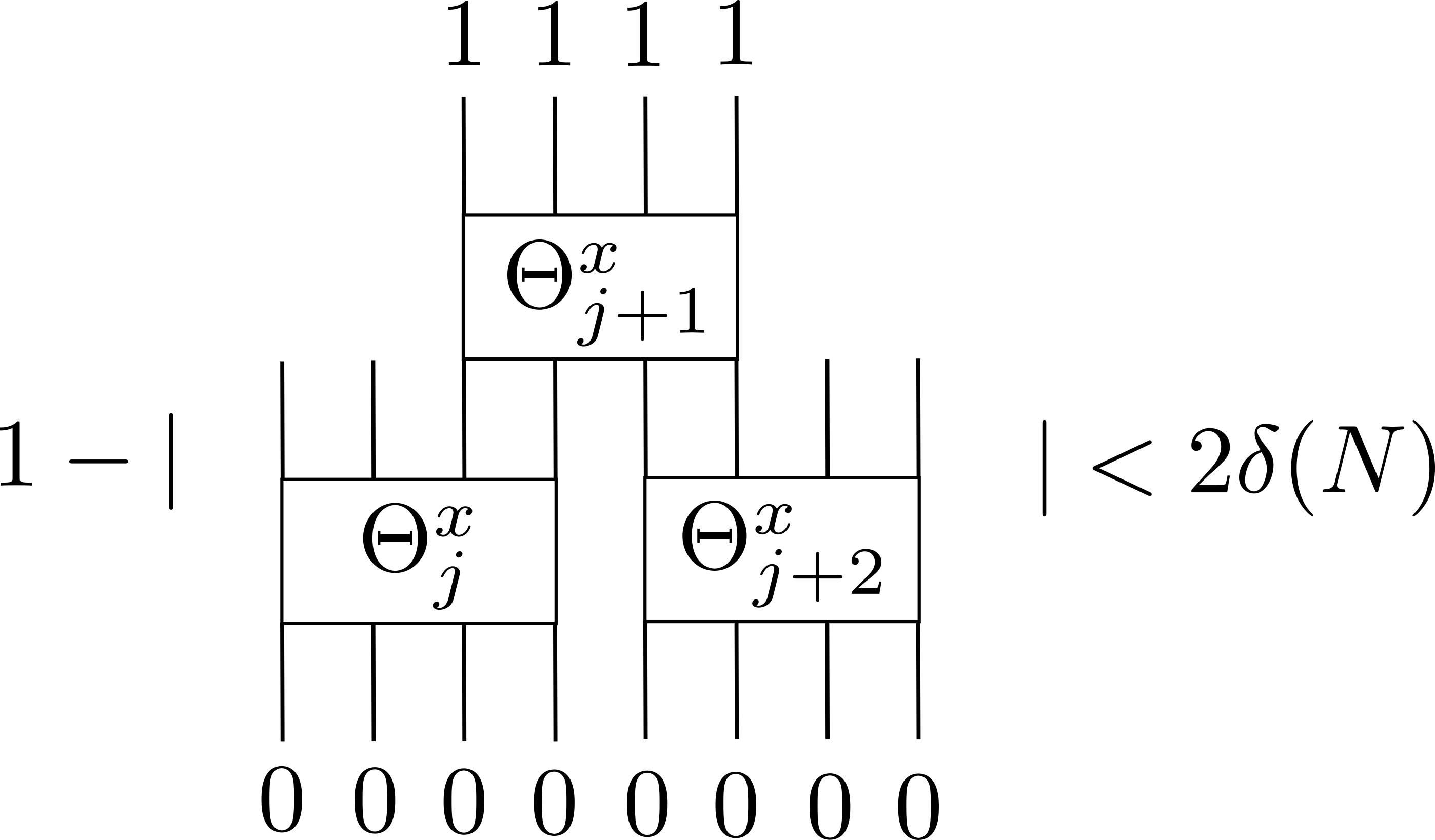}
 \end{aligned}. \label{eq:Theta10}  
\end{equation}
Note that the last two diagrams denote vectors, since there are open legs pointing upwards. 
Using Eqs.~\eqref{eq:Theta01} and~\eqref{eq:Theta10} repeatedly in an arbitrary order yields
\begin{equation}
\begin{aligned}
\includegraphics[width=0.48\textwidth]{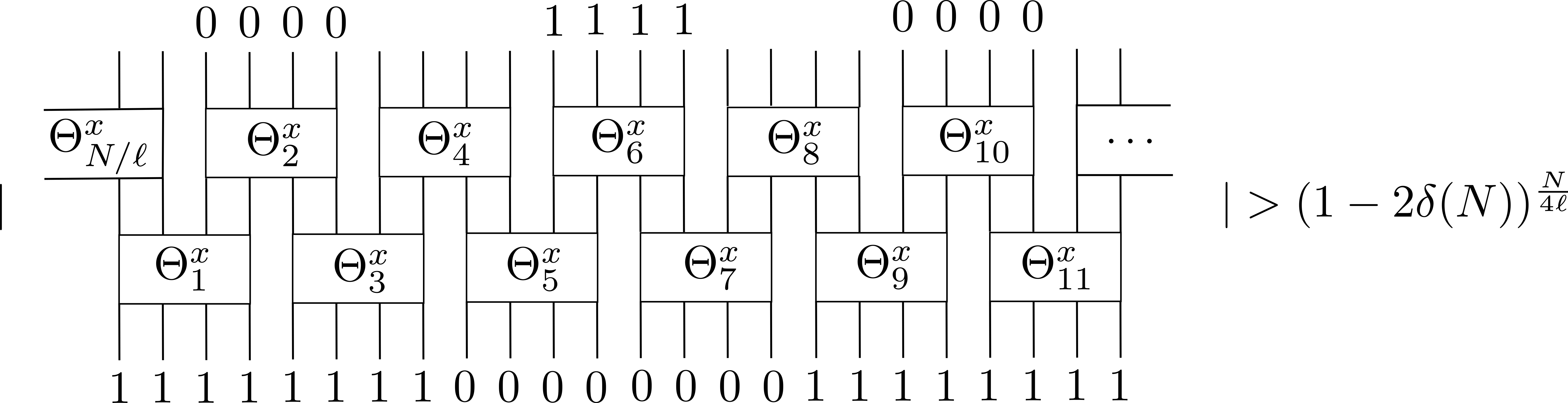}
 \end{aligned} \label{eq:Theta_sequence}  .
\end{equation}
There are $2^\frac{N}{4\ell}$ possible orderings in which Eqs.~\eqref{eq:Theta01} and~\eqref{eq:Theta10} can be combined. If we choose $\ell = b N^\beta$ with $b > 0$ and $1 - \mu < \beta \leq 1$,  the right hand side of Eq.~\eqref{eq:Theta_sequence} becomes 1 in the limit $N \rightarrow \infty$. Specifically for $\beta < 1$,  the above number of combinations $2^{\frac{N^{1-\beta}}{4b}}$ grows subexponentially with $N$. Since each of them corresponds to an off-diagonal element of $\Theta_x$, the number of exact energy degeneracies grows also at least subexponentially with $N$ leading to a proliferation of resonances and thus breakdown of FMBL in the thermodynamic limit.

\newpage

\bibliography{biblioMBL}{}

\end{document}